\documentclass[fleqn,usenatbib]{mnras}

\usepackage{newtxtext,newtxmath}

\usepackage{color}
\usepackage{ulem}

\usepackage[T1]{fontenc}

\usepackage{subfigure}
\usepackage{enumerate}

\DeclareRobustCommand{\VAN}[3]{#2}
\let\VANthebibliography\thebibliography
\def\thebibliography{\DeclareRobustCommand{\VAN}[3]{##3}\VANthebibliography}

\usepackage{graphicx}	
\usepackage{amsmath}	

\title[Local photoionization in simulations of Milky Way-sized galaxies]{The effect of local photoionization on the galaxy properties and the circumgalactic medium in simulations of Milky Way-sized galaxies}

\author[B.~Zhu and V.~Springel]{Bocheng Zhu$^{1,2}$ and Volker Springel$^{2}$\thanks{\href{mailto:vspringel@mpa-garching.mpg.de}{vspringel@mpa-garching.mpg.de}}
\vspace*{0.1cm}\\%
$^{1}$Key Laboratory for Research in Galaxies and Cosmology, Shanghai Astronomical Observatory, Chinese Academy of Sciences, \\ 80 Nandan Road, Shanghai 200030, People's Republic of China\\%
$^{2}$Max-Planck-Institut für Astrophysik, Karl-Schwarzschild-Straße 1, 85741 Garching, Germany}

\date{Accepted XXX. Received YYY; in original form ZZZ}

\pubyear{2024}

\begin{document}
\label{firstpage}
\pagerange{\pageref{firstpage}--\pageref{lastpage}}
\maketitle

\begin{abstract}
In this study, we investigate the impact of local stellar radiation in cosmological zoom simulations of the formation of Milky Way-sized galaxies. We include the radiation field as an additional feedback component that is computed alongside gravity with a tree code in an optically thin approximation. We resimulate the initial conditions of five Milk Way-like systems taken from the Auriga project with and without stellar radiation, and study the effects of local stellar radiation on several properties of the galaxies and the circumgalactic medium (CGM). Similar to previous findings, we observe with our current model that local stellar radiation can modify gas cooling in the CGM and thus suppress star formation and the surface densities of young stars and HI gas, while having little impact on the total gas content. In particular, it also suppresses the peak of the rotation curve and reduces the mass of the stellar bulge. In the CGM region, the young stellar radiation exceeds the external UVB and dominates the radiation field within the virial halo at all redshifts. Nevertheless, we find that the local stellar radiation, as implemented in the current study, has overall little impact on the radial density and temperature profile of the CGM gas.  However, for the ion species HI and MgII the column densities within $\sim 0.3\,R_{\rm vir}$ are reduced, while the OVI column density is hardly impacted by the radiation field due to a lack of soft X-ray components in our current model. Additional effects can be expected from the radiation of the central AGN during phases of quasar activity and from soft X-ray sources, which have not yet been included in the simulations of the present study.
\end{abstract}

\begin{keywords}
galaxies: formation -- methods: numerical -- galaxies: evolution -- hydrodynamics
\end{keywords}



\section{Introduction}

In the classical theory of galaxy formation,  gas follows the collapse of dark matter (DM) due to the pull of gravity and reaches virial equilibrium through shock-heating \citep{silk77, Rees77, white78, white91}. However, unlike DM, gas is also subject to radiative cooling. Upon dissipating its thermal support, it can further collapse into the center of a DM halo, eventually becoming dense enough for star formation, or becoming available to feed the growth of a  supermassive black hole (SMBH). On the other hand, SMBHs and stars will also inject energy due to the release of gravitational potential energy and as a result of  nuclear reactions, respectively. The released photons can heat and photoionize the gas, thereby in principal counteracting the cooling processes in the halos. 

The injected radiation energy plays a significant role in galaxy formation and evolution. The associated photons give rise to a background radiation field in the universe, an important part of which comprises the so-called ultraviolet background \citep[UVB, ][]{haardt96, faucher09, haardt12, faucher20}. In the early universe, the UVB is important for photoionizing the neutral hydrogen that fills the universe after recombination, converting the universe from a neutral state back to a highly ionized state. This transition is known as the epoch of reionization \citep[EoR, ][]{shapiro87,loeb01,haiman16}. Importantly, the UVB can also suppress the formation of  small structures through the heating associated with ionizing the intergalactic medium (IGM), making the gas  difficult to capture by low mass halos \citep{efstathiou92, cen92, babul92, haiman96, quinn96}. As a result, star formation in these small halos is suppressed, as the gas is no longer able to efficiently cool in them \citep{wang09, finlator11, borrow23}.

Current cosmological simulations  commonly implement the UVB as a spatially uniform, time-dependent radiation field \citep[e.g.][]{thoul96, quinn96, katz96, weinberg97, vogelsberger14, crain15, pillepich18}, with a time-dependent spectrum as modeled in several works \citep{haardt96, faucher09, haardt12, faucher20}. The photoionization rate and the radiative heating and cooling rates associated with the UVB are usually calculated with spectral synthesis codes like Cloudy \citep{ferland98}, assuming that the gas is in collisional and photionization equilibrium. The inclusion of an externally imposed homogeneous UVB in cosmological simulations makes them able to reproduce the average thermal history of the IGM, and helps to match the observed galaxy population by suppressing galaxy formation in very small dark matter halos \citep{cen93, benson02, okamoto08}. 

On the other hand, in cosmological simulations of Milky Way (MW)-like galaxies the UVB has been found to not significantly affect their evolution  \citep{thoul96, quinn96, weinberg97}. Since the virial temperature of MW-like galaxies is about $10^6~{\rm K}$, the effective UVB temperature is too low to significantly affect the gas in the halos of MW-like galaxies. However, at the center of such galaxies, the gas temperature is relatively low, suggesting that the gas may here be more prone to be affected. On the other hand, the gas density is so high there that the intensity of the UVB is likely not sufficient to significantly affect the high-density gas. 

But the UVB is in reality not a spatially uniform radiation field, rather it arises from the  collective light emitted by a discrete population of galaxies. It should therefore have a higher intensity when approaching the center of a DM halo, where the photon sources -- the stars and SMBHs  -- are located. Some  cosmological simulation projects try to account for this effect by considering the radiation emitted from each discrete SMBH via a $r^{-2}$-law  \citep{vogelsberger14, pillepich18, dave19}. These studies found that including such a local radiation contribution from SMBHs has non-negligible effects on the galaxy evolution. For example, \citet{dave19} reported that massive galaxies at $z=0$ become redder when accounting for the local radiation from SMBHs, which implies that this local radiation field can help quench massive galaxies.

There are also some theoretical works and simulation studies that considered the radiation feedback from stars. For example, \citet{hopkins12} performed several galaxy formation simulations with different halo masses, considering both the short-range and long-range radiation pressures as well as the photoionization heating from HII region by using cosmological zoom-in simulation. They found that radiation pressure plays a key role in generating galactic winds in massive galaxies. \citet{Roskar14} simulated  MW-sized galaxies with short-range radiation pressure on dust, and found that such radiative feedback can suppress star formation and the clumpiness of the cold gas in the disk. However, these works did not take the photoionization heating caused by the certain stellar populations into account. \citet{ceverino2014} have performed a cosmological zoom-in simulation of the progenitor of a MW-sized galaxy with radiation pressure, photoionization and photoheating calculated by a detailed stellar population description and found that radiation feedback is helpful for reducing and flattening the rotation curve. However, long-range effects of the local stellar radiation were not included in the study.

\citet{cantalupo10} investigated the impact of  X-ray radiation emitted by  hot winds from young stars, supernova remnants, and X-ray binaries on the gas cooling time. This differed from previous works which had only considered the radiation from young stars themselves. They found that this additionally included X-ray radiation  significantly affects the gas cooling time, and that the cooling time  increases with increasing star formation rate (SFR). This ``stars quench stars'' paradigm constitutes a form of preventive feedback, as the X-ray radiation prevents the gas from cooling and thus ceases star formation. 

Inspired by \citet{cantalupo10}, \citet{kannan14} studied the impacts of local stellar radiation on the evolution of MW-like galaxies, which is the first work to consider local stellar radiation in cosmological simulations. They separately considered the radiation of young and old stars, as well as the X-ray radiation from the hot gas components considered by \citet{cantalupo10}, then calculated the gas cooling and heating rate using Cloudy. They incorporated the local stellar radiation in the McMaster Unbiased Galaxy Simulations suite \citep[MUGS, ][]{stinson10}, which is a cosmological zoom-in simulation project performed with the {\small GASOLINE2} \citep{wadsley04} code. The radiation field was calculated by attaching the radiation information to the gravitational tree structure in  an optically thin approximation. \citet{kannan14} found that local stellar radiation can significantly affect the evolution of MW-like galaxies, suppressing star formation by about 30\% and lowering the central peak of the rotation curve. Based on these findings, they introduced the term local photoionization feedback (LPF) for the locally varying stellar radiation field.

\citet{obreja19} further improved the LPF implementation of \citet{kannan14} and applied it in the project Numerical Investigation of a Hundred Astrophysical Objects suite \citep[NIHAO, ][]{wang15}, which is another cosmological simulation project performed by {\small GASOLINE2}. \citet{obreja19} simulated a sub-$L^{*}$, an $L^{*}$, and a super-$L^{*}$ galaxies with and without the LPF, and confirmed the basic conclusions obtained by \citet{kannan14} for MW-like galaxies. Further, they found that the effects of the LPF become weaker in more massive galaxies, and that the LPF also affects the CGM properties. In independent work, \citet{hopkins18, hopkins20} implemented two different numerical schemes as part of the FIRE project to study the impact of stellar radiation feedback on galaxy evolution, both for dwarf- and MW-sized galaxies. Differently from the conclusions reached in \citet{obreja19}, they found that the local stellar radiation has no impact on the evolution of MW-like galaxies. 
 
Besides galaxy evolution, the local radiation field can potentially also affect the CGM properties \citep{stern16, Suresh17, Oppenheimer18, qu18, nelson18, fielding2020, strawn21}. However, most current CGM simulation studies consider a uniform UVB when studying the ionization state of the CGM, and there is so far comparatively little work including the local radiation. Notably, \citet{Suresh17} compare the OVI column density with and without the galaxies' central stellar radiation and found that it helps to increase the OVI column density in the inner region. \citet{Oppenheimer18} consider the flickering AGN radiation and non-equilibrium ionization of the CGM gas in EAGLE zoom simulations, finding that the AGN radiation can enhance the OVI column density allowing the COS-halos observations \citep{tumlinson11, werk14} to be matched. Finally, \citet{nelson18} studied the OVI column density in the TNG-100 simulation with several variations of the feedback prescriptions. They found that the OVI column density becomes lower when turning off the BH radiative feedback compared to a simulation with the fiducial TNG model, for halo masses in the range $10^{11.5}<M_{\rm halo}/{\rm M_{\odot}}<10^{12.5}$.

As mentioned above, there is no consensus in the field about how strongly local stellar radiation can affect the evolution of MW-like galaxies, and the impact of the local stellar radiation on the CGM is also not well understood. To further investigate these two questions, we here revisit simulations of MW-like galaxies that include local stellar radiation. We use the successful model of the Auriga project \citep{grand17} as the basis for our simulations, and implement the local stellar radiation with a scheme technically similar to the LPF treatment introduced by \citet{kannan14}. 

This paper is organized as follows. In Section~\ref{Sec:Methods}, we briefly introduce the Auriga project, describe the spectral energy distribution (SED) of the local stellar sources we used, and the numerical implementation of the local photoionization feedback (LPF) in the code. In Section~\ref{Sec:GalProperties}, we compare the simulated Auriga galaxies with and without LPF to study the impacts of LPF on the evolution, the gas as well as stellar disk properties, and the rotation curve of the simulated galaxies. In Section~\ref{Sec:CGM}, we investigate the impacts of LPF on the CGM gas properties, including the radial profile, the gas phase distribution, and the properties of some selected species. In Section~\ref{sec:discussion}, we compare our results with previous studies, discuss the potential effects of some radiation sources that are not considered in the present work, and test the non-linear coupling of the interaction between AGN feedback and local stellar radiation. Finally, we summarize our findings and conclude in Section~\ref{sec:conclusions}.

\section{Methodology}  \label{Sec:Methods}

\subsection{Simulations}\label{method:auriga}

This work is an extension of the Auriga project \citep{grand17}, which consists of a large suite of high-resolution magnetohydrodynamical simulations of the formation of MW-like galaxies, and whose data was recently been made publicly available \citep{Grand2024}. The computations utilize the `zoom-in' technique to follow galaxies within a full cosmological context. The project has been carried out using the {\small AREPO} code \citep{springel10}, which is a gravitational N-body and moving-mesh magnetohydrodynamical code. Due to its quasi-Lagrangian nature, the moving-mesh approach can provide an automatically adaptive resolution and manifest Galilean invariance at the discretized level of the fluid equations. These properties are particularly useful for the often highly supersonic flows occurring in galaxy formation. 

The galaxy formation physics model we employ in this work is based on that of IllustrisTNG \citep{weinberger17, pillepich18}. This in turn employs a star formation and stellar feedback model which is based on the two-phase ISM and non-local stellar wind subgrid prescription described in \citet{springel03}. The gas in the simulation is eligible for star-formation gas when its density is higher than $\sim 0.1\, {\rm cm^3}$ and the temperature is equal or less than $\sim 10^4~{\rm K}$. This specific density threshold value is calculated in the SN feedback-regulated two-phase ISM model as a function of the overall gas consumption timescale, with the latter matched to observations. When a gas cell becomes star-forming, its thermal state is assumed to follow an effective equation of state, and the star formation rate (SFR) is taken to be proportional to the dynamical time of the gas. As a central feedback channel, stellar winds  from the star-forming region are invoked. Since the multi-phase structure of the ISM is not spatially resolved, the numerical implementation of this wind feedback is realized in terms of temporary wind particles that are decoupled from the gas until they leave the star-forming region, thereby mediating a non-local kinetic feedback just outside the star forming region. For a more detailed description of the star formation, stellar wind, and stellar yield subgrid model and its parameters, we refer to \citet{pillepich18}.

Black holes (BH) are modelled as sink particles in the centers of galaxies. They can accrete gas with a rate estimated with the Bondi-Hoyle formula using the local gas properties at the BH's position. We employ the two-mode BH feedback model introduced in the TNG galaxy formation model \citep{weinberger17}, which includes both a thermal and a kinetic mode. When the black hole accretion rate (BHAR) is higher than a critical value (i.e.~for high Eddington ratio), the BH feedback is assumed to be in a purely thermal form (`quasar mode'), whereas for low accretion rate it occurs in a kinetic form (sometimes called `radio mode'). The critical BHAR where the transition occurs is scaled with the BH mass itself, making it easier for massive BHs to transition into the kinetic mode. This two-mode AGN feedback model can quite successfully produce a galaxy stellar mass function and a bimodality in the galaxy color distribution at $z = 0$ consistent with observations.

The Auriga project selected 30 halos with mass $M_{\rm halo}=1-2\times 10^{12}\,{\rm M_{\odot}}$ from the dark matter only simulation of the EAGLE project \citep{crain15}, and resimulated them with the the zoom-in technique keeping the original box length of 100 cMpc. The adapted cosmological parameters for the simulations were taken from the \citet{Planck2014} analysis.

In this work, we select a subset of five typical halos from the Auriga galaxy set to study the impacts of the local stellar radiation on the galaxy and CGM properties. The selected halos correspond to  Au3, Au5, Au6, Au15, Au16 from Auriga, as presented in \citet{grand17}. However, note that we do not employ the original Auriga galaxy formation model, but rather we simulate the halos with a fiducial, slightly adjusted TNG galaxy formation model in which we omit  AGN radiative feedback. In addition, we resimulate them again by accounting for a local stellar radiation field as described below. For reference, the primary galaxy properties of these five halos at $z=0$, with and without the local stellar radiation, are given in Table~\ref{table:galaxies}.  

\begin{table*}
 \caption{The properties of the simulated galaxies at $z=0$. The columns are (1) model name; (2) halo mass defined by the FoF algorithm; (3) halo virial radius; (4) total stellar mass; (5) BH mass; (6) half-mass radius of young stars; (7) half-mass radius of total stars; (8) gas fraction within the halo.}
 \label{tab:galaxies}
 \begin{tabular}{lrcccrcc}
  \hline
 Run & $\frac{M_{\rm halo}}{10^{10}~{\rm M_{\odot}}}$ & $\frac{R_{200}}{\rm kpc}$ & $\frac{M_{\star}}{\rm 10^{10}~M_{\odot}}$ & $\frac{M_{\rm BH}}{\rm 10^{7}~M_{\odot}}$ & $\frac{R_{\rm young~star,~1/2}}{\rm kpc}$ & $\frac{R_{\rm total~star,~1/2}}{\rm kpc}$ & $\frac{f_{\rm gas}}{\%}$\\
 \hline
 Au3NoRad & 135.77 & 185.69 &  5.96 & 4.78 & 10.06 & 2.48 & 4.40\\
 Au3StarRad & 132.54 & 184.25 & 4.49 & 9.13 & 14.71 & 1.60 & 3.53\\
 Au6NoRad & 98.79 & 167.06 & 3.14 & 3.41 & 13.43 & 4.53 & 6.78\\
 Au6StarRad & 97.50 & 166.33 & 2.48 & 2.78 & 7.75 & 4.01 & 7.10\\
 Au16NoRad & 133.70 & 184.79 & 5.16 & 2.71 & 14.61 & 5.98 & 7.21\\
 Au16StarRad & 136.29 & 185.97 & 4.19 & 4.41 & 18.50 & 6.15 & 8.51\\
 Au5NoRad & 110.63 & 173.48 & 4.10 & 3.57 & 2.65 & 1.50 & 4.16\\
 Au5StarRad & 111.54 & 173.96 & 3.17 & 4.68 & 2.84 & 1.37 & 5.67\\
 Au15NoRad & 93.18 & 163.84 & 2.21 & 2.07 & 6.86 & 1.60 & 7.20\\
 Au15StarRad & 91.97 & 163.12 & 1.87 & 2.73 & 4.12 & 1.79 & 6.48\\
 \hline
 \end{tabular}
 \label{table:galaxies}
\end{table*}

\subsection{Radiation and photoionization from stellar populations}\label{radiationfield}

In this section, we specify the photoionization feedback model that we introduced as a physics extension in our simulations. 

In general, the photoionizing radiation in the universe comes from AGN and stars. For a single galaxy, the radiation sources can be further divided into local radiation sources and radiation from the other galaxies. The cumulative radiation of AGN and stars from other galaxies results in the well-known cosmic UV background (UVB). This UVB radiation has been extensively studied \citep{faucher09,haardt12,faucher20} and has been incorporated into many cosmological simulations. Here, we follow the previous work but use the updated UVB spectrum produced by \citet{faucher20}. 

For representing local radiation sources, one should aim to include both AGN and stellar radiation. A simple model for AGN radiation has been incorporated in the Auriga simulations following \citet{vogelsberger14}. However, our technical implementation of stellar radiation, as described in the following section, is not readily compatible with this AGN radiation module. More importantly, since the aim of this paper is to study the impact of local stellar radiation, we prefer to consider this first in isolation for clarity. We thus turn off the AGN radiation in the following and focus on the stellar radiation alone, deferring an investigation of the joint impact of both AGN and stellar radiation to future work.

\subsubsection{Stellar radiation}\label{method:starrad}

Our setup for the radiation treatment from stars follows the approach of \citet{kannan14} but with several modifications. In the following, we concisely summarize our approach and refer readers to \citet{kannan14} for additional details.

The stellar radiation is divided into two parts -- radiation from star forming regions and from old stellar populations. For radiation from ongoing star formation, \citet{kannan14} consider the young stellar population and the X-ray emission from SNe remnants. \citet{kannan16} further account for the X-ray emission from high-mass X-ray binaries (HMXB), whose companions are O or B type stars. We combine these three kinds of radiation in our model in a single channel. For the radiation from the young stellar component, following again \citet{kannan14}, we do not consider the time evolution of the SED. Instead, the SED of the young stellar radiation is fixed to the SED of a 10 Myr old stellar population for simplicity. We use the Binary Population and Spectral Synthesis v2.3 \citep[BPASS, ][]{eldridge17} model to generate the spectrum of a stellar population at 10 Myr for a given constant star formation rate. 

The young stellar radiation will be partly absorbed by hydrogen and dust when it leaves the star-forming region. Here, we follow the approach of \citet{kannan14}, which is similar to \citet{cantalupo10}, and describe the frequency-dependent escape fraction $f^{\nu}_{\rm esc}$ as
\begin{equation}
f^\nu_{\rm esc} = f^{\rm LL}_{\rm esc}+(1-f^{\rm LL}_{\rm esc})\,{\rm e}^{-\tau_{\nu}}.
\end{equation}
The escape fraction at the Lyman limit frequency $f^{\rm LL}_{\rm esc}$ is set to 5\%. Here, only absorption of the neutral hydrogen is taken into account,  thus $\tau_{\nu}=\sigma_{\nu} N{\rm (H^{0}})$. $\sigma_{\nu}$ is the cross-section of neutral hydrogen and $N{\rm (H^{0}})$ is the column density of neutral hydrogen in the star-forming region. The value of $N{\rm (H^{0}})$ is set to $10^{20}~{\rm cm^{-2}}$, as in \citet{kannan14}.

For the X-ray emission from SNe remnants, we follow the description presented in \citet{cervino02}. They reported that the total energy released in the {\small EINSTEIN} band ($0.1 - 3.4 \, {\rm keV}$) of an individual SNe is $\sim1.07\times 10^{40} {\rm erg}$. Based on this value, we calculate the SNe event rate from a young stellar population with a \citet{chabrier03} IMF and constant $1~{\rm M_{\odot}~yr^{-1}}$ SFR and normalize the SED accordingly. The shape of the SED is as follow:
\begin{equation}
    f_{\nu} = 0.65 f_{\rm RS}(0.76 {\rm keV}) + 0.175 f_{\rm RS}(0.23 {\rm keV}) + 0.175 f_{\rm RS}(1.29 {\rm keV})
\end{equation}
where $f_{\rm RS}(kT)$ is the Raymond-Smith model \citep{raymond1977}.

The HMXB emission in this work is assumed to follow a $\Gamma=2$ power law. Based on \citet{anderson13}, the total SED of HMXB is in the range from 0.5-8.0 keV band, and the scaling relation of SFR and total X-ray luminosity is:
\begin{equation}
L_{\rm HMXB}\approx 1.4\times 10^{39} \frac{\rm SFR}{\rm M_{\odot}~yr^{-1}}~{\rm erg~s^{-1}}.
\end{equation}
For the radiation from the evolved stellar population, both radiation from old stars and low-mass X-ray binaries (LMXB) is included. Similar to the radiation from young stellar populations, we do not consider the time evolution of the radiation of old stellar populations, for simplicity. Instead we fix the SED of an old stellar population at an age of 2 Gyr under the simple stellar population (SSP) assumption, i.e., the stars are assumed to have formed at the same time. We also use the BPASS v2.3 model and assume that the escape fraction of the old stellar radiation is unity. 

The energy of the LMXBs comes from black holes or neutron stars accreting material from their low-mass companion stars. We assume that the shape of the SED from LMXBs is a $\Gamma=1.5$ power law. Following the description in \citet{anderson13}, the SED is in the range from 0.3 keV to 8 keV, and the LMXB luminosity as a function of total stellar mass scales as
\begin{equation}
L_{\rm LMXB} = 10^{40}\frac{M_{\star}}{10^{11}{\rm M_{\odot}}}~{\rm erg~s^{-1}}.
\end{equation}

\begin{figure}
   \includegraphics[width=0.45\textwidth]{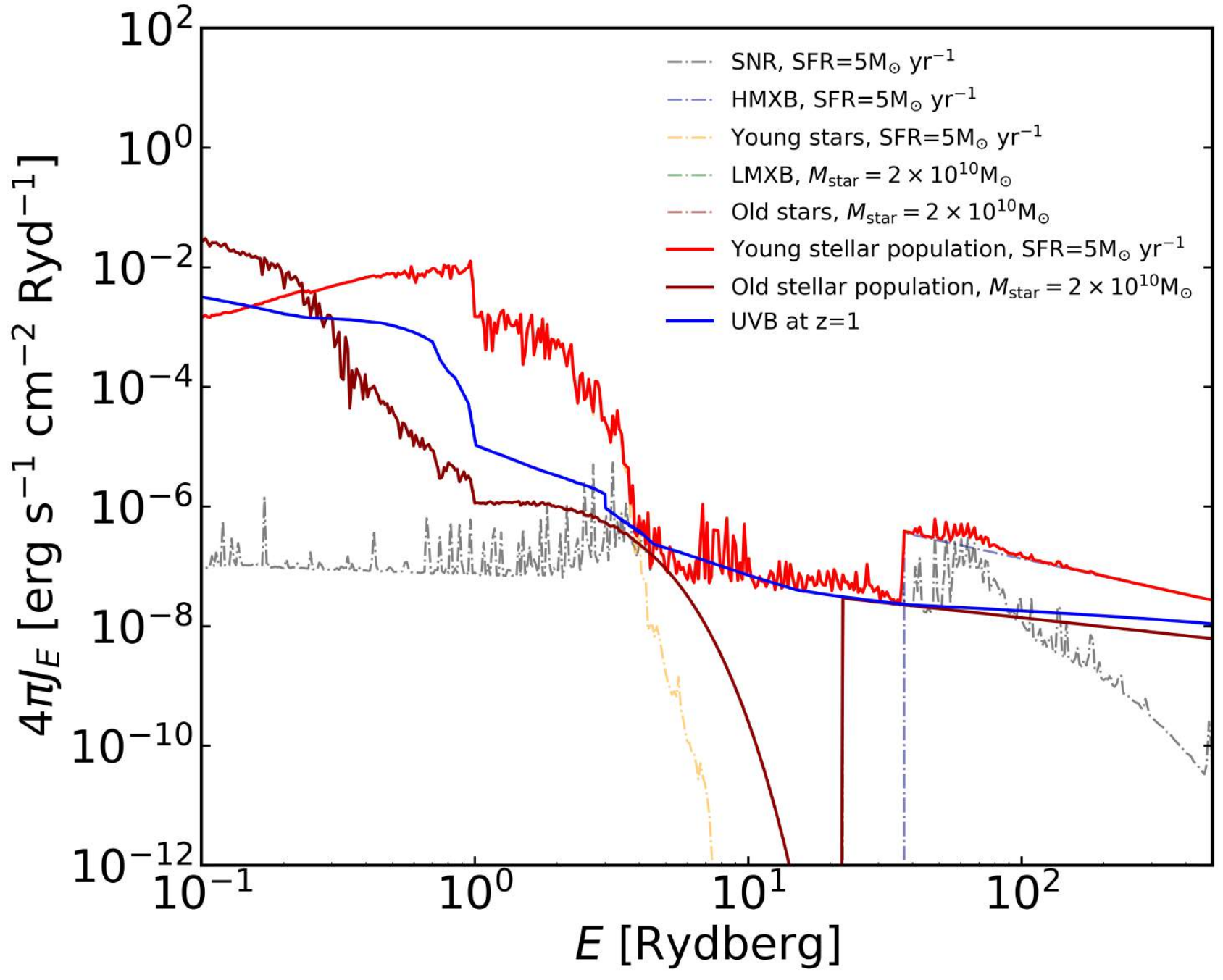}
   \caption{The intensity of the young stars, SNR, and HMXB with SFR equal to $5~{\rm M_{\odot}~yr^{-1}}$, and old stars and LMXB with mass ${\rm M_{\mathrm{star}}=2\times10^{10}\,\mathrm{M_{\odot}}}$, adopting a distance of $50\,{\rm kpc}$ away from the sources, which is around $0.5\, R_{\rm vir}$ of the simulated halos at $z = 1$. The total SEDs from the young stellar region and the old stellar population are also presented. The SED from the UVB at $z=1$ is included as well for comparison.}
    \label{sed}
\end{figure}

The intensities of the different components mentioned above are shown in Figure \ref{sed}. The radiation intensity is calculated by assuming a SFR or $1~{\rm M_{\odot}~yr^{-1}}$ for young stellar populations, $M_{\mathrm{star}}=10^{11}~{\rm M_{\odot}}$ for the old stellar population, and adopting a distance of 200 kpc away from the galaxy.

The hard truncation of the SED in the LMXB and HMXB regime is not a particularly accurate physical model since there are certainly some photons emitted below 0.3~keV. For example, \citet{US2018} use an empirical extrapolation of the HMXB SED below 0.3~keV based on a power law index of $-0.7$. The corresponding emission is uncertain and missing in our present analysis. We will examine the potential impact of these lacking photons in future work.

\subsubsection{Gas heating and cooling with irradiation}\label{method:cool&heat}

Having specified the SEDs of the radiation components as described above, we can calculate the gas heating and cooling rates in the presence of the radiation field.
For the heating and cooling of the primordial gas, Compton cooling due to the interaction between the cosmic microwave background (CMB) and the gas, and other cooling/heating processes are considered separately in the simulation. The treatment of the heating and cooling of the primordial gas, and the gas cooling due to CMB, follows the calculation in the IllustrisTNG galaxy formation model with small modifications. The primordial gas is assumed to be always in an ionization equilibrium state since the simulation time span (Hubble time) is much larger than the relaxation time ($\sim10^{7}\,{\rm yr}$). This assumption is valid for most of the time except at the epoch of reionization.

The number density of different species of the primordial gas can be calculated by the following equilibrium rate equations \citep{katz96}:
\begin{equation}\label{eq:ion1}
\alpha_{\rm H~II}~n_{\rm H~II}~n_{e}-\Gamma_{e{\rm H~I}}~n_{e}~n_{\rm H~I}-\Gamma_{\gamma{\rm H~I}}~n_{\rm H~I} = 0,
\end{equation}
\begin{equation}\label{eq:ion2}
(\alpha_{\rm He~II}+\alpha_{\rm d})~n_{\rm He~II}~n_{e}-\Gamma_{e{\rm He~I}}~n_{e}~n_{\rm He~I}-\Gamma_{\gamma{\rm He~I}}~n_{\rm He~I} = 0,
\end{equation}
\begin{equation}\label{eq:ion3}
\begin{aligned}
&\alpha_{\rm He~III}~n_{\rm He~III}~n_{e}+\Gamma_{e{\rm He~I}}~n_{e}~n_{\rm He~I}+\Gamma_{\gamma{\rm He~I}}~n_{\rm He~I}\\
&- (\alpha_{\rm He~II}+\alpha_{\rm d})~n_{\rm He~II}~n_{e} -\Gamma_{e{\rm He~II}}~n_{e}~n_{\rm He~II}-\Gamma_{\gamma{\rm He~II}}~n_{\rm He~II} = 0,
\end{aligned}
\end{equation}
\begin{equation}\label{eq:ion4}
\alpha_{\rm He~III}~n_{\rm He~III}~n_{e}
-\Gamma_{e{\rm He~II}}~n_{e}~n_{\rm He~II}-\Gamma_{\gamma{\rm He~II}}~n_{\rm He~II} = 0, 
\end{equation}
where  $\alpha_{i}$ is the radiative recombination coefficient, $\alpha_{\rm d}$ is the dielectric recombination coefficient, $\Gamma_{ei}$ is the collisional ionization rate for different species, and $\Gamma_{\gamma i}$ is the photonionization rate for different species. The values of the radiative recombination coefficient, the dielectric recombination coefficient, and the collisional ionization rate are given in \citet{katz96}. With the UVB and stellar radiation field, $\Gamma_{\gamma i}$ can be calculated by
\begin{equation}\label{eq:ion5}
\Gamma_{\gamma i} = \int^{\infty}_{\nu_{Ti}}\frac{4\pi}{h\nu}\sigma_{\nu i} (J_{\nu,~{\rm UVB}}(z)+J_{\nu,~\star}) {\rm d}\nu ,
\end{equation}
where $J_{\nu,~{\rm UVB}}(z)$ is the redshift-dependent, frequency-dependent, and solid-angle-averaged intensity of UVB, and $J_{\nu,~\star}$ is the frequency-dependent, solid-angle-averaged intensity of the total radiation from the stellar population. As described in Section~\ref{radiationfield}, the UVB radiation is described by the tabulated results in \citet{faucher20}, and the radiation from the stellar populations is obtained by summing up the total radiation of the stellar populations in the galaxies.

Based on equations (\ref{eq:ion1})-(\ref{eq:ion5}) and the number conservation equations, the abundance of different species can be solved. Gas cooling and heating can then be calculated. The gas cooling due to collisional excitation, collisional ionization, recombination, dielectric recombination and free-free collision can be calculated using Table~1 in \citet{katz96}. The photoionization heating is calculated by
\begin{equation}
H = n_{\rm H~I}\varepsilon_{\rm H~I}+n_{\rm He~I}\varepsilon_{\rm He~I}+n_{\rm He~II}\varepsilon_{\rm He~II}  
\end{equation}
where 
\begin{equation}
\varepsilon_{i} = \int_{v_{T_i}}^{\infty}\frac{4\pi}{h\nu}\sigma_{\nu i}(h\nu-h\nu_{T_i})(J_{\nu,~{\rm UVB}}(z)+J_{\nu,~\star}){\rm d}\nu~{\rm erg~s^{-1}}
\end{equation}
$\varepsilon_{i}$ due to UVB is based on the tabular results from \citet{faucher20}, while $\varepsilon_{i}$ due to the stellar radiation is obtained by using {\small CLOUDY}.

\begin{figure*}
   \subfigure{\includegraphics[width=0.45\textwidth]{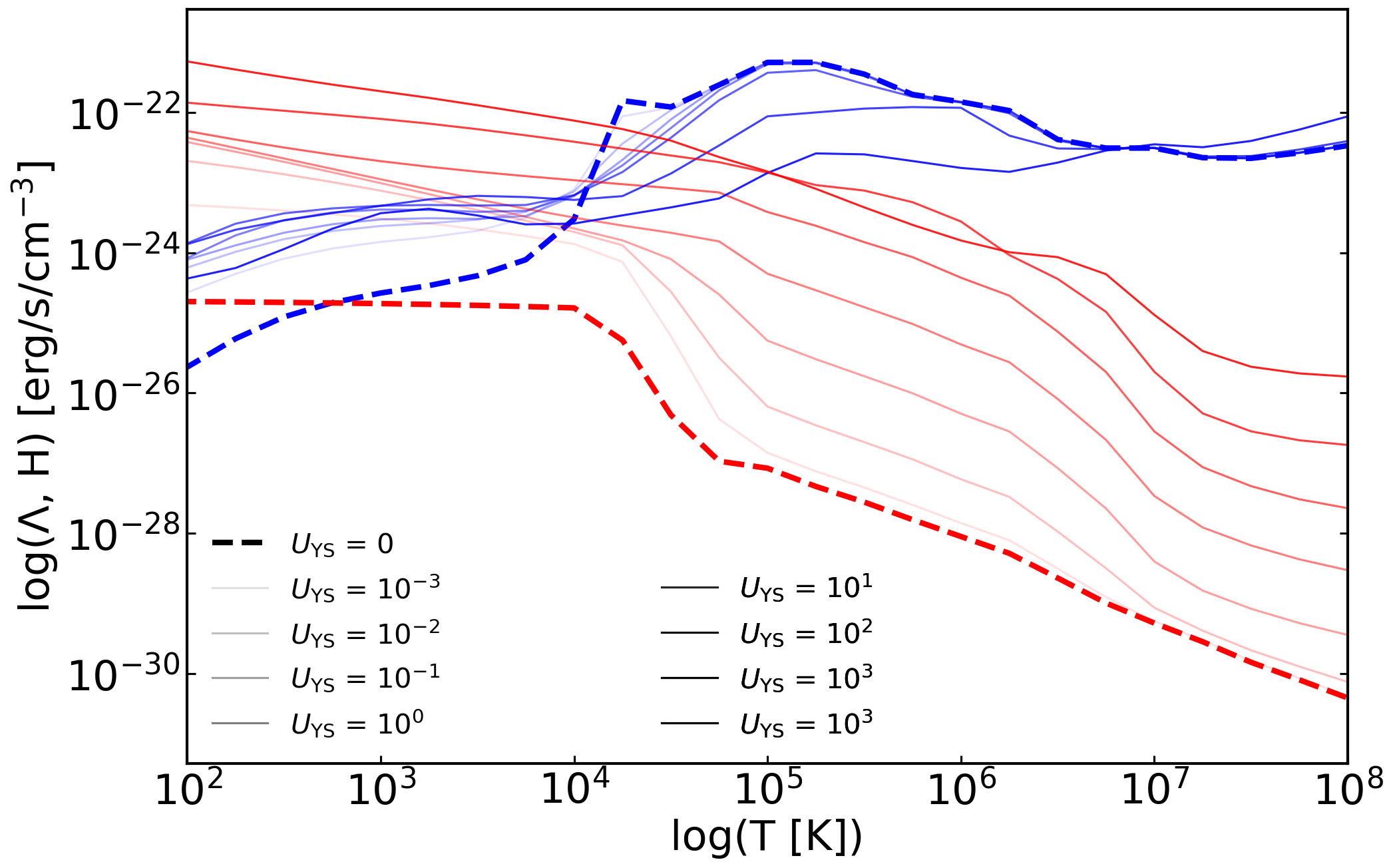}}
   \subfigure{\includegraphics[width=0.45\textwidth]{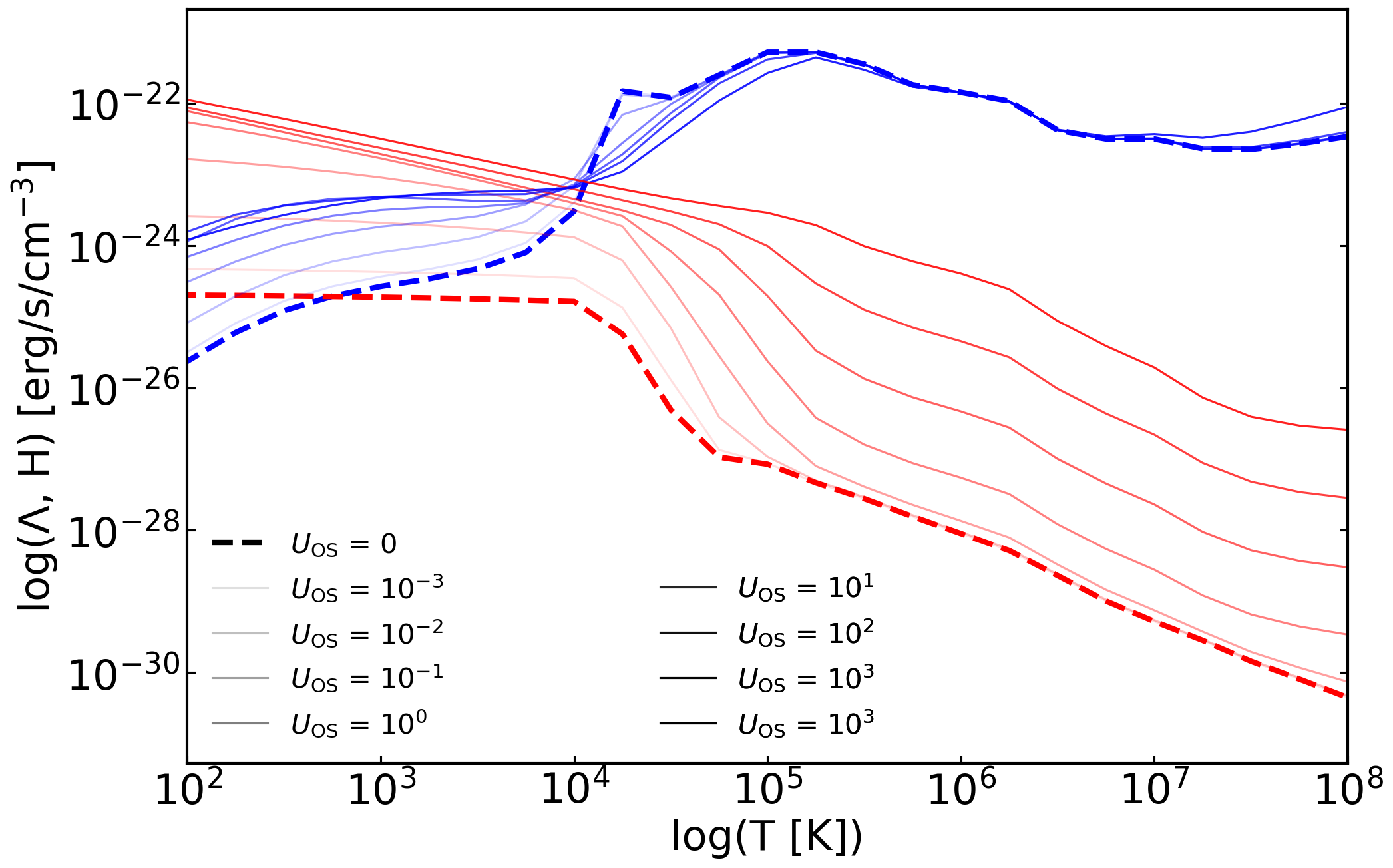}}\\
   \subfigure{\includegraphics[width=0.45\textwidth]{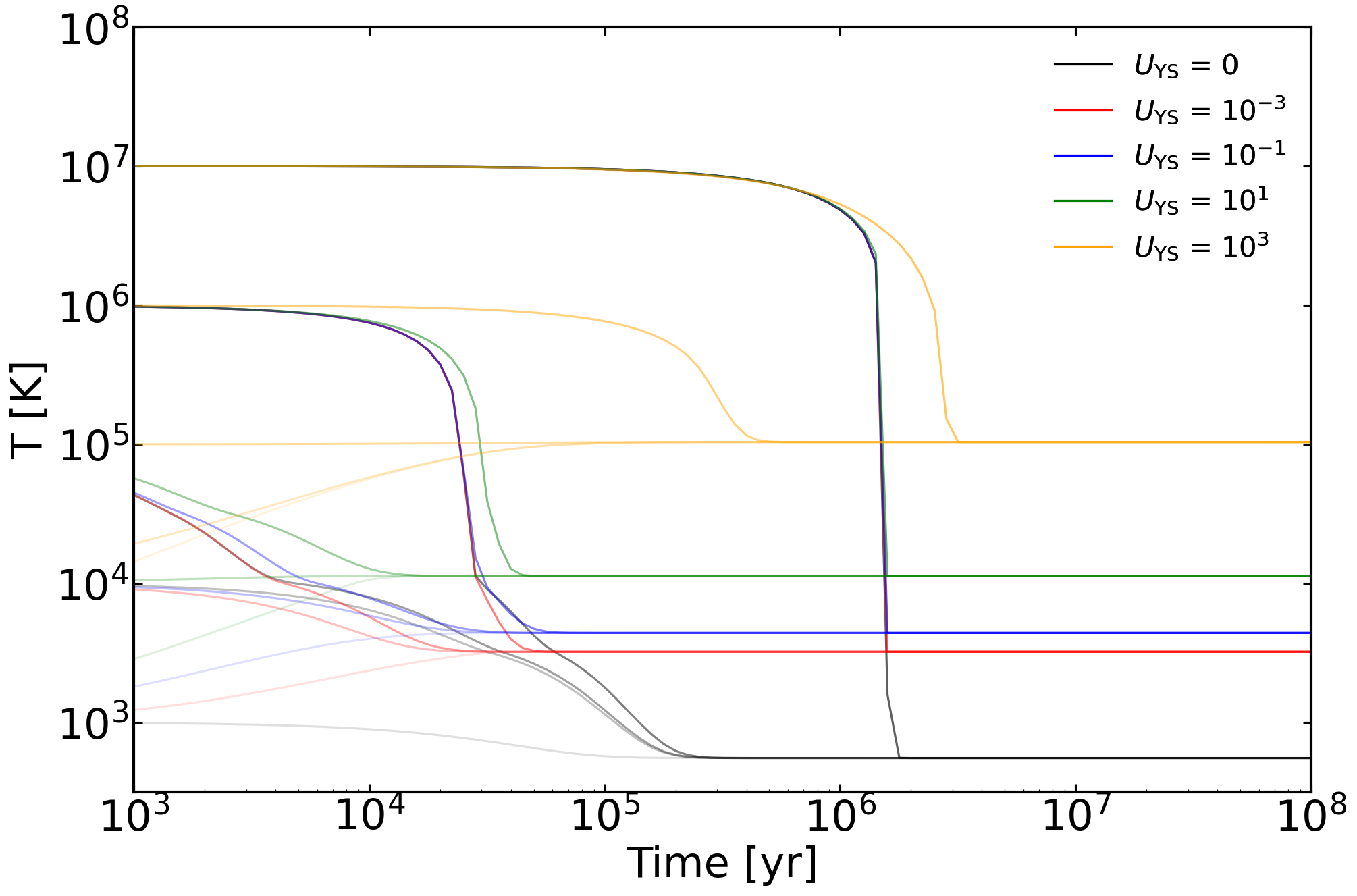}}
   \subfigure{\includegraphics[width=0.45\textwidth]{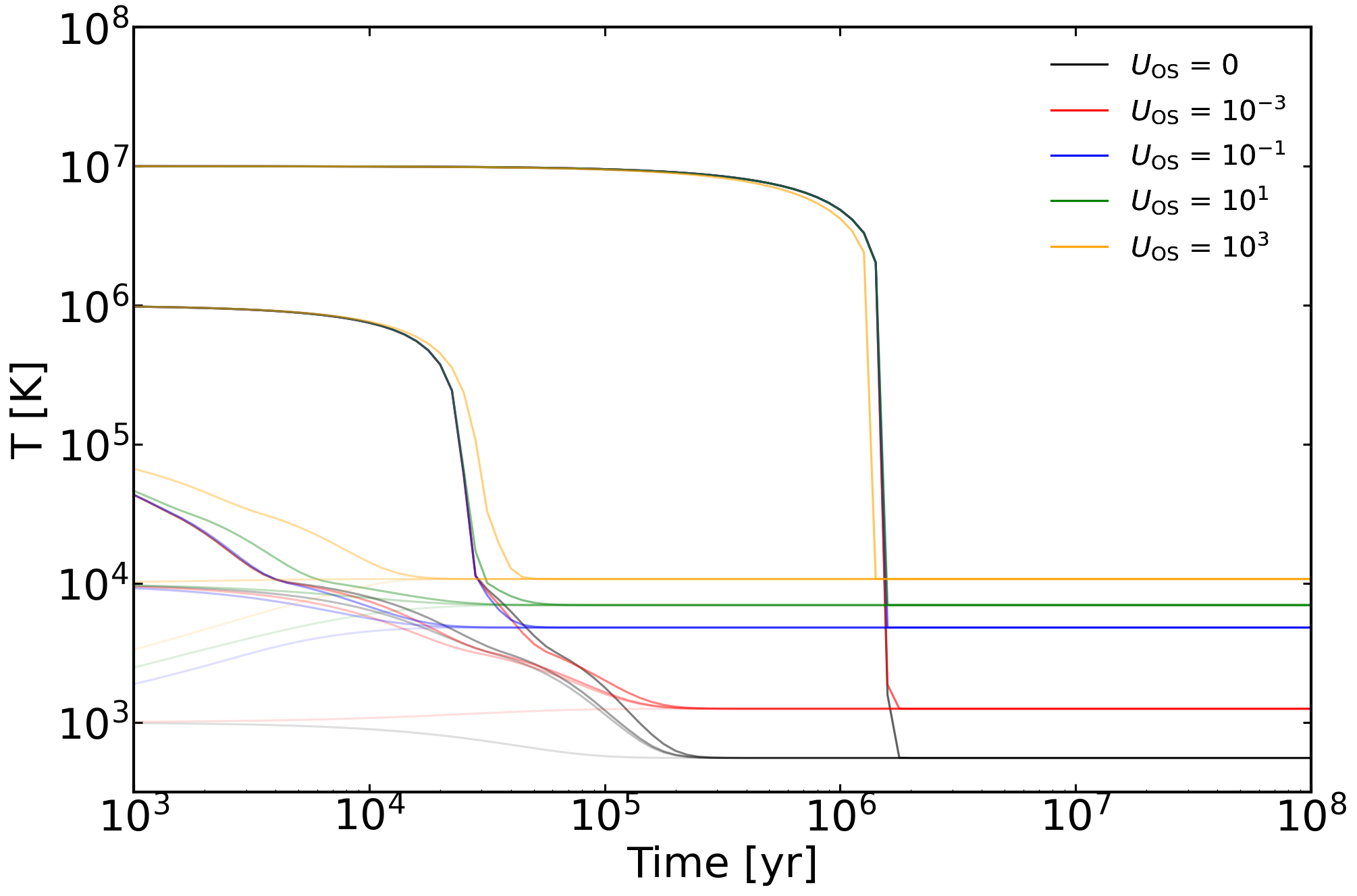}}
   \caption{{\it Upper panels:} Cooling curves of the gas for different ionization parameters. The left panel shows the cooling curve when the ionization parameter $U_{\rm YS}$ of young stellar radiation  (as labeled, in units of ${\rm erg\,s^{-1}\,cm^{3}}$) is varied, while the right panel shows the cooling curve for different ionization parameters $U_{\rm OS}$ due to old stellar radiation. {\it Lower panels:} The time evolution of the temperature for a gas cell with different initial temperature and ionization parameters, and a prescribed gas density corresponding to a number density of $1\,{\rm cm^{-3}}$. The left panel shows the time evolution of the temperature for different $U_{\rm YS}$, while the right panel gives the temperature decline for different $U_{\rm OS}$. The lines with different colors represent the time evolution of the temperature of the gas cells with ionization parameters equal to 0, $10^{-3}$, $10^{-1}$, $10^1$, and $10^3$ ${\rm erg~s^{-1}~cm^{3}}$. The different lines with the same color represent the time evolution of the temperature of gas cells with  initial temperatures equal to $10^3$, $10^4$, $10^5$, $10^6$, and $10^7\,{\rm K}$.}
    \label{coolingfunciton}
\end{figure*}

Other cooling/heating processes, including metal cooling/heating and Compton cooling due to the stellar radiation field, are calculated in the form of an input table for the simulations with {\small CLOUDY} under the ionization equilibrium assumption. Combined with all the processes described above, the total cooling/heating rate $\Lambda$ in the simulation can then be obtained as
\begin{equation}
\Lambda = \Lambda_{\rm p}+\Lambda_{\rm C}+\Lambda_{\rm o},
\end{equation}
where $\Lambda_{\rm p}$ is the cooling/heating rate of the primordial gas, $\Lambda_{\rm C}$ is the Compton cooling due to CMB, and $\Lambda_{\rm o}$ is the cooling/heating rate due to other cooling/heating processes. The resulting cooling/heating curves for different ionization parameters of young and old stellar radiation $z=0$ are shown in the first row of Figure~\ref{coolingfunciton}.

Since the cooling/heating rate due to other processes, $\Lambda_{\rm o} = \Lambda(n,T,J_{\rm ys},J_{\rm os},Z,z)$, is a function of gas density $n$, temperature $T$, radiation field from young stellar population $J_{\rm ys}$, radiation field from old stellar population $J_{\rm os}$, metallicity $Z$ and redshift $z$, we generate a 6-dimensional cooling/heating table and store it in a \textsc{HDF5} file. The redshift dependence of the function is due to the variation of the UVB spectrum. Because metal cooling and heating due to the stellar radiation only occur after stars have formed, we ignore the metal cooling/heating from the UVB at $z>10$.

For the table of the cooling/heating function, the gas number density ranges from $10^{-8}-10^{2}~{\rm cm^{-3}}$ in spacing of 1 dex in log space. The temperature ranges from $10^{1}-10^{9}~{\rm K}$ with 0.5 dex steps in log space. The mean intensity of the radiation from young and old stellar radiation ranges from $10^{-7}$ to $10^{7}~{\rm erg~s^{-1}~cm^{-2}}$ with 1 dex in log space. The metallicity ranges from $10^{-3}-10^{1}~{\rm Z_{\odot}}$ with steps of 1 dex in log space. The redshift covers $0-10$, again  with 1 dex in log space. These bins are relatively coarse due to the high memory cost of more densely sampled parameter space. To investigate the validity of our produce table, we have randomly chosen $10^6$ particles with different states $(n,T,J_{\rm ys},J_{\rm os},Z,z)$ and evolved them with different resolutions of the cooling/heating table and calculated the relative difference of the results. Reassuringly, we found relative difference less than 10\% in $99\%$.

\subsection{Numerical implementation of the radiation field}\label{method:implementation}

As described in the previous subsection, the gas cooling/heating rate is a function of the local radiation field. Thus, to implement the photoionization feedback into the simulation, we need to obtain the radiation intensity where the gas grid is located. 

For the ambient UVB radiation, we follow the treatment in previous work, i.e.~the UVB radiation is approximated as a redshift-dependent homogeneous radiation field. The UVB intensity value we adopt follows the results of \citet{faucher20}, as mentioned above. 

In contrast, we calculate the radiation from the local stellar population similar to the method described in \citet{kannan14}. First of all, to get the strength of the radiation emitted from a young stellar population, the luminosity is scaled with the total mass in young stars. We identify star particles borne more recently than 100 Myr as young stars. The total luminosity and spectrum of young stars are obtained using the method described in Section~\ref{method:starrad}. We assume that star particles whose age is older than 100 Myr belong to the old stellar population. Again, the luminosity of the old stellar population is scaled in proportion with the stellar mass and its luminosity can be calculated easily. We note that our criterion for the young stellar population is less strict than \citep{kannan14}, who only identified star particles borne more recently than 10 Myr as young stars. However, as the typical age of stars on the main sequence with mass $\sim8~{\rm M_{\odot}}$ is $\sim 30-50$ Myr, we adopt 100 Myr as an upper bound in the current work to safely include them. We also have run various test simulations with different criteria for the age cut, finding that the value of the age criterion for young stars does affect the results at a quantitative level, for example, smaller value of the age criterion for young stars lead to weaker effects of the local star radiation. However, our conclusions are not affected qualitatively by the value for the age cut. Hence, for the rest of this study we adopte the 100 Myr value for  our study of the effects of the local stellar radiation. 

Our approximate radiation transfer method for the local stellar emission follows the methods described in \citet{kannan14} and \citet{woods15}. The radiation intensity impinging on a gas cell from local radiation sources is calculated using an octree by assuming that the gas outside the star-forming region is optically thin. Therefore, the radiation intensity from a single source at any location is calculated by 
\begin{equation}
 j = \frac{L}{4\pi r'^2},
\end{equation}
where $L$ is the luminosity of the single source and $r'$ is the distance between a given location and the radiation source. The luminosity of a single source in turn can be obtained by the methods described in the previous subsections. The total radiation intensity at any location can then be simply calculated by adding the radiation from all local sources, with the gravitational tree algorithm being repurposed for this task, allowing the corresponding summation to be computed efficiently. 

\begin{figure*}
   \includegraphics[width=\textwidth]{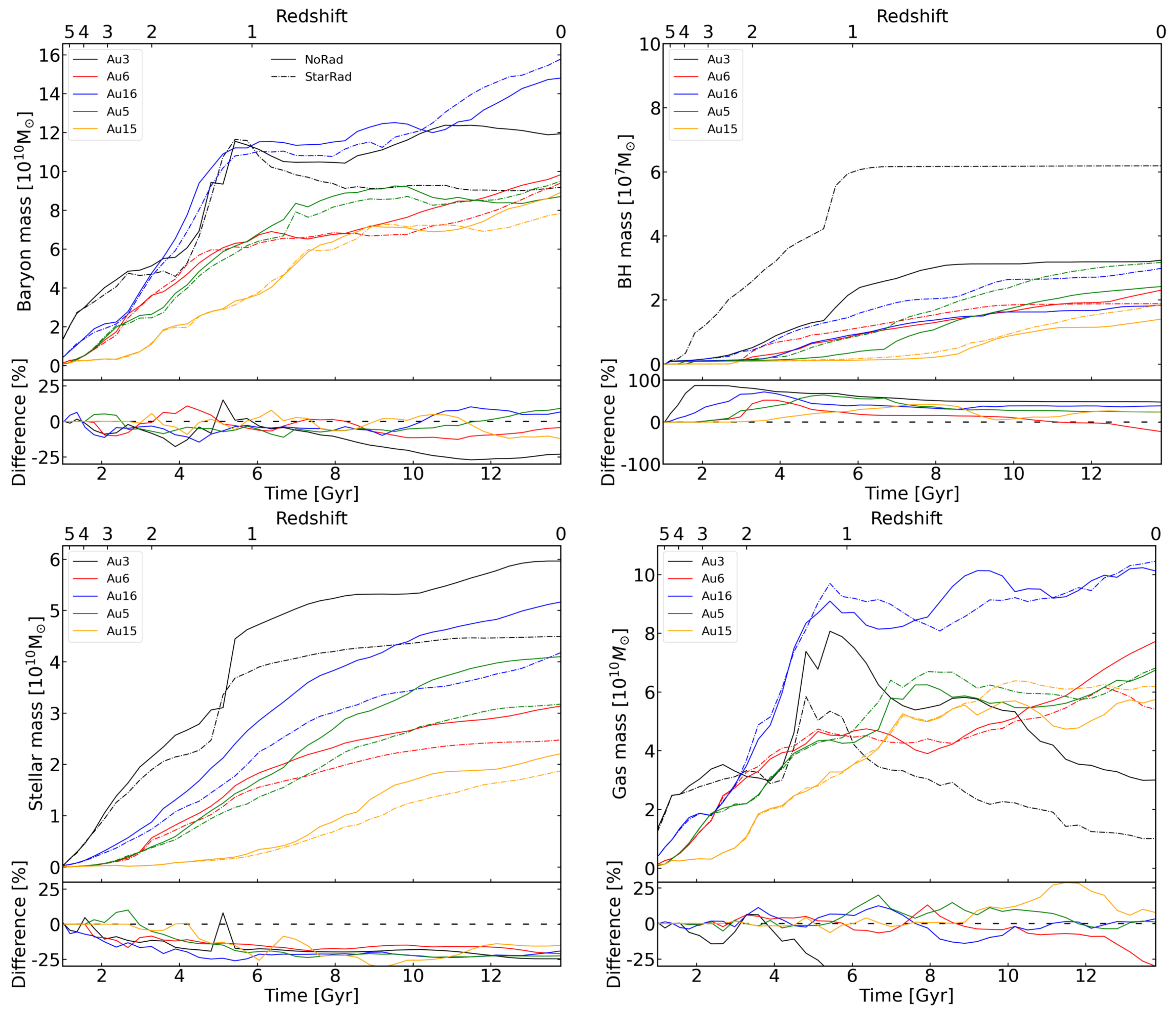}
   \caption{{\it Upper panels}: The time evolution of the total baryonic mass, the BH mass, the stellar mass, and the gas mass from $z=5.5$ to $z=0$, as labelled, in our five selected galaxies. The solid lines give the results from the NoRad simulations, while the dashed lines represent the results from the StarRad simulations. {\it Bottom panels}: The relative difference between the simulations with and without stellar radiation. The NoRad simulations are treated as here as fiducial reference simulations.}
    \label{time_evo}
\end{figure*}

\begin{figure}
   \includegraphics[width=0.45\textwidth]{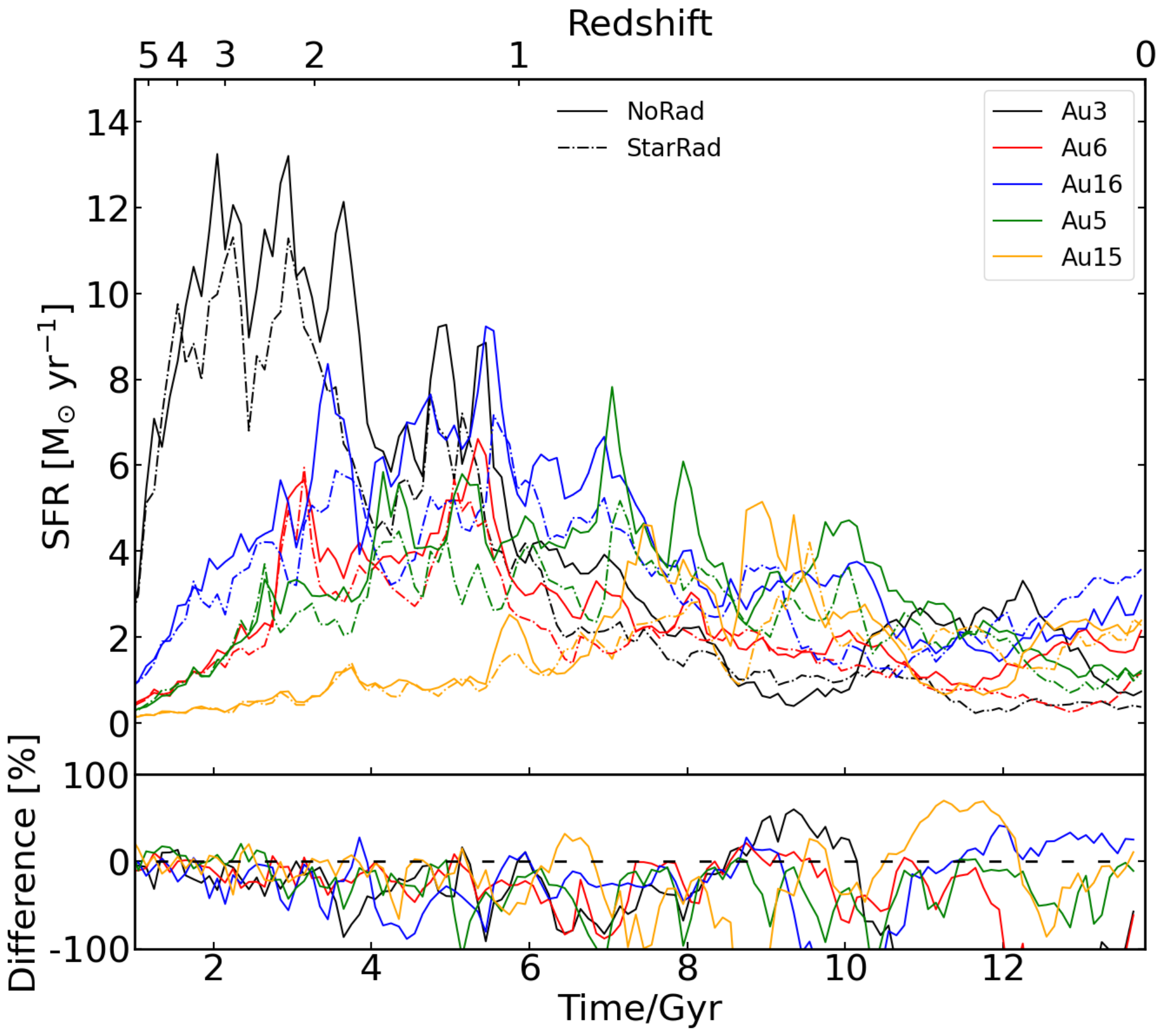}
   \caption{{\it Upper panels}: The time evolution of the 100 Myr-averaged SFR from $z=5.5$ to $z=0$ in our five selected galaxies. The solid lines give the results from the NoRad simulations, while the dashed lines represent the results from the StarRad simulations. {\it Bottom panels}: The relative difference between the simulations with and without stellar radiation. The NoRad simulations are treated as here as fiducial reference simulations.}
    \label{sfr}
\end{figure}

\begin{figure*}
   \includegraphics[width=\textwidth]{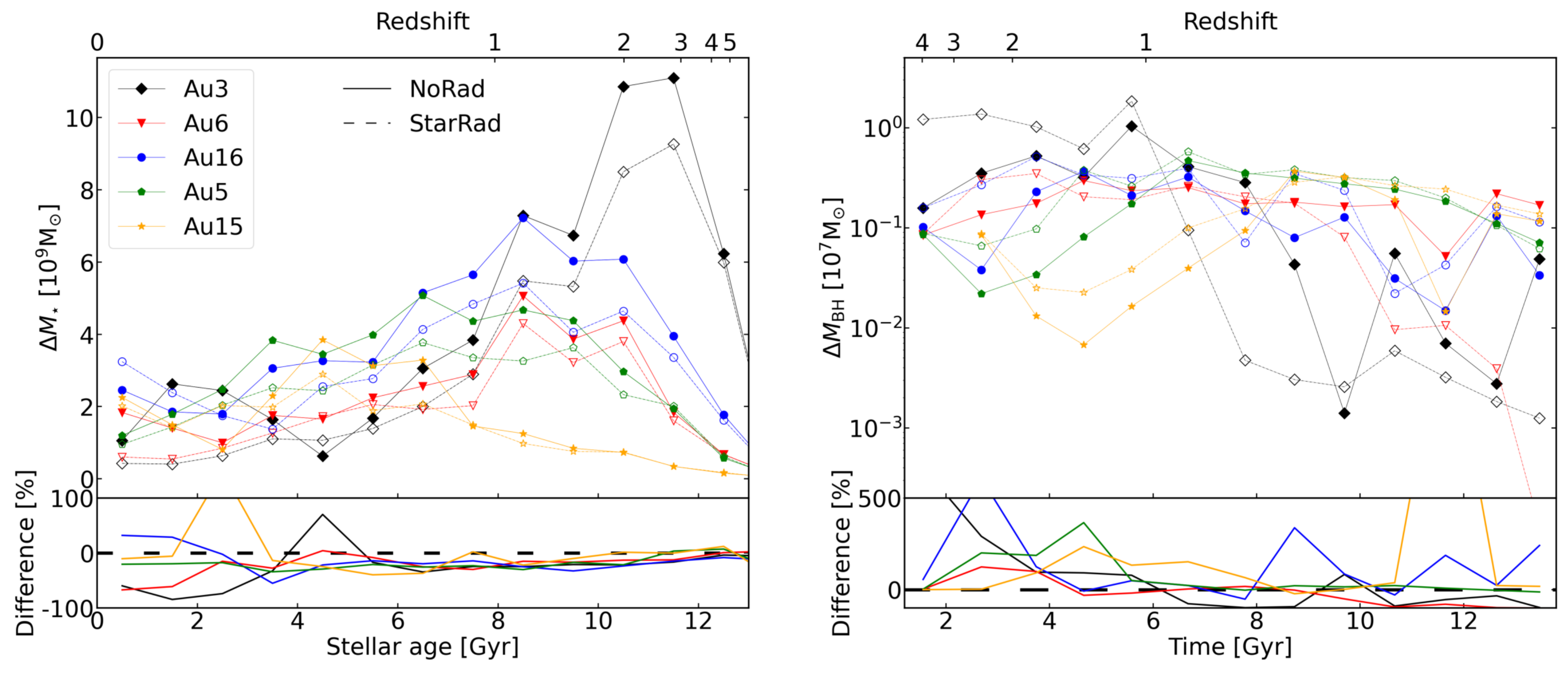}
   \caption{{\it Left panel}: The stellar mass growth for different subsequent time bins in five selected Auriga galaxies at $z=0$. The time interval for each bin is 1~Gyr, and this is used here to obtain a time-averaged version of the star formation rate history in order to smooth out temporal fluctuations in the SFR. {\it Right panel}: The BH mass growth for different time bins in five selected Auriga galaxies. Again, we use time intervals equal to 1~Gyr  in order to smooth the BHAR. Solid lines in both subfigures represent the results from the NoRad simulations, while the dashed lines give the results from the StarRad simulations. The relative difference of the NoRad and StarRad simulations is shown in the bottom panels of each panel.}
    \label{deltaM}
\end{figure*}

\begin{figure}
   \includegraphics[width=0.45\textwidth]{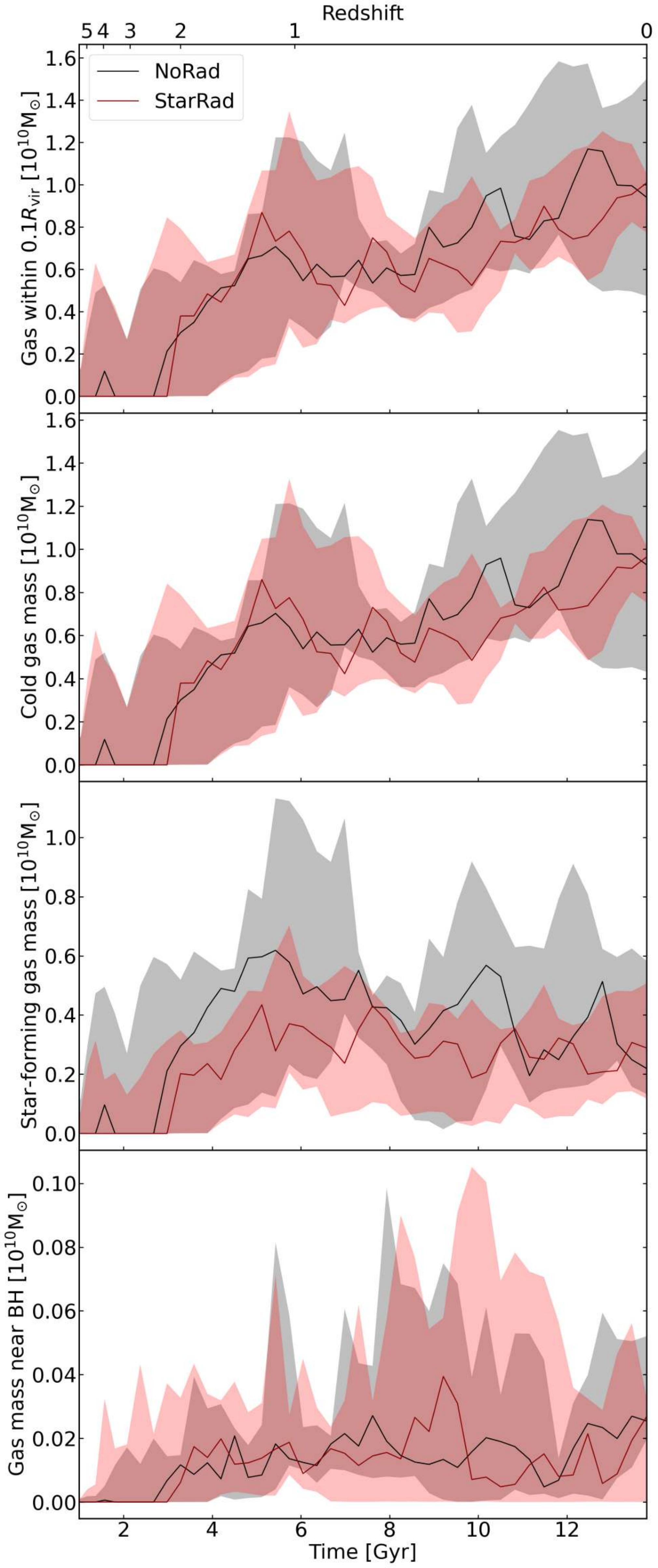}
   \caption{The time evolution of the gas mass selected according to different criteria. From the upper to the bottom panels we show: all gas within $0.1\, R_{\rm vir}$, cold gas ($T<10^{5.5}\,{\rm K}$) in the same region, star-forming gas in the same region, and finally the gas mass within a 1~kpc aperture around the central BH. The solid lines represent the median values of the results of our five simulated galaxies, while the shaded regions give the corresponding maximum and minimum values over this set of galaxies. Both simulations with and without local radiation are shown, as labelled.}
    \label{0.1Rvir_evo}
\end{figure}

\section{The impacts of local ionization feedback on galaxy properties} \label{Sec:GalProperties}

In this section, we analyze and discuss the differences in the time evolution, in the properties of gas and stellar disks, and in the rotation curve between simulated galaxies with and without local stellar radiation.

Before we consider our results for the simulated galaxies, we investigate the time evolution of the temperature for several fiducial gas states with different initial temperatures and different ionization parameters of young stellar radiation $U_{\rm YS}\equiv J_{\rm YS}/n$ and old stellar radiation ${U_{\rm OS}}\equiv J_{\rm OS}/n$. By analyzing the temperature evolution of these illustrative test situations, we can develop a first rough understanding of the heating effects of the radiation on the expected evolution of gas thermal states.

The second row of Figure~\ref{coolingfunciton} shows the time evolution of the temperature of test gas cells with  initial temperatures $T=10^{3}$, $10^{4}$, $10^{5}$, $10^{6}$, and $10^{7}\,{\rm K}$, and $U_{\rm YS}$ or $U_{\rm OS}$ equal to $0$, $10^{-3}$, $10^{-1}$, $10^{1}$, $10^{3}\,{\rm erg\,s^{-1}\,cm^{-2}}$, respectively. At first glance, we can infer from the figure that the temperature will eventually reach an equilibrium temperature when radiation is included, which corresponds to the point where the cooling rate equals the heating rate. This equilibrium temperature is sensitive to the strength of the young stellar radiation. It can increase from $\sim10^{3.5}\,{\rm K}$ at $U_{\rm YS}=10^{-3}\, {\rm erg\,s^{-1}\,cm^{-2}}$ to $\sim10^5\,{\rm K}$ at $U_{\rm YS}=10^{-3}\, {\rm erg\,s^{-1}\,cm^{-2}}$. Note that the equilibrium temperature due to the old stellar radiation is always close to or even lower than $10^{4}\,{\rm K}$, and in general, it is lower than the equilibrium temperature due to the young stellar radiation.

The stellar radiation can also delay the cooling when the gas temperature is mildly higher than the equilibrium temperature, and this effect is more pronounced for the old stellar radiation. For example, for both the young and old stellar radiation, the gas temperature of the test gas configuration with an initial temperature equal to $10^{5}~{\rm K}$ stays longer higher at any given time for a higher ionization parameter, before the equilibrium temperature is reached. This conclusion still holds in the presence of young stellar radiation for gas with an initial temperature equal to $10^{6}$ or $10^7\,{\rm K}$, but for high ionization parameter. For the old stellar radiation, this effect becomes weaker,
but it can still be found for $U_{\rm OS}\ga 10^{3}\,{\rm erg~s^{-1}\, cm^{-2}}$ in gas with  initial temperature equal to $10^{6}$ or $10^7\,{\rm K}$.

\subsection{Global time evolution of different baryonic components}

\begin{figure}
   \subfigure{\includegraphics[width=0.45\textwidth]{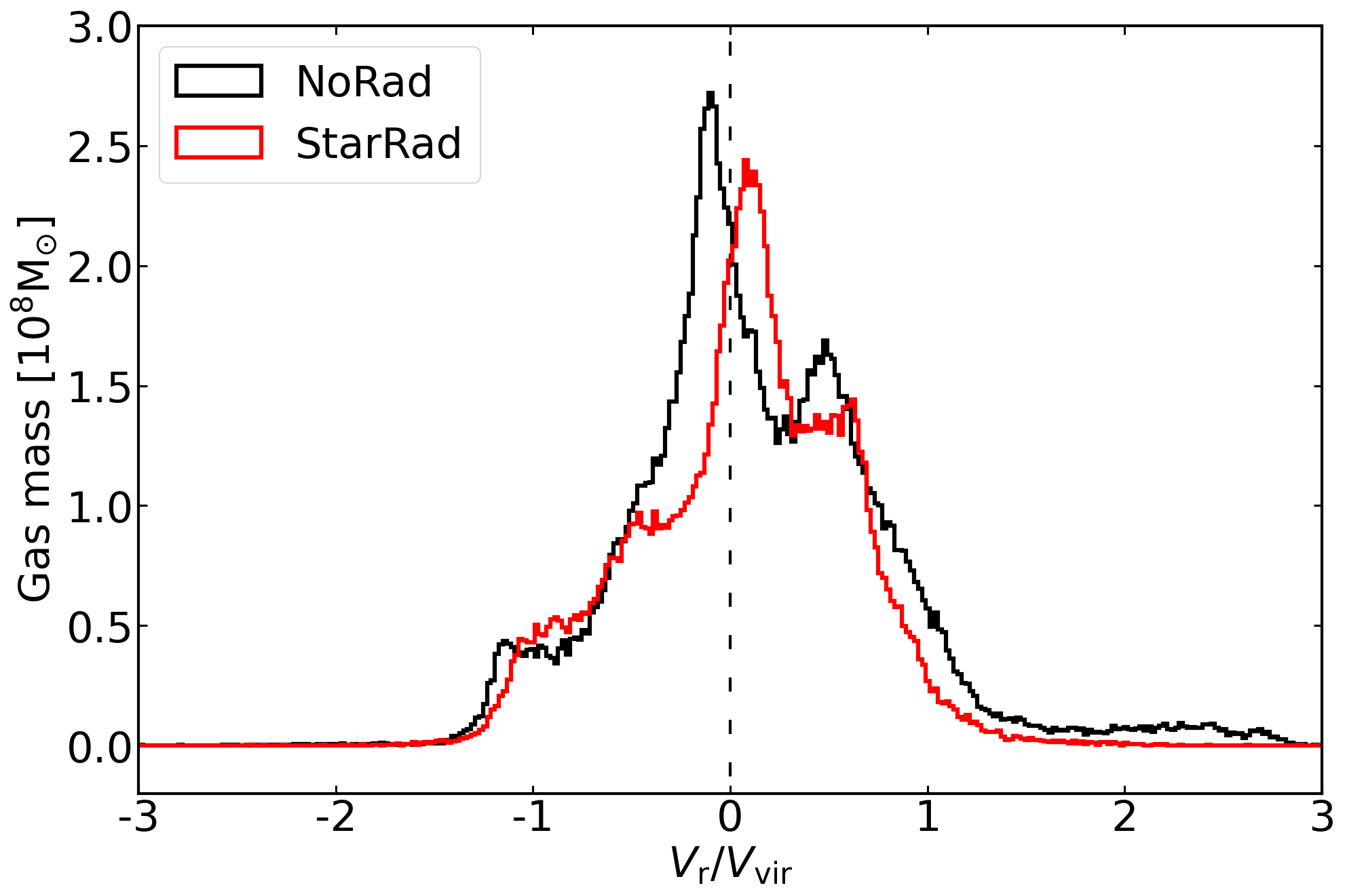}}
   \caption{The distribution of the radial velocity of gas within $0.1\, R_{\rm vir}$ for times $z<0.1$. The data is averaged over five selected Auriga galaxies. The black line represents the simulation results from the NoRad runs, while the red line is for the corresponding StarRad runs.}
    \label{GasVelHist}
\end{figure}

Figure~\ref{time_evo} shows the time evolution of the total baryonic mass, the BH mass, the stellar mass, and the gas mass in five selected halos from $z=5.5$ to $z=0$. The first panel gives the evolution of the baryonic mass and the relative difference of five simulated halos with the NoRad and StarRad galaxies. In all but the Au3 galaxies, the relative differences between the NoRad and StarRad galaxies  are $< 20\%$ and do not have a clear trend, implying that the current implementation of local photoionization sources has little impact on the gas accretion of the halos. The only exception is Au3, where the total baryonic mass decreases in the StarRad galaxies compared to the NoRad galaxies after 6~Gyr. The third panel of Figure \ref{time_evo} also shows the signal of the large variation in the time evolution of the stellar mass of Au3. The stellar mass of Au3 doubles around 6 Gyr, which implies that this system encounters a major merger. Combined with the third panel of Figure~\ref{time_evo}, the large difference in the Au3 galaxy may be caused by the AGN feedback during the major merger at 6 Gyr. Since AGN feedback will cause strong disturbances of the gas and nonlinearity of the fluid behavior, combined with the local stellar radiation, the subsequent evolution may experience strong differences. A detailed analysis of the nonlinearly combined influence of AGN feedback and local stellar radiation during and after the major merger of the Au3 galaxy is presented in Section~\ref{sec:combined}.

To further confirm this conclusion, we present the time evolution of the 100 Myr-averaged star formation rate in Figure~\ref{sfr}. We focus on the Au3 and Au16 galaxies. The first panel of Figure~\ref{time_evo} shows that the time evolution of the baryonic mass is similar in Au3 and Au16 at $z\ga1$. However,  at $z\sim1$ the growth rate of the baryonic mass rapidly decreases in the Au3 StarRad model. However, we can observe in Figure~\ref{sfr} that the SFR in Au3 at $z\sim 1$ does not show significant differences compared to Au16. This confirms our conclusions mentioned above that our model of local stellar radiation does not have a significant impact on the gas accrection of the halos. The differences of  the evolution of the baryonic mass in Au3 at $z\la1$ is due to the AGN feedback during the major merger events.

The second panel of Figure~\ref{time_evo} shows the growth of the BH mass and the relative differences between the NoRad and StarRad galaxies in all simulated halos. In the figure, most of the simulated galaxies have a positive value for the relative differences, which indicates that the BH is larger in the StarRad galaxies than in the NoRad galaxies. The difference grows quickly and can be greater than $50\%$ in the first 2 Gyr, but then becomes lower at later times. This is also reflected in the right panel of Figure \ref{deltaM}, which shows the growth of the BH mass over time intervals of 1 Gyr. In the first 6 Gyr, the relative difference of the BH mass is in general larger in the StarRad model than in the NoRad model. The BH mass growth can even become six times higher in some simulated galaxies. After 6 Gyr, the difference becomes small, however, and the relative difference can even become negative, thus reversing the trend.

These results indicate that local stellar radiation can strongly enhance the growth of BH at high redshift, while having only a minor impact at later time. This may be caused by the suppression of star formation in dense gas when local stellar radiation is considered. Since our model prevents gas cells from becoming star-forming when the gas ionization parameter is larger than 1, these gas cells can stay at relatively low temperature because their equilibrium temperature is lower than the effective equation of state temperature calculated by the star formation model in TNG.

We similarly find that the relative differences in the growth of the stellar mass between the NoRad and StarRad galaxies are higher at earlier times.  In general, the relative differences increase with time in the first 6 Gyr and then reduce to $\sim 25\%$ after 6 Gyr. This result implies that the local stellar radiation can suppress star formation preferentially at high redshift, which is a robust conclusion for the Auriga galaxies with our photoionization feedback described above. At late times, the stellar mass can be suppressed by around $25\%$, which is still stronger than the results reported in \citet{obreja19} and \citet{hopkins20}.  We will discuss the comparison with these two works in Section~\ref{sec:comparison}.

Note, however, that these trends are not always very clear at $z\geq2$. In fact, models Au5 and Au15 have actually larger stellar mass with the StarRad model than with the NoRad model at high redshift. Given the more robust trend at low redshift, this may be because local stellar radiation is only effective in suppressing star formation once a disk has formed. The results in \citet{kannan14}, where star formation was found to be suppressed by $\sim30\%$ in a disk galaxy, are consistent with this conjecture. In contrast, at high redshift, the galaxies are still in an intense formation phase, where their virial temperatures are lower, the gas infall rates are very high, and stable disks have not yet formed. The impact of local stellar radiation in this environment is apparently weaker and likely more complicated to understand. In addition, we also expect a larger system to system variation at these early times. A further discussion of the high-redshift evolution is beyond the scope of this work, but we intend to investigate it in more detail in future work.

\begin{figure*}
   \includegraphics[width=\textwidth]{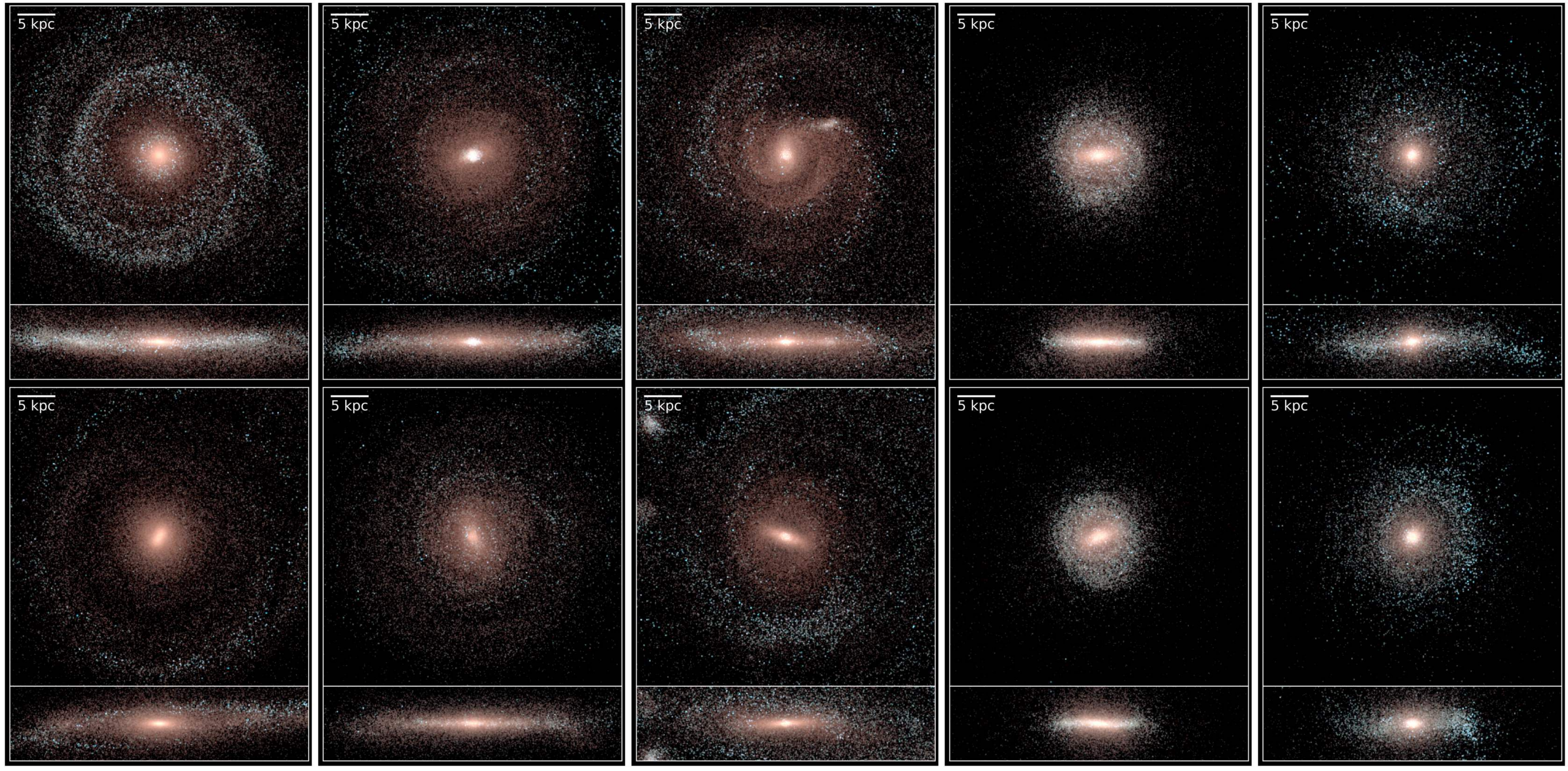}
   \caption{Stellar images of different Auriga galaxies, stacked from $z=0.1$ to $z=0$, a timespan still short compared to the bar rotation period. From the left to the right columns we show Au3, Au6, Au16, Au5 and Au15. The top row shows the stellar distributions of the NoRad simulations, while the bottom row gives the stellar images of the corresponding StarRad simulations.}
    \label{StellarImage}
\end{figure*}

\begin{figure}
   \includegraphics[width=0.45\textwidth]{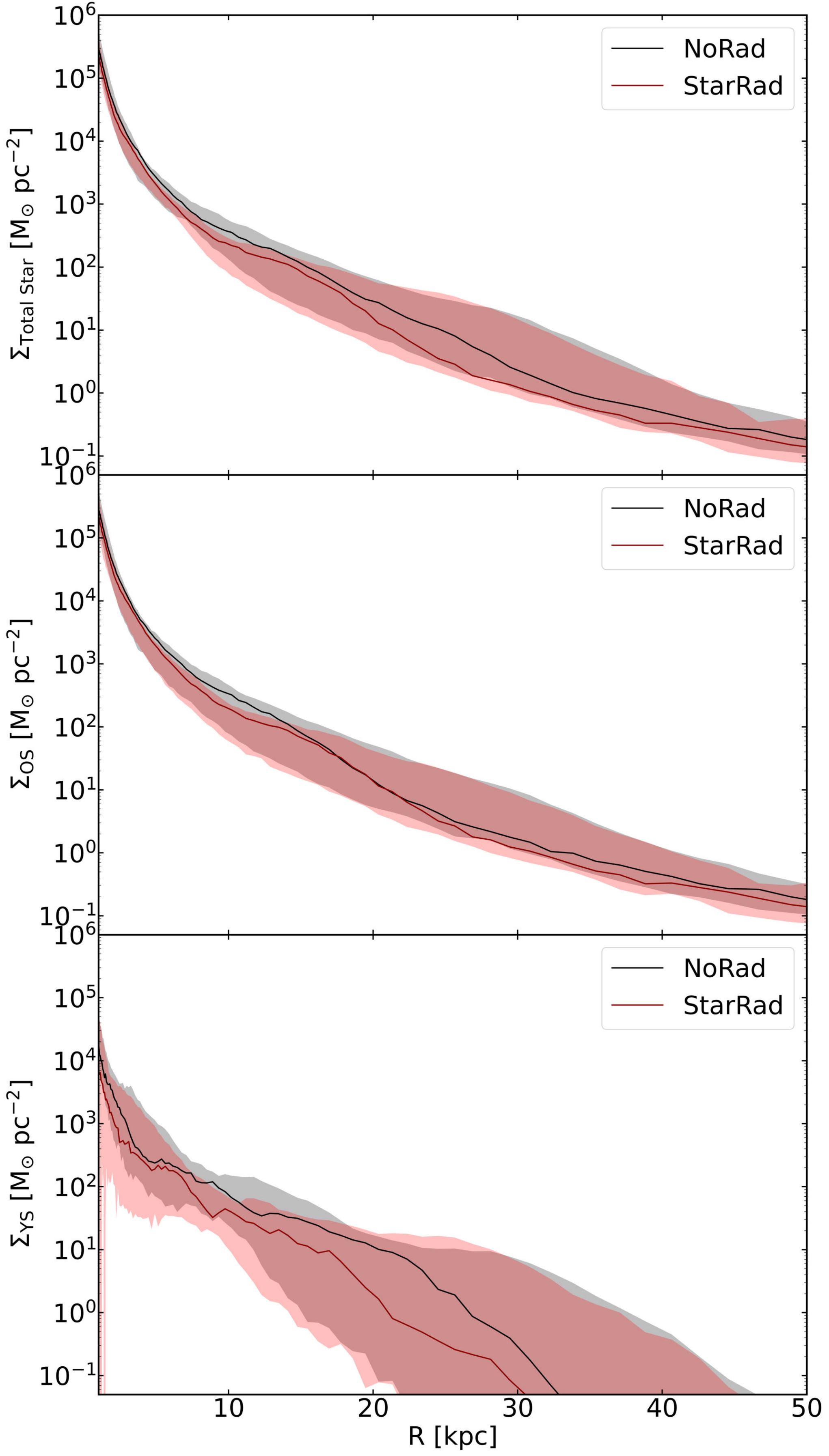}
   \caption{Stacked radial stellar surface density profiles of different Auriga galaxies, averaged from $z=0.1$ to $z=0$. We show results for simulations both with and without local stellar radiation, as labelled. The shaded areas illustrate the 10 to 90 percentiles of the five galaxies in our simulation set, while the solid lines are the medians.}
    \label{StarSurf1D}
\end{figure}

We further investigated the stellar age distribution of simulated galaxies at $z=0$. The left panel of Figure \ref{deltaM} shows the stellar age distribution of the NoRad and StarRad galaxies at $z=0$ and their relative differences. The differences indicate that the stellar mass formed at $z\sim2$ to $z\sim0.7$ is systematically lower in the StarRad galaxies than in the NoRad galaxies. At $z\ga2$, the value of the relative differences is slightly higher than zero in Au5 and Au15, which is consistent with the results in the third panel of Figure~\ref{time_evo}. At $z\la0.7$, the relative differences exhibit large scatter, which may be related to the growing importance of AGN feedback at late times. Since the BH mass becomes large there, the cumulative energy released by the AGN grows as well, and the impact of the associated energy feedback becomes significant \citep{zinger20}. Note that although the stellar mass also increases with time, the impact of the radiation from old stars on the gas is much weaker than that of radiation from young stars. Given also the declining star formation rate at low redshift, it is then not surprising that the impact of the AGN feedback becomes more important than the local stellar radiation. Thus, the impact caused by local stellar radiation becomes subdominant to the impact of AGN feedback and cannot be distinguished separately any more.

The final panel of Figure~\ref{time_evo} shows the time evolution of the total gas mass and the relative differences between the NoRad and StarRad galaxies in all simulated halos. Similar to the time evolution of the baryonic mass, the figure shows good convergence in the first 8 Gyr but highlights larger relative differences between the NoRad and StarRad galaxies at later time. Since the stellar mass exhibits a systematic decrease with radiation included, the gas mass shows a corresponding
increase in its proportion relative to the stellar mass and the the total baryon mass later than $\sim 6$ Gyr. This result again confirms that less gas mass is converted to stars in the StarRad galaxies.

The time evolution of the total baryon mass in both the NoRad and StarRad galaxies indicates that the local stellar radiation will not affect the total gas accretion into the halo. To further investigate whether gas accretion into galaxies remains equally unaffected, we show the time evolution of the gas mass within $0.1\,R_{\rm vir}$ in both the NoRad and StarRad galaxies in the first panel of Figure~\ref{0.1Rvir_evo}. We directly put the five NoRad/StarRad galaxies together for a statistical robust mean trend. Similarly to the total baryon mass, the difference in the gas within $0.1\, R_{\rm vir}$ in the first 8~Gyr is small, indicating that the feedback processes are still weak and do not provide enough energy to push the gas out. After 8~Gyr, the median difference between the two is still small. However, the range between the 10 and 90 percentile values in the StarRad galaxies becomes lower and smaller than in the NoRad galaxies. This is because local stellar radiation can heat the gas, either preventing it from flowing in, or even pushing it to larger radii through thermal pressure.

The second panel of Figure~\ref{0.1Rvir_evo} shows the time evolution of the cold gas. The cold gas is here defined as gas with a temperature lower than $10^{5.5}\, {\rm K}$. The result is almost the same as seen for the time evolution of the total gas within $0.1\, R_{\rm vir}$, indicating that the dominant gas component within $0.1\, R_{\rm vir}$ is cold gas. The reduction of the cold gas at low redshift can be ascribed to the heating effect of local stellar radiation. The third panel of Figure~\ref{0.1Rvir_evo} shows the time evolution of the star-forming gas mass within $0.1 R_{\rm vir}$. Unlike the time evolution of the total gas mass shown in the first panel of Figure~\ref{0.1Rvir_evo}, the total amount of star-forming gas in the StarRad galaxies is systematically lower than in the NoRad galaxies from high redshift. This follows the trend of the star formation history shown in the first panel of Figure~\ref{deltaM}, again implying that local stellar radiation affects star formation by reducing the amount of star-forming gas, but not the total gas mass. 

The final panel of Figure~\ref{0.1Rvir_evo} gives the time evolution of the gas mass within 1~kpc around the central BH. From the figure, we can see that the median value of the gas mass evolves similarly in both the NoRad and StarRad galaxies. However, the variation of the gas mass in StarRad galaxies is much lager than in NoRad galaxies, especially in the first 5~Gyr. A higher gas mass in principle implies that the BHs have more gas to accrete from, so this result can partly explain why the BH mass can grow faster in the StarRad galaxies at high redshift. 

Figure~\ref{GasVelHist} provides another view on the impact of the stellar radiation by showing the distribution of radial velocities of the gas, in units of the virial velocity. From the figure we can infer that the amount of gas with a negative inflow velocity as well as with a positive outflow velocity (wind feedback) is reduced in the StarRad galaxies compared to the NoRad galaxies. This implies that the local stellar radiation effectively reduces the amount of material available for star formation, and thus indirectly the feedback strength. Note that the local stellar radiation thus acts more as a preventive process that keeps gas from flowing in in the first place.

\subsection{Disk properties at $z=0$}

The previous section examined the impact of local ionization feedback on the formation history of the simulated galaxies. In this section, we analyze the impact of local ionization feedback on both stellar and gas disk properties $z=0$.

\subsubsection{Stellar disk}

Figure~\ref{StellarImage} shows stacked stellar images of five simulated galaxies with and without local stellar radiation, where the surface density of each simulation is averaged from $z=0.1$ to $z=0$. The upper row of panels gives the simulated galaxies without local stellar radiation, while the bottom row of panels shows the corresponding simulated galaxies with local stellar radiation. At first glance, the NoRad galaxies have slightly higher surface brightness in the face-on view. The NoRad galaxies display more blue light outside 5~kpc, and more red light at the center. Except for the surface brightness, both the red-light and blue-light of NoRad galaxies are also more extended. These results imply that local stellar radiation can also affect the stellar disk properties by influencing the spatial location of star formation. Similar conclusions can also be drawn from the edge-on views.

To compare the stellar distribution quantitatively in both the NoRad and StarRad galaxies, we show the one-dimensional surface density of total stars, old stars, and young stars in Figure~\ref{StarSurf1D}. Young stars and old stars are separated by their age. When the stellar age is younger than 3~Gyr, the stars are identified as young stars, otherwise as old stars. Again, we stack all of the NoRad/StarRad galaxies from $z=0.1$ to $z=0$. The solid line represents the median value of the surface density, while the shaded area in the figure represents the 10 to 90 percentile range of the distribution of measurements.

The first panel of Figure~\ref{StarSurf1D} gives the surface density of the total stars of the stacked NoRad and StarRad galaxies. We find that the total stellar surface density in the NoRad galaxies is slightly higher than in the StarRad galaxies. The difference of the median values between the StarRad and NoRad galaxies is relatively small at the center, but becomes larger outside 10 kpc. For determining the 10 to 90 percentiles, note that the coverage within 10 kpc is generally lower for StarRad galaxies than for NoRad galaxies. Outside 10 kpc, the coverage is roughly the same except that the lower limit in the StarRad galaxies is slightly smaller.

To investigate the relative contributions of old and young stars to the surface density, we also show the surface density of old and young stars in the second and third panels of Figure~\ref{StarSurf1D}, respectively. As for the total, the solid lines in the second panel represent the median values of the surface density of old stars, while the shaded area indicates the 10 and 90 percentile values. For the median value, we find that the surface density of the NoRad and StarRad galaxies is similar to each other within 50 kpc. The median and 90 percentile values of the surface density in the StarRad galaxies are slightly lower than the NoRad galaxies within $\sim 15$ kpc, which is similar to the surface density of total stars. This is consistent with the finding that the NoRad galaxies are brighter in the red band at the center of the stellar images. However, in general, the differences in the surface density of the old stars between the NoRad and StarRad galaxies are rather small.

This is different for the surface density of young stars, shown in the third panel of Figure~\ref{StarSurf1D}. Here the median value for the StarRad galaxies is lower than for the NoRad galaxies over the entire 50 kpc range. This relative difference becomes larger when the radius increases. This result can explain the higher surface brightness in the blue band in the NoRad galaxies outside of 5~kpc in Figure~\ref{StellarImage}. The large spread between the 10 and 90 percentile values is here primarily induced by the large size variations of the five simulated galaxies. Since the stellar mass distribution in the outer regions is disk-like, these results imply that local stellar radiation is comparatively effective in suppressing the star formation in disk-like configurations. 

\begin{figure*}
   \subfigure{\includegraphics[width=0.9\textwidth]{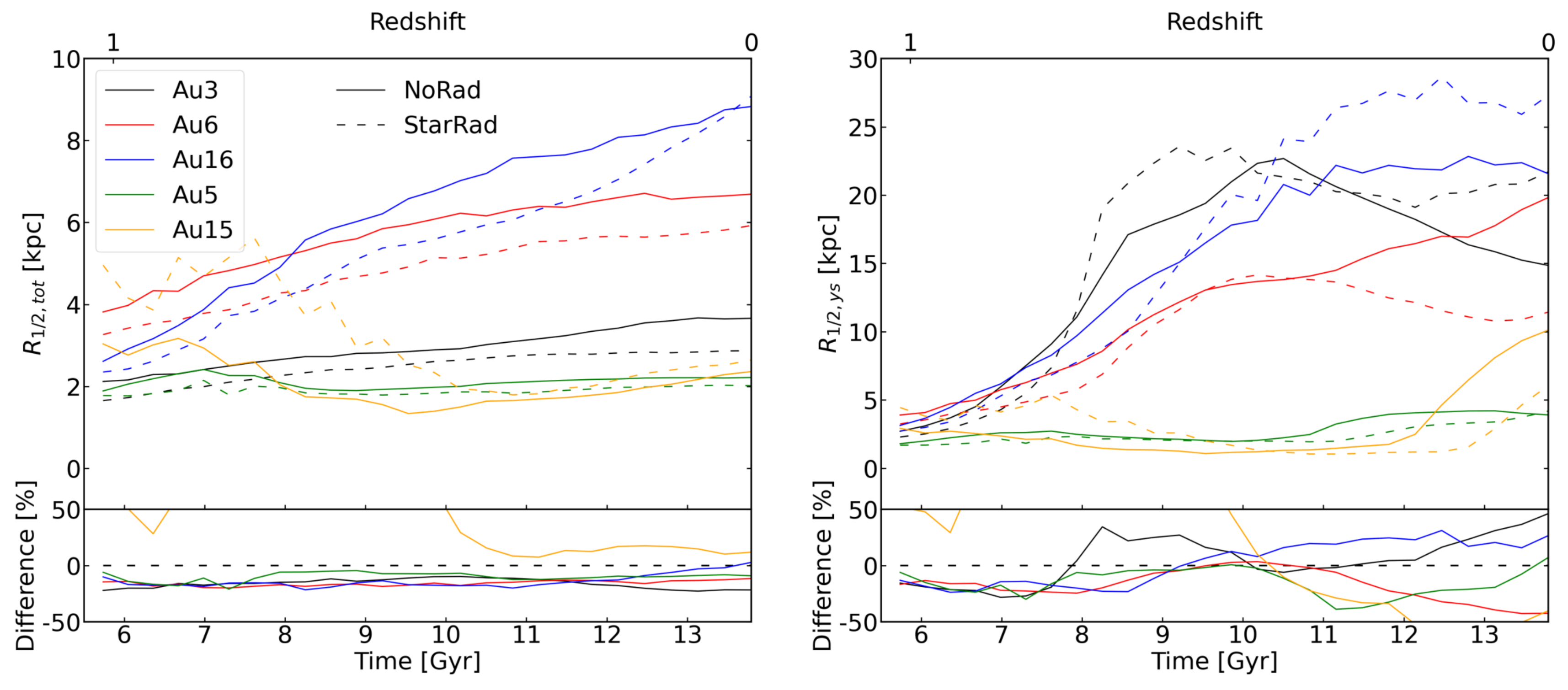}}
   \caption{Time evolution of the half-mass radius of total stars (left panel) and young stars only (right panel). The solid lines represent the simulation results for our NoRad runs, while the dashed lines give the results for StarRad simulations. The relative differences of the corresponding NoRad and StarRad runs is shown in the bottom panels.}
    \label{r0.5}
\end{figure*}

The above results confirm that local stellar radiation can efficiently suppress the stellar surface density. We further investigate its impacts on the concentration of the stellar distribution. To this end we have fitted the stellar surface density with Sersic profiles, using separate fits for the surface density of old and young stars. We note, however, that the fitting parameters can sometimes be sensitive to details of the fitting procedure, calling for caution in interpreting their precise values. This is also reflected in relatively noisy time evolutions of the Sersic index and the effective fit radius, which obscures clear trends. Instead of using a parametric fit, we therefore in the following characterize the concentration of the stellar mass through the stellar half-mass radius. This non-parametric measure avoids a sensitivity to details of the fitting procedure and thus tends to be more robust.

Figure~\ref{r0.5} shows the time evolution of the half-mass radius of all stars and  of the young stars, and of the relative difference in our five NoRad and five StarRad galaxies at $z<1$. The left panel of the figure gives the time evolution of the half-mass radius of all stars. Each of the simulated galaxies has a smaller half-mass radius of all stars in the StarRad variant compared with the NoRad variant, except for Au15. Recall that in the first panel of Figure~\ref{deltaM} we found that the formation history of Au15 is different from those of the other galaxies, making this galaxy a bit of an outlier. In particular, the stellar age of Au15 is younger than that of the other galaxies, and its star formation reaches its peak at $z<1$. So aside from Au15, the smaller half-mass radius of all stars in the other StarRad galaxies implies that local stellar radiation can increase the concentration of the galaxies. On other hand, through its different formation history Au15 corroborates again the result that the impact of local stellar radiation differs in the early- and late-stages of disk galaxy formation.

The second panel of Figure~\ref{r0.5} shows the time evolution of the half-mass radius of young stars. Unlike the half-mass radius of the total stars, the half-mass radius of young stars does not yield a consistent trend over the entire period. At 6 to 8~Gyr, it is smaller in the StarRad than in the NoRad galaxies -- except for Au15. After 8 Gyr, it does however not show an obvious and clear systematic trend any more for the full sample of simulated galaxies. Here the half-mass radius of young stars in the StarRad galaxies can be found to be both larger or smaller than in the NoRad galaxies. This may be due to the AGN feedback becoming important after 8~Gyr and beginning to affect the gas, especially in the inner regions.

\subsubsection{Gas disk}\label{sec:gasdisk}

\begin{figure*}
   \includegraphics[width=\textwidth]{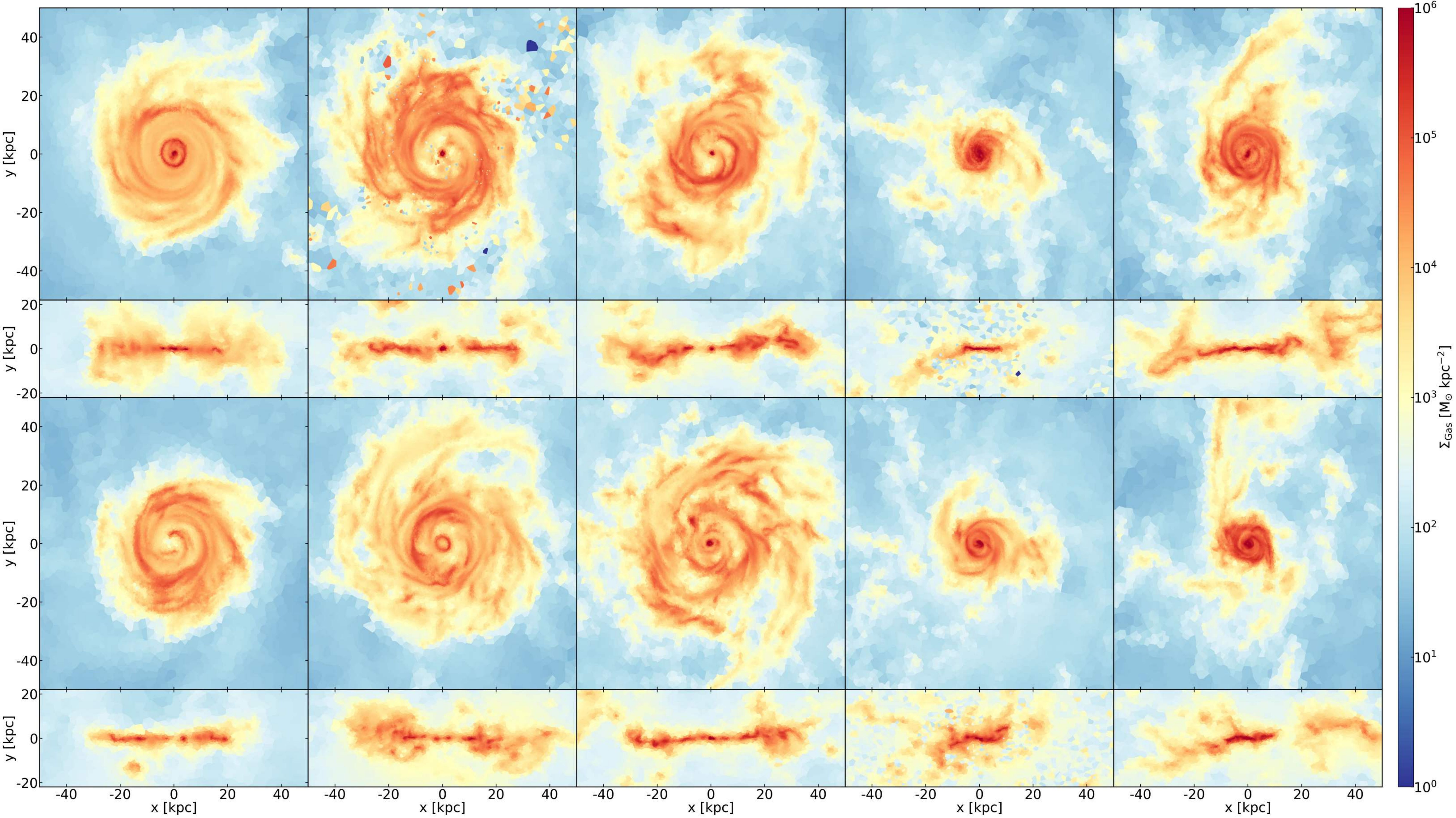}
   \caption{Maps of the total gas surface density of different Auriga galaxies at $z=0$. From the left to the right columns, we display  Au3, Au6, Au16, Au5, and Au15. The top row gives results for the  NoRad simulations, both in face-on and edge-on projections (large and small subpanels, respectively), while the bottom row shows the corresponding StarRad simulations.}
    \label{GasImage}
\end{figure*}

\begin{figure}
   \includegraphics[width=0.45\textwidth]{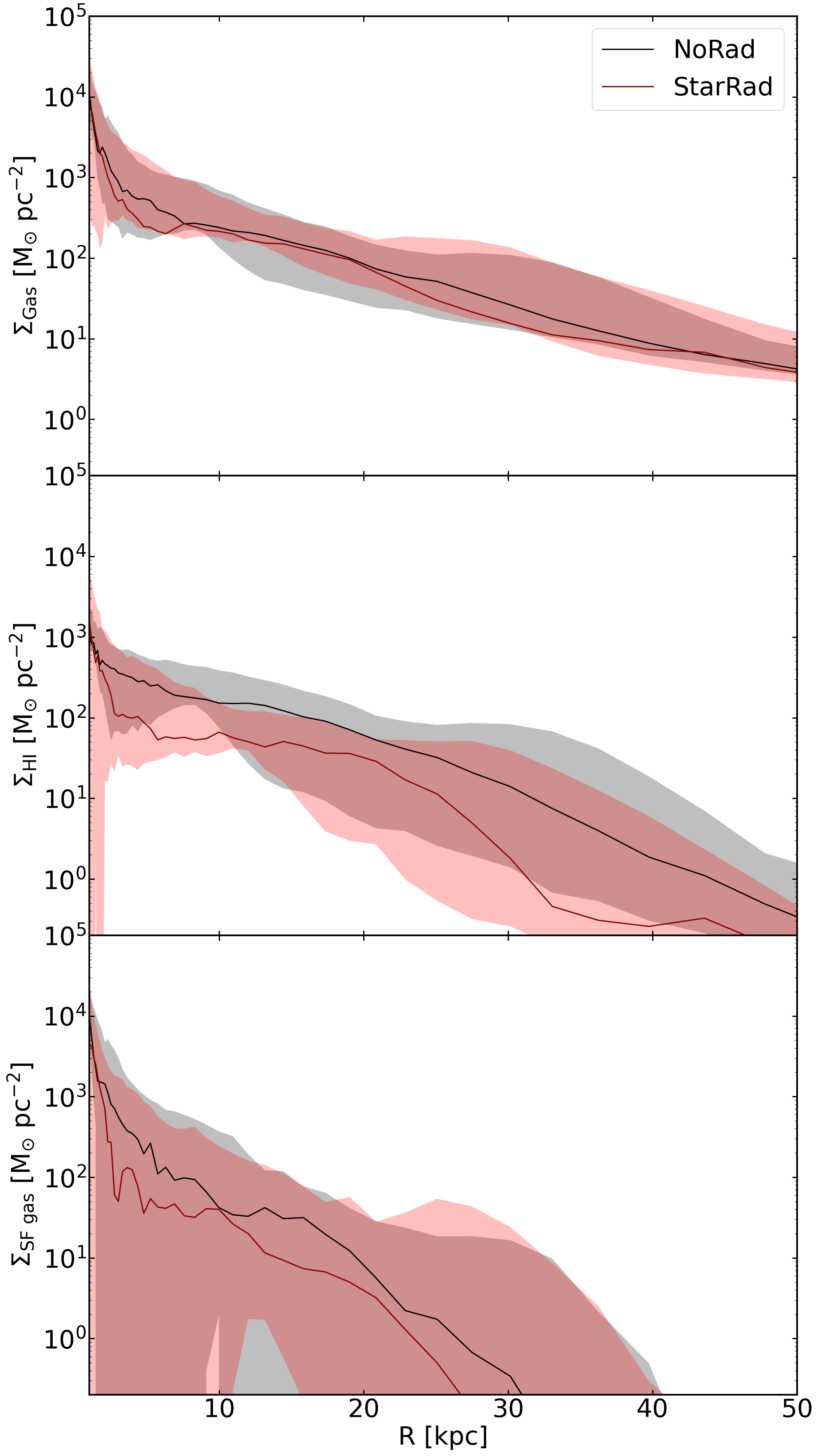}
   \caption{Radial surface density profiles of total gas (top panel), HI gas (middle panel) and just star-forming gas (bottom panels) of our sample of five Auriga galaxies, averaged for the cases with and without local stellar radiation (as labelled), and over time from $z=0.1$ to $z=0$. The shaded area shows the 10 to 90 percentiles of the measured values for the stacked simulation set, while the solid lines are the medians.}
    \label{GasSurf1D}
\end{figure}

\begin{figure*}
   \includegraphics[width=\textwidth]{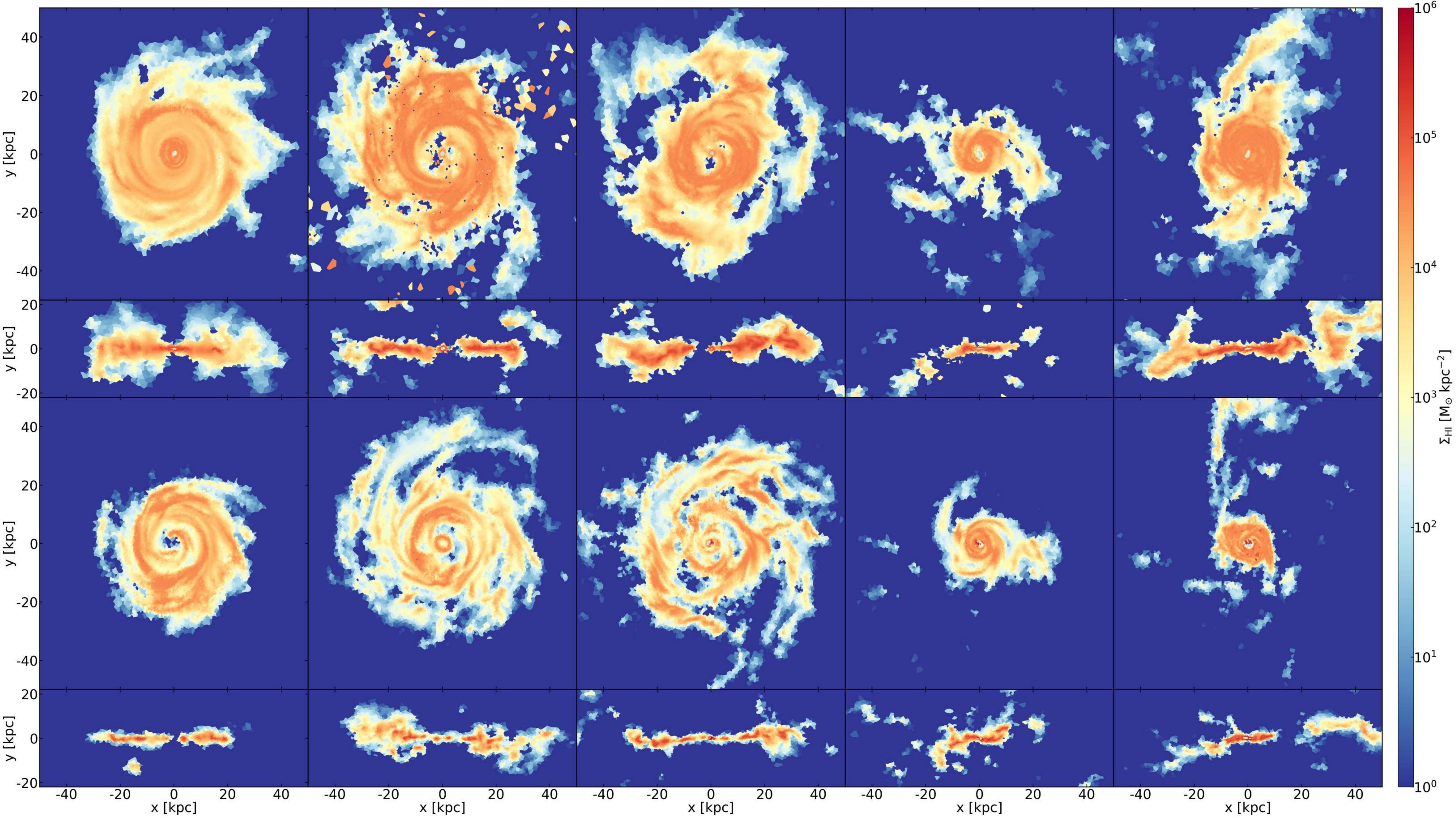}
   \caption{Same as Figure~\ref{GasImage}, but for the HI gas surface density.}
    \label{HIGasImage}
\end{figure*}

Figure~\ref{GasImage} shows the total gas surface density in different simulated galaxies at $z=0$. The first and second panels display the face-on and edge-on views of the total gas surface density in our five NoRad galaxies, while the third and fourth panels give the corresponding information for our five StarRad galaxies. The face-on views of the gas disk in the NoRad and StarRad galaxies highlight that they have similar morphology and size. The disk heights also do not indicate significant differences, based on comparing the edge-on views of both NoRad and StarRad galaxies. This result is consistent with the finding in the first panel of Figure~\ref{0.1Rvir_evo}, which shows that the median value of the total gas within $0.1\,R_{\rm vir}$ is similar in the NoRad and StarRad galaxies. Since the gas disks in both models have similar morphology and size, the angular momentum of the inflowing gas should also be similar, again providing indirect evidence that the local ionization feedback does only weakly affect the inflow and is not an overly effective preventive feedback process. 

To further investigate the differences of the gas disks in the NoRad and StarRad galaxies, we compare the radial profile of the total gas surface density in the first panel of Figure~\ref{GasSurf1D}. The profiles are stacked and averaged from $z=0.1$ to $z=0$. In general, the total gas surface density in the NoRad galaxies is higher than in the StarRad galaxies, although the difference is very small. This confirms the conclusion obtained by eye from the 2D projections of the total gas surface density. The slightly lower median value in the StarRad runs within $10\,{\rm kpc}$ occurs because local stellar radiation can heat the cold gas. As the gas temperature increases, the gas density needs to reduce to satisfy the pressure equilibrium.

Figure~\ref{HIGasImage} shows the HI gas surface density of different simulated galaxies at $z=0$. Similarly to Figure~\ref{GasImage}, the top row of panels gives the face-on and edge-on views of five NoRad galaxies, and the bottom row of panels is for our five StarRad galaxies. Unlike for the total gas surface density, the HI gas surface density exhibits significant differences between the NoRad and StarRad cases. The face-on view indicates that the HI disk of the NoRad galaxies features  a relatively smooth distribution. However, the HI disk in the StarRad galaxies appears  clumpy. In addition, the HI disk in the StarRad galaxies has a smaller size. From the edge-on view, we see that the HI disk in the StarRad galaxies has also a lower disk height. These results are consistent with the findings in \citet{obreja19}.

These differences are quantified in the stacked radial profiles of the HI gas surface density shown in the second panel of Figure~\ref{GasSurf1D}. From the figure we see that the HI gas surface density in the StarRad galaxies is systematically lower than in the NoRad galaxies. The median value at 5-25 kpc in the StarRad galaxies is around $10^2\,{\rm M_{\odot}\,pc^{-2}}\sim 10^{21}\,{\rm cm^{-2}}$, which is one order of magnitude lower than in the NoRad galaxies. Another difference is that the HI disk surface density drops rapidly at $\sim 25-35\,{\rm kpc}$ in the StarRad galaxies. These differences in HI morphology and surface density between NoRad and StarRad galaxies indicate that local stellar radiation can effectively ionize the HI gas in the disk, making the HI disk smaller, clumpier, and giving it a lower surface density. 

The final bottom panel of Figure~\ref{GasSurf1D} shows the stacked surface density of  star-forming gas in the NoRad and StarRad galaxies. We find that the surface density of star-forming gas is systematically lower in the StarRad galaxies compared with the NoRad galaxies, reflecting the results for the HI gas disk. This lower star-forming gas surface density of the StarRad galaxies can also explain the lower disk surface density of young stars seen in Figure~\ref{StarSurf1D}.

\subsection{Rotation curve}

\begin{figure}
   \subfigure{\includegraphics[width=0.45\textwidth]{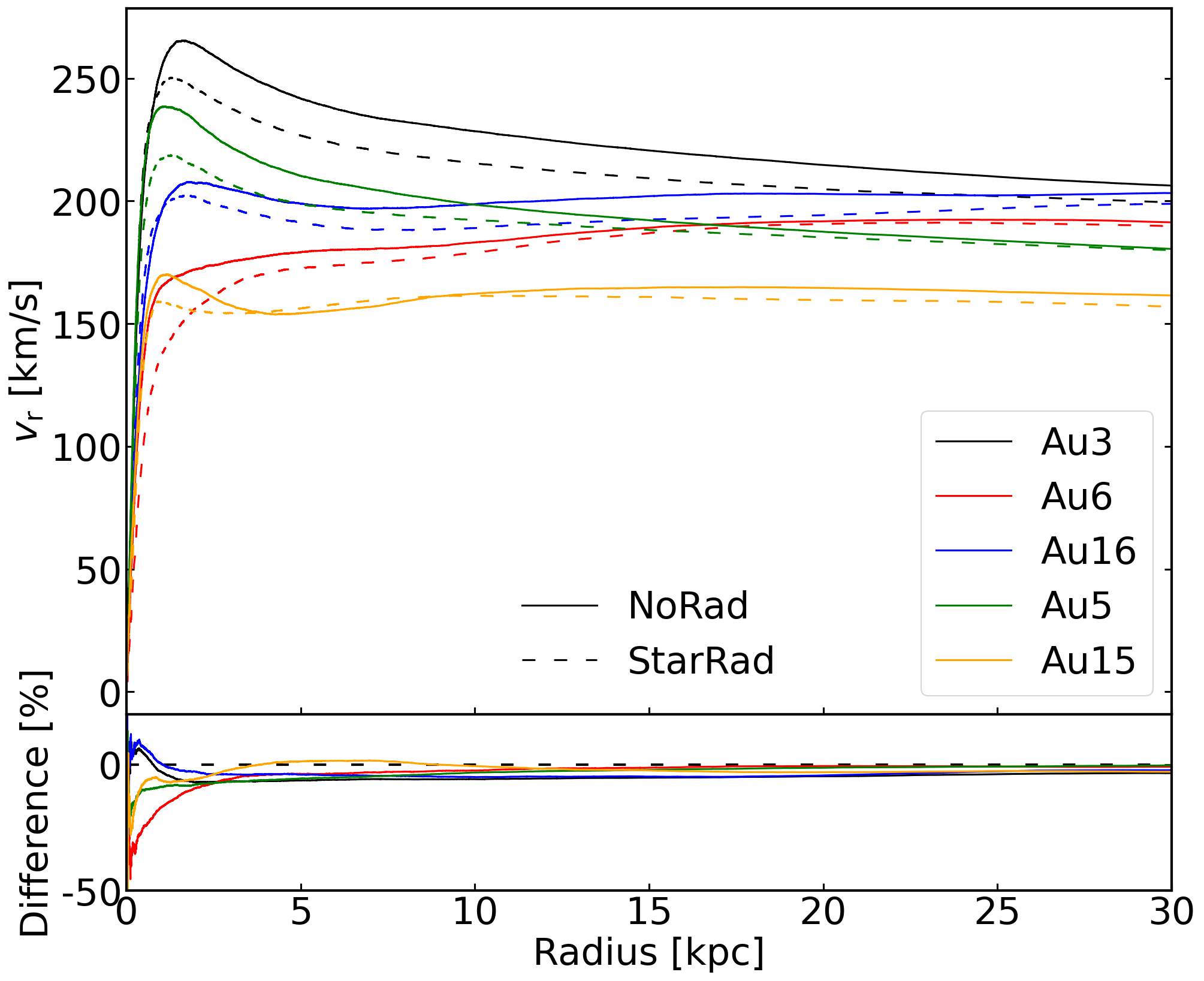}}
   \caption{Comparison of the $z=0$  rotation curve of our five simulated Auriga galaxies. Solid lines give the results for the NoRad runs, while dashed lines are for the corresponding StarRad simulations where local stellar radiation is included. The bottom panel shows their difference relative to the NoRad simulations.}
    \label{rc}
\end{figure}

Arguably one of the most notorious difficulties in simulating MW-like galaxies is the overcooling problem \citep{balogh01}. The low angular-momentum gas accreted into the DM halos at the early stages of galaxy formation needs to be removed to avoid forming too many stars in a central bulge \citep{navarro91, governato2004, scannapieco12}. One of the symptoms of the overcooling problem, if not sufficiently prevented, is the existence of a central peak in the rotation curve. \citet{kannan14} found in their simulations that local stellar radiation can effectively suppress this peak, indicating that local stellar radiation is a process that can contribute to solving the overcooling problem. Furthermore, \citet{obreja19} implemented local ionization feedback scheme in a comparable way to \citet{kannan14} in the NIHAO simulations, finding a similar effect on the rotation curve as in \citet{kannan14}. We here therefore also check the effect of the local stellar radiation on the rotation curve in our simulations.

Figure~\ref{rc} shows the averaged  $z=0$ rotation curve of our five NoRad and five StarRad galaxies, respectively. Compared to the NoRad case, the peak of the rotation curve shows some suppression. However, this suppression is relatively small compared to the results in \citet{kannan14} and \citet{obreja19}. The relative difference in the peak of the rotation curve between the NoRad and StarRad galaxies is less than $10\%$. The reason why the local stellar radiation in our work shows an overall weaker effect may be related to the star formation model. In \citet{kannan14} and \citet{obreja19}, the multiphase gas is in part directly resolved  even at high density. Their temperature and density thresholds for the star formation model are $15000~{\rm K}$ and $10~{\rm cm^{-3}}$, and the gas can still be affected by radiation even if it is star-forming. The density threshold for the star formation model in our work is instead $\sim 0.1~{\rm cm^{-3}}$, and the thermal state of the denser gas is set by an effective EOS, suppressing resolved multi-phase structure in the star-forming gas. While gas with temperature lower than $\sim10^{4.5}~{\rm K}$ is more vulnerable to local stellar radiation, once the gas becomes star-forming, its effective temperature is no longer affected by local stellar radiation in our model, except for our added criterion that gas with high ionization parameter will not be allowed to form stars. Especially at the center of galaxies, the gas density is relatively high, and we expect only mild effects from this ionization parameter criterion. This may be the main reason why the suppression effects of local stellar radiation on the peak of the rotation curve are relatively weak in our simulations.

\section{Structure of the circum-galactic medium} \label{Sec:CGM}

\begin{figure}
   \includegraphics[width=0.45\textwidth]{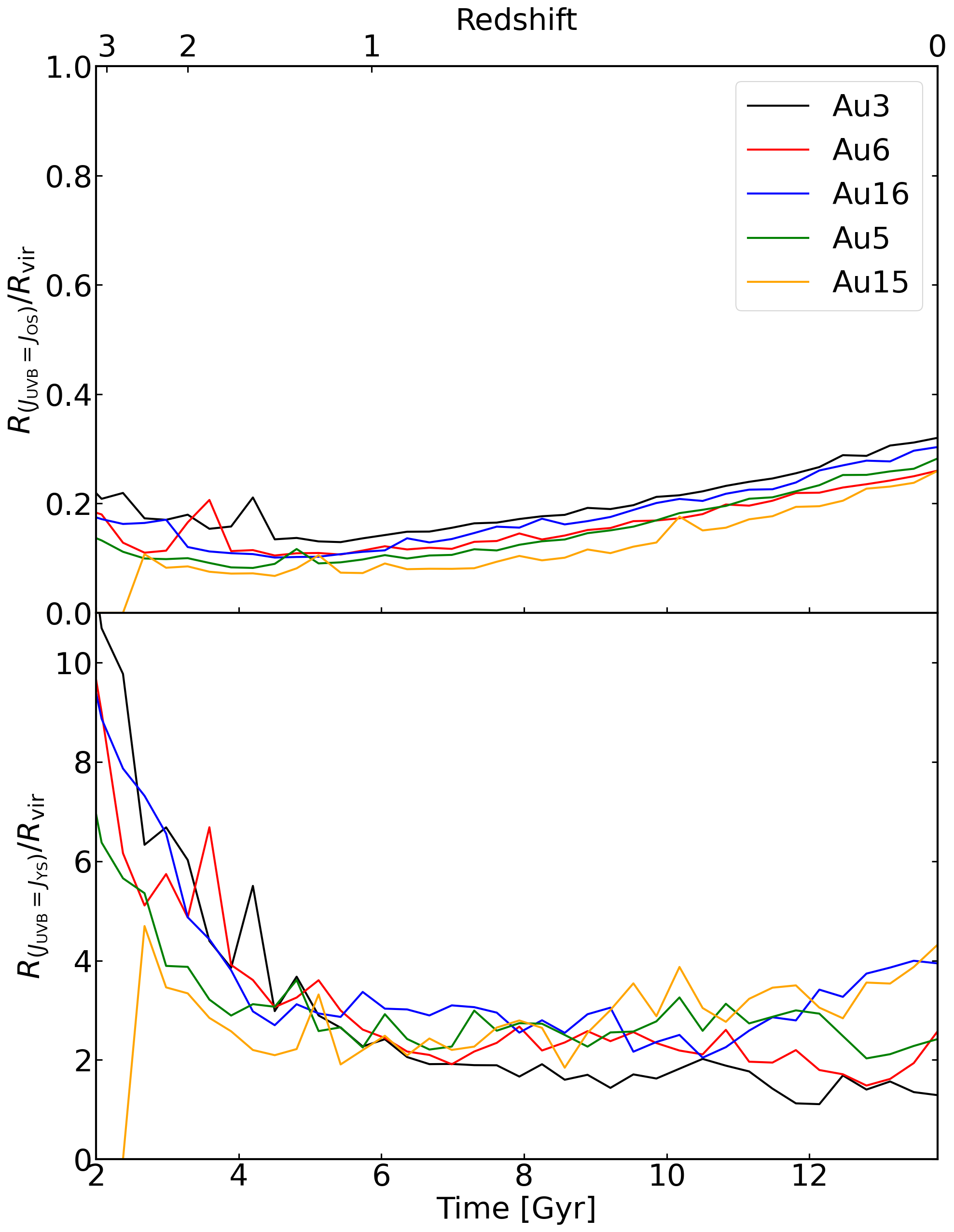}
   \caption{Time evolution of the {\it physical} radius in units of the virial radius where the old and young stellar radiation intensity equals the ambient UVB intensity, separately shown for our five simulated Auriga galaxies. The bottom panel is for the old stellar component, whereas the bottom panel is for the young stars. Note the different vertical scale in the two panels. Unlike the radiation field from the old stars, the young stellar component can dominate the UVB throughout the halo, and even beyond.}
    \label{UVBRadius}
\end{figure}

Previous work studying the impact of radiation on CGM gas usually considered the influence of a UV radiation background, whereas only few studies considered the radiation from the galaxies themselves \citep[e.g., ][]{Suresh17, Oppenheimer18, nelson18}. However, the radiation from nearby stars may play a significant role for the CGM region.

To illustrate this point, Figure~\ref{UVBRadius} shows the time evolution of the radius where the young/old stellar radiation is equal to the UVB intensity, in units of the virial radius. For old stars as sources, this radius gradually increases with time, due to the monotonic growth of the stellar mass and the decrease of the UVB intensity at $z<3$. The radius varies between $0.1-0.4\,R_{\rm vir}$ over time, which means that the stellar radiation of old stars is comparable of higher than the UVB only in the innermost regions of the CGM. Furthermore, since the older stellar populations emit few ionizing photons, this significantly limits the overall impact this radiation component can have on the CGM gas.

For younger stars, on the other hand, the radius where their influence dominates gradually decreases with time. It evolves from $\sim 10$ times the virial radius to $\sim 2$ times the virial radius from $z=3$ to $z=0$, which reflects that  the star formation history reaches a peak at $z\sim3$ and then decreases with time. This result implies that within the whole CGM gas the radiation field is actually dominated by the young stellar radiation from the central galaxies instead of the UVB. Unlike for the  radiation from old stars, the fraction of ionizing photons in the young stellar radiation can reach 23\%, indicating that this radiation component can potentially have a significant impact on the CGM gas, even outside the halo itself. In this section, we will thus analyze the influence  of the local stellar radiation on the CGM properties. 

\subsection{Radial profiles of gas properties}

\begin{figure}
   \includegraphics[width=0.45\textwidth]{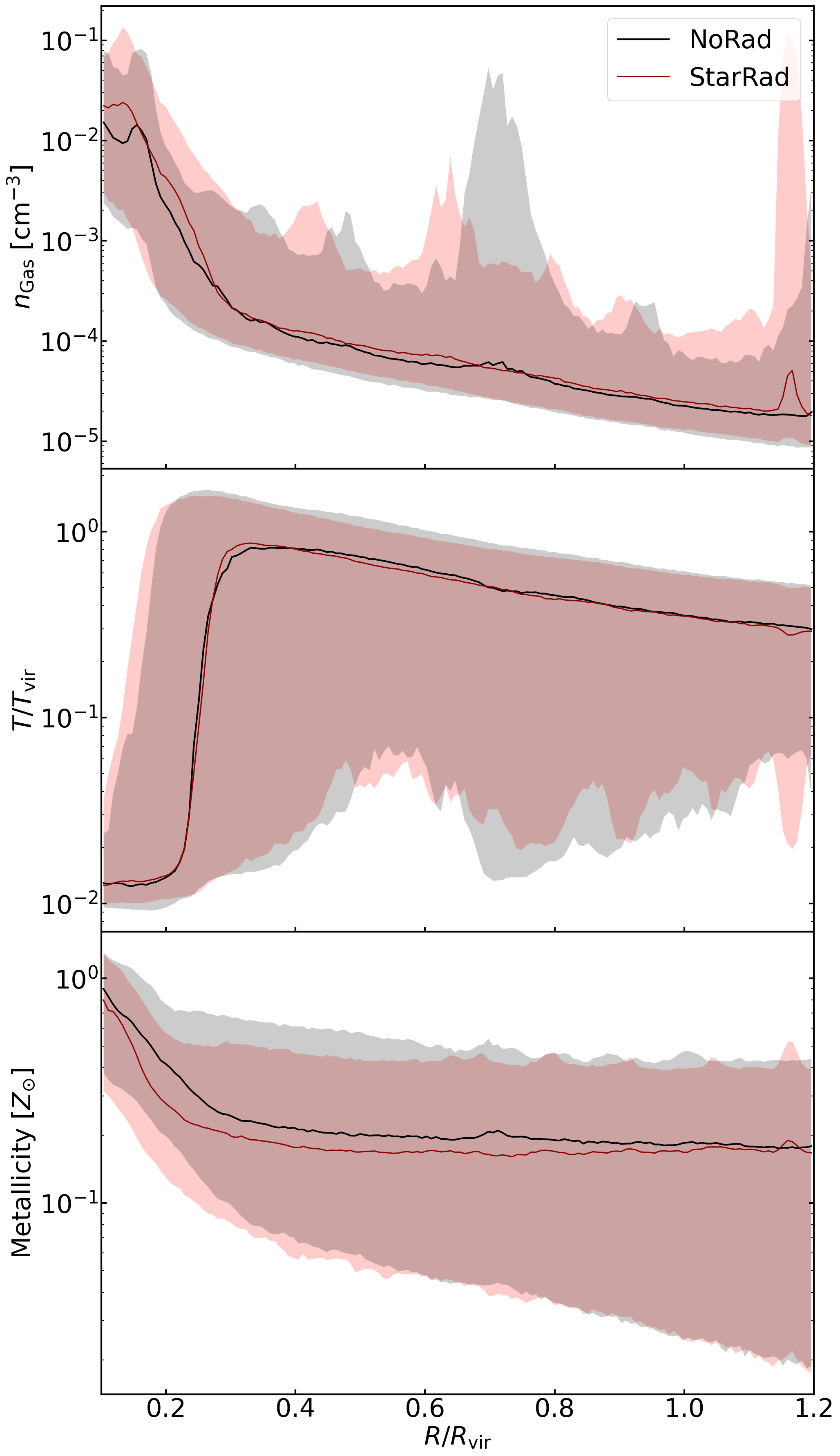}
   \caption{Radial profiles of number density (top panel), temperature (middle panel) and metallicity (bottom panel) of our Auriga galaxies, stacked for the different models simulated with and without radiation (as labelled) and averaged from $z=0.1$ to $z=0$ to reduce the influence of temporal and stochastic fluctuations. The shaded areas show the 10 to 90 percentile over the simulation set, while the solid lines give median values.}
    \label{CGM1D}
\end{figure}

We begin by studying the impact of local stellar radiation on the global structure of the gaseous halos. Figure~\ref{CGM1D} gives the stacked radial profiles of the gas number density, temperature, and metallicity, in both the NoRad and StarRad galaxies. Only non-starforming gas is included, i.e.~the ISM component is excluded. The data are stacked for all galaxies with the same physical model, and are averaged over $z=0.1$ to $z=0$ to eliminate temporal fluctuations. The solid line represents the median value of the individual measurements, while the shaded area indicates the 10\% to 90\% percentiles.

The number density profile, shown in the first panel of Figure~\ref{CGM1D},
can evidently be separated into two characteristic regions. Within $0.3 \,R_{\rm vir}$, the number density is still comparatively high, reflecting contributions from the extended gas disk. Outside $\sim 0.3 \, R_{\rm vir}$, the number density then quickly drops to $10^{-4}-10^{-5}\, {\rm cm^{-3}}$. Outside this radius of $\sim 0.3\,R_{\rm vir}$, the differences in the number density profiles between the NoRad and StarRad galaxies appear negligible. Some of the discrete peaks in the shaded region are due to cold gas debris or satellites, and these features would be expected to disappear when averaging over a still larger galaxy sample. In the radial range $0.1-0.3\,R_{\rm vir}$, the StarRad galaxies have a slightly higher number density than the NoRad galaxies. Note that this corresponds to radii of $\sim 20-60\,{\rm kpc}$ in our simulated galaxies. For the total gas surface density shown in the first panel of Figure~\ref{GasSurf1D}, the StarRad galaxies have a slightly lower value than the NoRad galaxies in the corresponding region. This difference arises from the omission of the star-forming gas in this profile. Since the local stellar radiation prevents some gas from becoming eligible for star-formation, the star-forming gas itself will have a lower density, and the median value of the gas number density in the StarRad galaxies is expected to be  lower than in the NoRad galaxies.

The second panel of Figure~\ref{CGM1D} shows the radial profile of the temperature. Unlike the number density, the median values of the temperature profiles show good consistency for the entire CGM region. While this result suggests that the local stellar radiation does not have a strong heating effect in the CGM region, it should be noted that the temperature tends to closely track the virial temperature of the halo even in the presence of strong heating, and it is set ultimately by gravity. Since the virial temperature of our simulated halos is $\sim10^6\,{\rm K}$, the coolants are dominated by ${\rm Ne^{\rm 5+}}$ and ${\rm Fe^{\rm 8+}}$ \citep{wiersma09}. Note that to photoionize these ions, the energy of the photons should be $\sim100-300\,{\rm eV}$, which are lacking in our model of the local stellar radiation. Nevertheless, there are still some indications in our result that local stellar radiation influences some aspects of the CGM gas. In particular, the lower limit of the shaded region is  slightly higher the StarRad galaxies than for the NoRad galaxies. This can be understood due to the ability of the local stellar radiation to heat  gas with temperature lower than $\sim10^{4.5}~{\rm K}$.

Finally, the third panel of Figure~\ref{CGM1D} shows the radial profile of the metallicity. From the figure we can see that the metallicity in the StarRad galaxies is lower than that in the NoRad galaxies over the entire CGM region. The difference decreases with increasing radius and disappears at $\sim R_{\rm vir}$. This radially dependent difference suggests that it originates in the center of the galaxy. Recall that Figure~\ref{GasVelHist} has shown that local stellar radiation damps the gas outflow, which is due to a suppression of star formation and thus a reduction of feedback energy injection. Because the gas outflows become weaker, they carry fewer metals into the CGM region.

\subsection{CGM phases by mass}

\begin{figure}
   \includegraphics[width=0.45\textwidth]{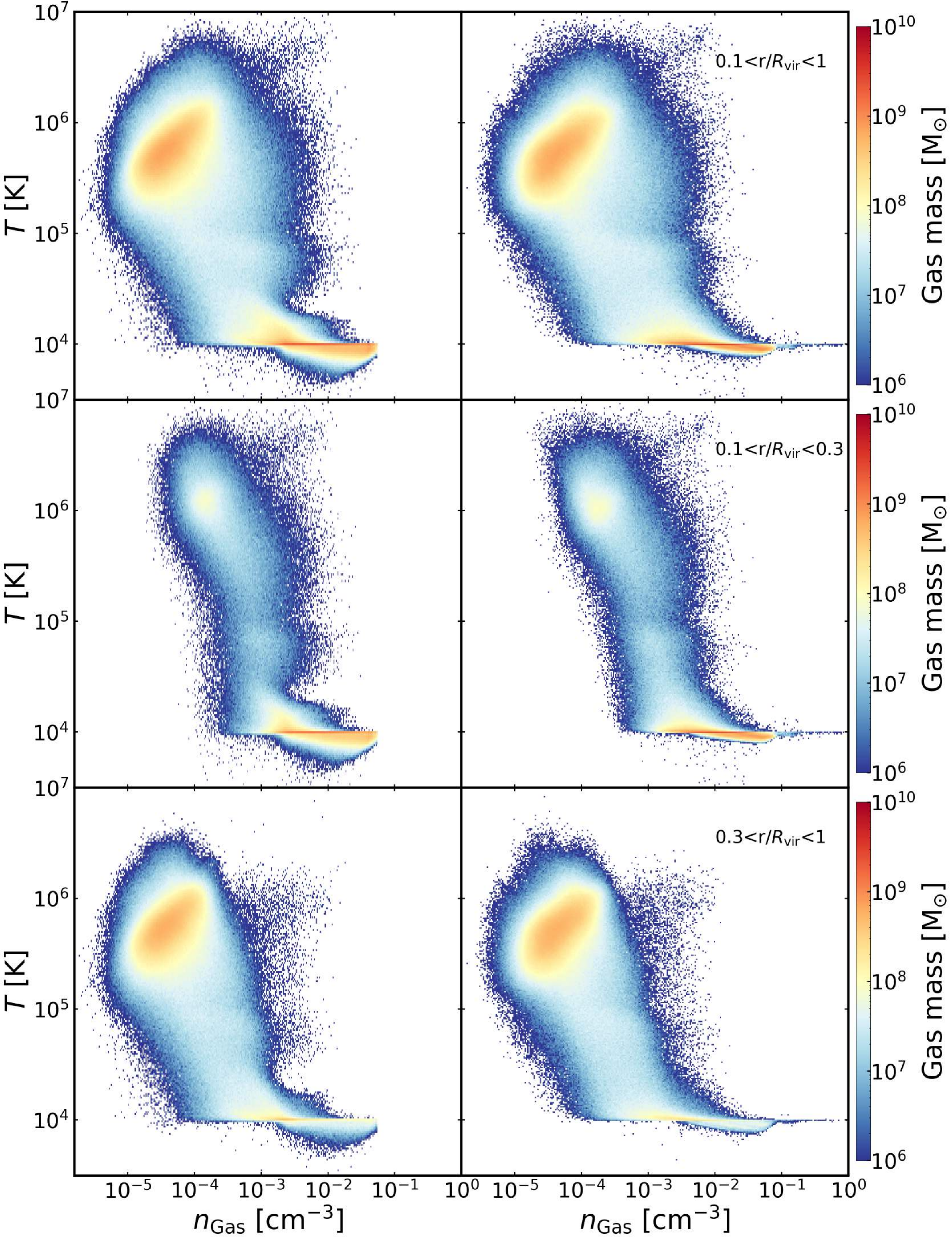}
   \caption{The temperature--number density distributions of CGM gas ($0.1<r/R_{\rm vir}<1$) in the upper row, inner halo gas  ($0.1<r/R_{\rm vir}<0.3$) in the middle row, and outer halo gas ($0.3<r/R_{\rm vir}<1$) in the bottom row. The left column represents the results from the NoRad simulations, while the right column gives the results of the StarRad simulations that include local stellar radiation. The data is stacked for all five simulated Auriga halos, and over outputs from $z=0.1$ to $z=0$.}
    \label{CGMGasPhase}
\end{figure}

\begin{figure}
   \includegraphics[width=0.45\textwidth]{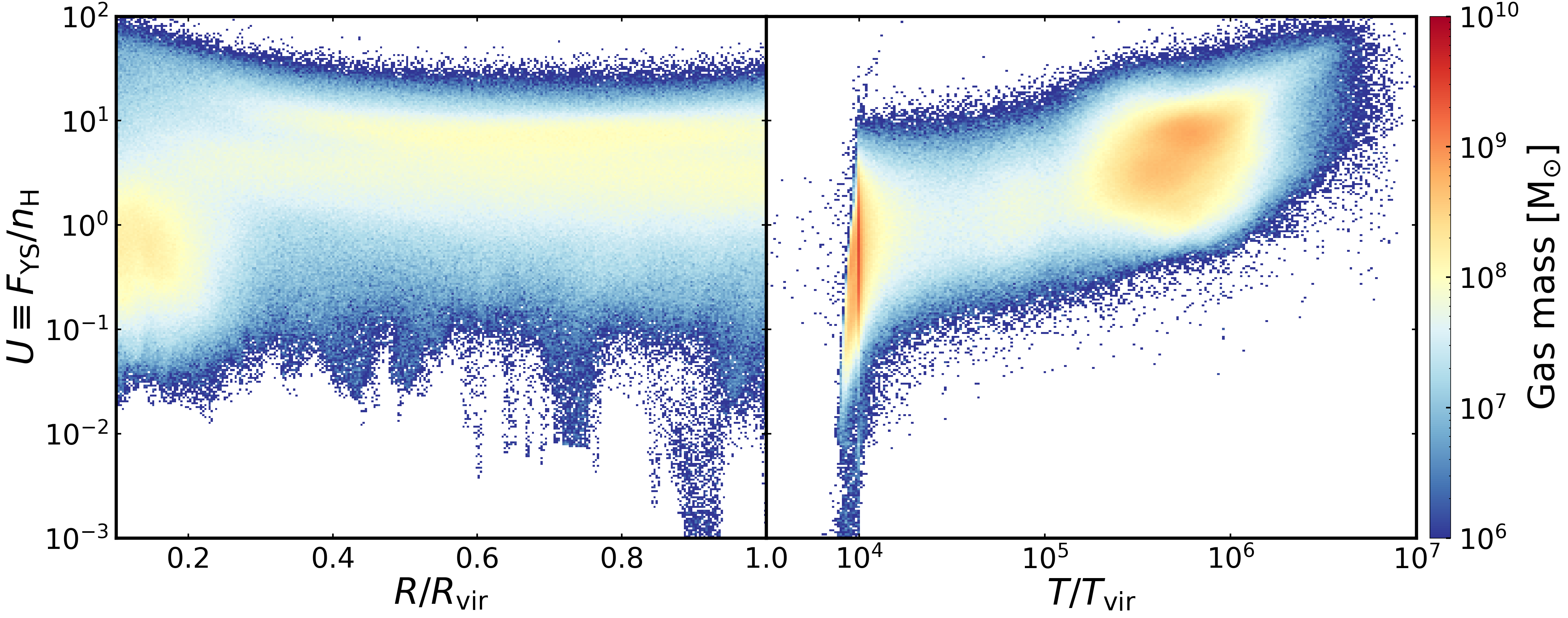}
   \caption{{\it Left panel}: Distribution of the ionization parameter due to the young stellar radiation as a function of radius in the StarRad simulations. {\it Right panel}: The distribution of the ionization parameter of the CGM gas due to young stars as a function of temperature. The color key indicates the amount of gas mass in the corresponding region of the distribution.}
    \label{CGMUTPhase}
\end{figure}

In the following we discuss the amount of gas in different CGM phases. To this end we divide the CGM into two regions: the inner halo ($0.1<r/R_{\rm vir}<0.3$) and the outer halo ($0.3<r/R_{\rm vir}<1$).  Figure~\ref{CGMGasPhase} shows the gas number density vs temperature distribution of the whole halo, and separately for the inner and outer halo, both in the NoRad and StarRad galaxies. As done earlier, we again combine the data of all galaxies with the same physics model, and average over the time span $z=0.1$ to $z=1$ to obtain statistically more robust results. The star-forming gas of the ISM is excluded.

In the top row of Figure~\ref{CGMGasPhase}, we show the gas number density-temperature distribution of the full CGM gas mass. Two distinct regions are readily apparent. In the low-density, high-temperature phase, the gas mass distribution is similar in both the NoRad and StarRad galaxies. The number density $n$ and temperature $T$ are related as $n\propto T^{3/2}$, which is an isentropic line. This happens because gas, after passing through the virial shock,  is adiabatically compressed when flowing to the halo center. The similar gas mass in this region for the NoRad and StarRad galaxies indicates that our model of local stellar radiation does not have a significant impact on this virialized gas component. On the other hand, the high-density, high-temperature region shows a more obvious difference between the NoRad and StarRad galaxies. The gas in the NoRad galaxies can reach temperatures of $\sim 10^{3.5}\,{\rm K}$, while this phase disappears in the StarRad galaxies. Note that gas with number density $n\la 0.1\,{\rm cm^{-3}}$ is missing in the NoRad galaxies (because this is all star-forming), whereas non-starforming gas denser than this threshold value can still be present in the StarRad galaxies, since gas with ionization parameter higher than unity is not eligible for star-formation in our model. 

We now divide the gas into inner and outer halo regions and investigate the gas distribution in the $n - T$ plane separately. The panels of the second and third row show the gas $n$-$T$ diagram in the inner halo and outer halo, respectively. Both phases exhibit a wide temperature range. The density in the inner halo is larger than $10^{-4}\,{\rm cm}^{-3}$, and most of the gas mass actually  resides in the low temperature regime. For the outer halo gas, the density is significantly lower, and a significant amount of the gas resides at the virial temperature. Although the total amount of cold gas in this region is low, it still shows a similar behavior as in the inner region, implying that local stellar radiation can still affect the gas even outside $0.3\,R_{\rm vir}$.  

The ionization parameter $U$ is a physical quantity that can measure the strength of the radiation effects. To understand the impact of local stellar radiation on different phases of the gas, we show the ionization parameter of the gas at different radii and for different temperatures in Figure~\ref{CGMUTPhase}. The left panel of Figure~\ref{CGMUTPhase} gives the gas mass distribution for different ionization parameters and radii in the StarRad galaxies. In the inner region, the gas mass experiences mainly  $U=0.1-1$, even though the gas is close to the radiation sources, but this effect is counter acted by the high gas density in this region. For radiation from the young stars, its heating effect becomes important  when the gas ionization parameter is greater than 1, and the temperature is lower than $10^{4.5}\,{\rm K}$. This result implies that local stellar radiation will have a mild heating effect on the inner halo gas. For the outer halo, it is interesting that the ionization parameter of the gas is around 10 and is approximately independent of radius. This is because the radiation flux decreases with $r^{-2}$ in the outer region, and the density happens to  decrease with a similar trend, making the ionization parameter roughly constant. 

The right panel of Figure~\ref{CGMUTPhase} shows the gas in the $T-U$ diagram. In the figure, most of the hot gas has a typical ionization parameter $U\sim1-10$, while most of the cold gas resides at $U\sim 0.1-10$. Although the ionization parameter of the hot gas is higher, local stellar radiation does not have a significant heating effect on it. Thus, the high ionization parameter does not modify the thermal state of the hot gas. The cold gas, on the other hand, is much more vulnerable to radiation. The presence of cold gas with high ionization parameter indicates that there is a significant amount of cold gas that is affected by local stellar radiation. Compared to the left panel, the total cold gas mass with a high ionization parameter is higher than that of the inner halo, implying that a certain amount of radiation-dominated cold gas resides throughout the entire CGM region.

\subsection{The abundance of different ion species}

Although our model of local stellar radiation does not have a significant heating effect on the hot gas, it may still affect its ionization state. In this section we therefore investigate the impact of local stellar radiation on different ionic species. HI, MgII, and OVI are three particularly important ions in CGM observations. Since they trace the cold gas within the CGM, they probe the outflow of cold gas from galaxies, as well as the associated metal enrichment. We thus focus on them in the following. 

\subsubsection{Column density}

\begin{figure}
   \includegraphics[width=0.45\textwidth]{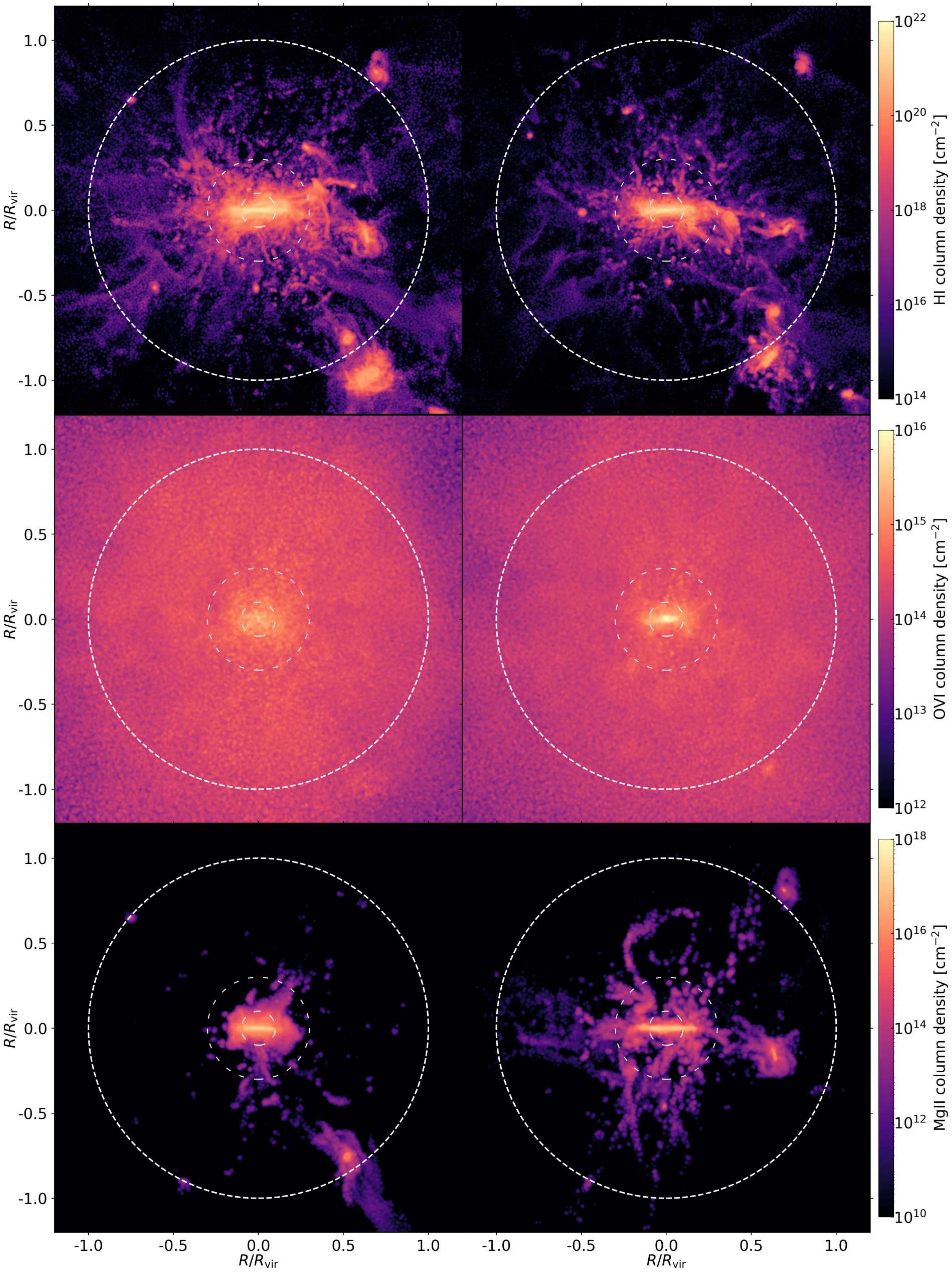}
   \caption{Stacked column density maps of HI (top row), MgII (middle row) and OVI (bottom row) gas of differently simulated Auriga galaxies. The left column gives results from our NoRad simulations, while the right column shows the corresponding results for our StarRad simulations which included local stellar radiation.}
    \label{IonColDens2D}
\end{figure}

\begin{figure}
   \includegraphics[width=0.45\textwidth]{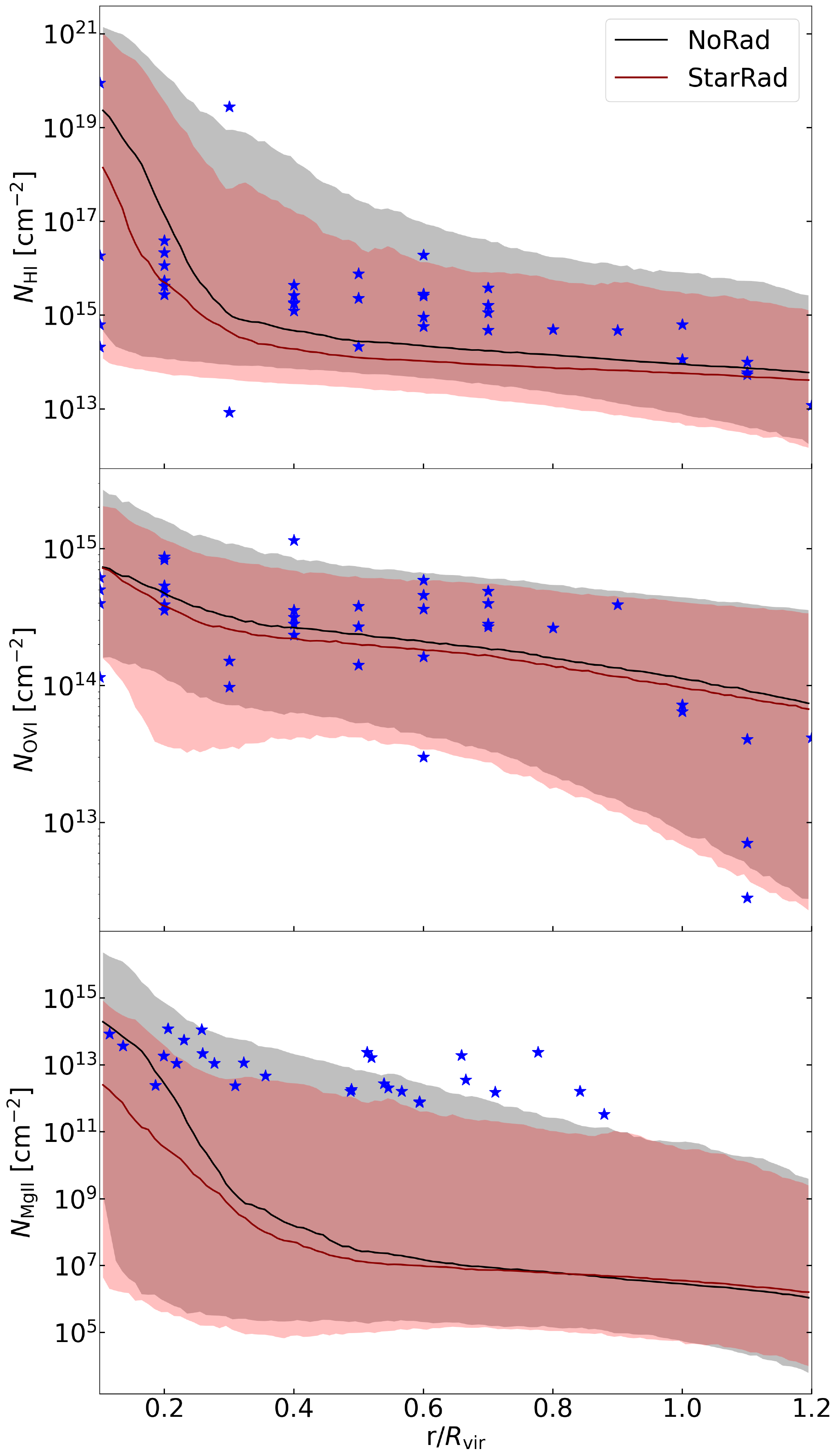}
   \caption{The radial profile of stacked HI, MgII and OVI column density distributions of our different Auriga galaxies, averaged from $z=0.1$ to $z=0$. The blue star symbols represent the observational results, coming from \citet{Johnson2015} for HI and OVI column densities, and
from \citet{Werk2013} for observed MgII column densities.}
    \label{IonColDens1D}
\end{figure}

Figure~\ref{IonColDens1D} shows stacked images of the HI, OVI and MgII column density in the galaxies with the same physics model, averaged in time over the outputs from $z=0.1$ to $z=0$. The three dashed lines in the figure indicate radii equal to 0.1, 0.3 and $1.0\, R_{\rm vir}$. The left column gives the result for the NoRad galaxies, while the right column is for the StarRad galaxies. Compared to the NoRad galaxies, the column density of HI gas is systematically lower in the StarRad galaxies. Since local stellar radiation can easily photoionize the HI gas, this result is to be expected. For the OVI gas, the column density in both the NoRad and StarRad galaxies is however quite similar, apart from a difference in the central column density, which is slightly higher in  the StarRad galaxies than in the NoRad galaxies. For the MgII gas, the NoRad galaxies feature a uniform distribution within $0.3\, R_{\rm vir}$, whereas the StarRad galaxies exhibit a more fragmented distribution. This can be explained by the patchy photoionization effects from the local stellar radiation.

To compare the differences in the column density in the NoRad and StarRad galaxies more quantitatively, we show the radial profile of the column density of different ions in Figure~\ref{IonColDens1D}. The first panel gives the radial profile of the HI column density. From the figure, we can see that the HI column density of the NoRad galaxies is higher than that of the StarRad galaxies, which confirms the conclusion we obtained by eye from Figure~\ref{IonColDens2D}. Inside $\sim$ 0.3 $R_{\rm vir}$, the simulation matches the observations well. However, recent 21-cm observations \citep{das2024} show that the HI gas column density within ~0.5 $R_{\rm vir}$ declines from $\sim5\times10^{18}~{\rm cm^{-2}}$ at $r\sim 0.1R_{\rm vir}$ to $\sim 10^{17}~{\rm cm^{-2}}$ at $r\sim 0.5R_{\rm vir}$, which is higher than in our simulations. In addition, outside $\sim$ 0.3 $R_{\rm vir}$, the observational data lies within the 10 and 90 percentile regions for both the NoRad and StarRad galaxies, but their median value is lower than the observations. Both lower estimates of the HI column density may be due to selection effects, or it may reflect resolution limitations of the simulations \citep{hummels19,peeples19, nelson20,vdv19}. 

The second panel shows the OVI column density in both the NoRad and StarRad galaxies. Similar to our observation in Figure~\ref{IonColDens2D}, the OVI column densities in both the NoRad and StarRad galaxies do not display a significant difference. The OVI column density in the StarRad galaxies is slightly lower than that of the NoRad galaxies, but the relative difference is only around 20\%. The OVI shows an overall good consistency with observations.

Finally, the third panel gives the MgII column density in both the NoRad and StarRad galaxies. As for HI, the MgII column density in the StarRad galaxies is lower than in the NoRad galaxies. MgII also traces the cold gas, and since local stellar radiation can heat or photoionize the cold gas, it can modify the abundance of this ion. The observational data resides in the upper range of the shaded region, indicating a  tension between the observations and the simulations. However, since current instrumentation can only observe luminous MgII sources, the observational data may be affected by a significant selection biases. From this point of view, it appears reasonable that the observational data reside in the upper parts of the shaded region. But it is also quite plausible that out results are affected by limited resolution, preventing us from resolving cool outflows accurately \citep{vdv19, hummels19, peeples19, nelson20}. Also note that \citet{ji20} found that including cosmic rays in the CGM can increase the MgII column density substantially, which may be another way to reproduce the observed MgII column densities.

\subsubsection{Ion mass phase diagram}

\begin{figure*}     \subfigure{\includegraphics[width=0.48\textwidth]{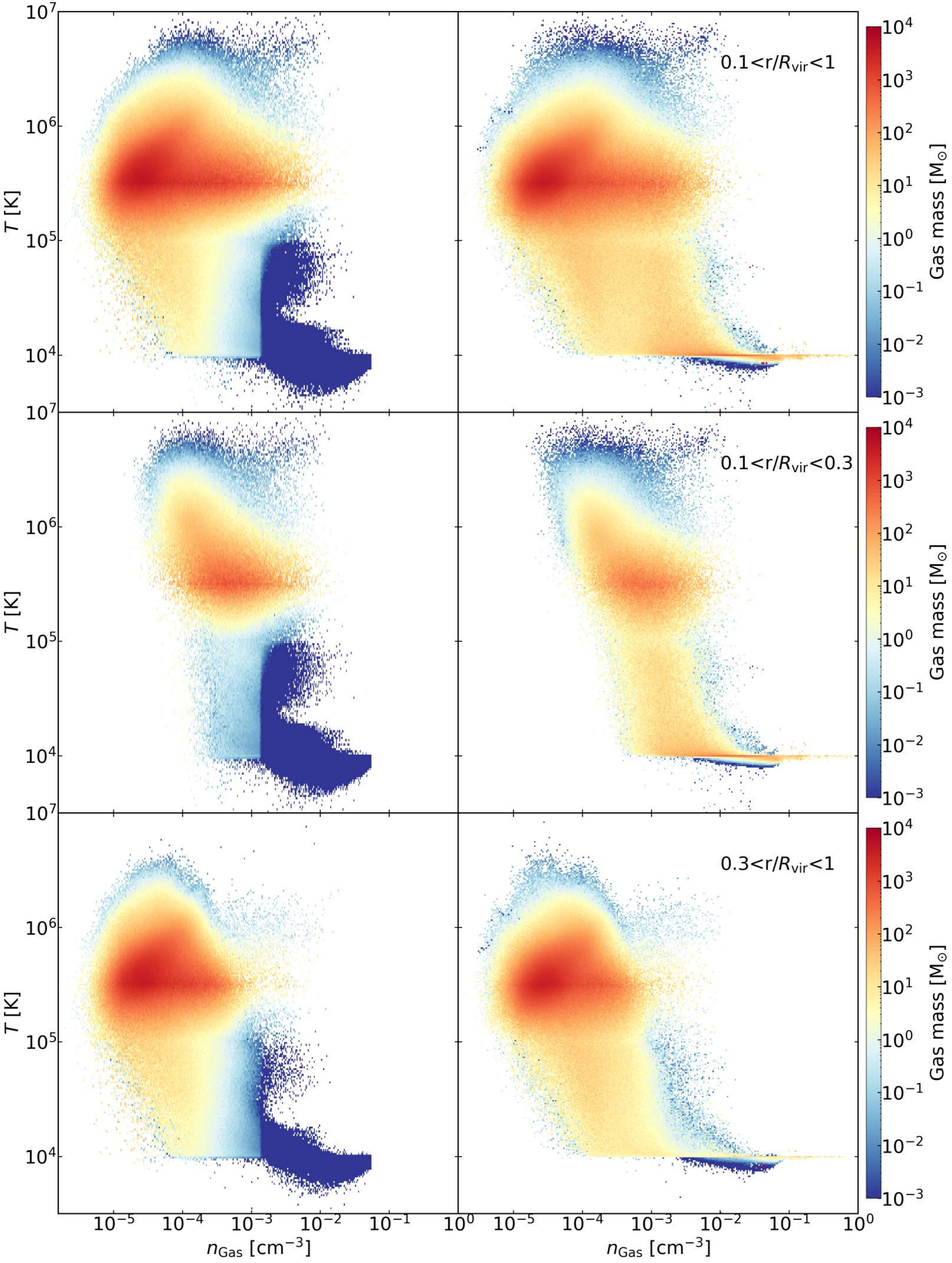}}%
   \subfigure{\includegraphics[width=0.48\textwidth]{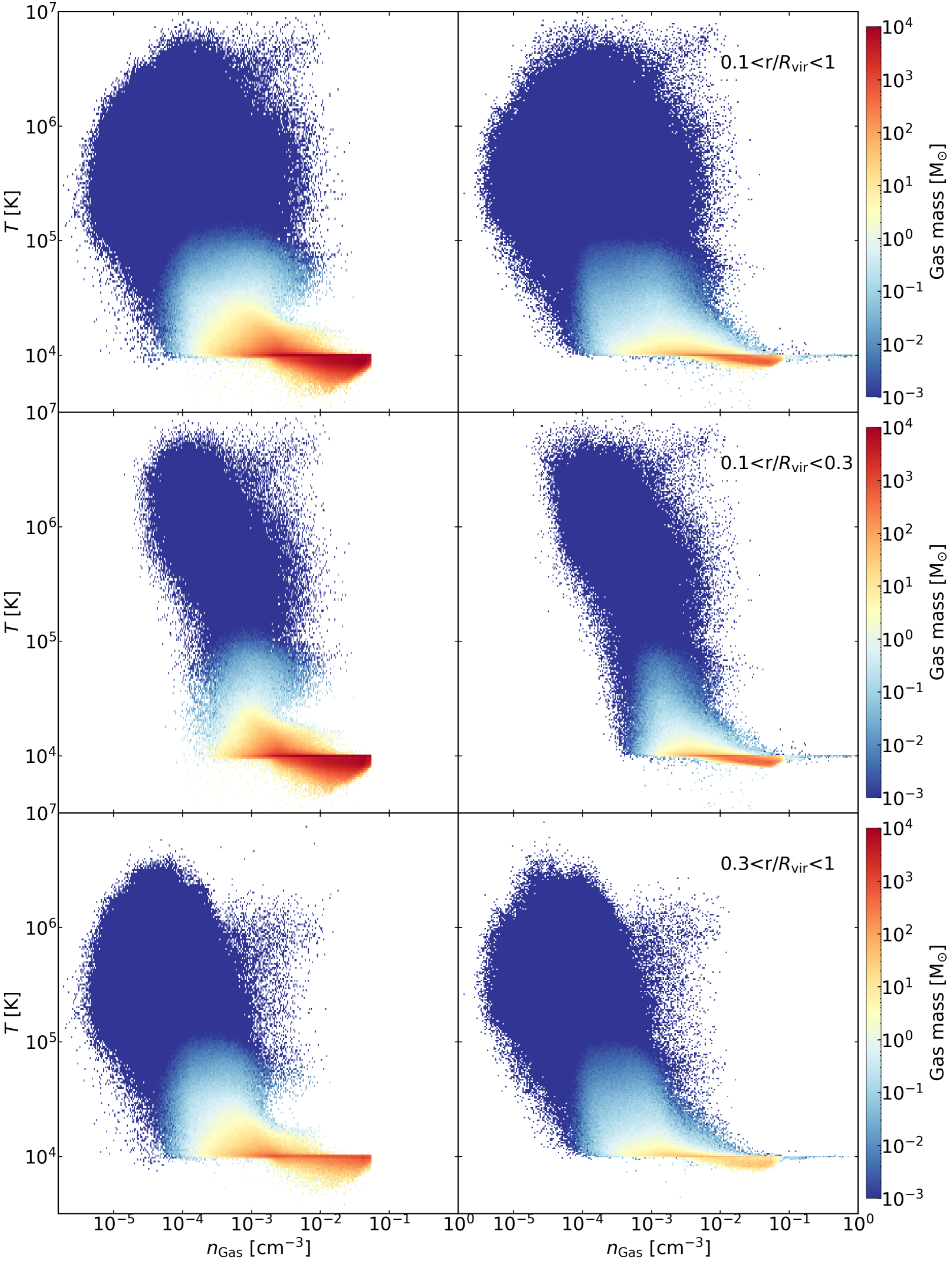}}
   \caption{The temperature--number density distribution for OVI gas (left) and MgII gas mass (right). In both cases, the upper, middle and bottom rows show the phase distribution of the CGM gas ($0.1<r/R_{\rm vir}<1$), inner halo gas ($0.1<r/R_{\rm vir}<0.3$) and outer halo gas ($0.3<r/R_{\rm vir}<1$), respectively, while the left column is for our NoRad simulations and the right column for the StarRad simulations that include local stellar radiation. The corresponding distribution for HI gas shows only minor differences, and is therefore left out for clarity.}
    \label{IonNTPhase}
\end{figure*}

Ions can be produced or destroyed by photoionization or collisional ionization. The collisional ionization process becomes more important as the temperature increases, while the strength of the photoionization process depends on the strength of the radiation field, which is characterized by the ionization parameter. Local stellar radiation can affect the ions both by photoionizing and by heating the gas to enhance collisional ionization. To further investigate the effects of local stellar radiation on different phases of the gas, we show the $n-T$ phase diagram of the HI, OVI, and MgII gas mass in the whole halo, inner halo, and outer halo in Figure~\ref{IonNTPhase}. The first panel of the figure gives the $n-T$ phase diagram of the HI gas mass. The left column shows the results for the NoRad galaxies, while the right column is for the StarRad galaxies. Compared to the NoRad galaxies, it can be seen that both, the area that the HI gas occupies and the total HI gas with the same density and temperature, are reduced in the StarRad galaxies. The region where $T\la10^{4}\,{\rm K}$ disappears and the amount of  HI gas mass at $T\sim10^4\,{\rm K}$ and $10^{-3}~{\rm cm^{-3}}\la n\la10^{-3}~{\rm cm^{-1}}$ also becomes smaller. This result indicates that local stellar radiation affects the HI gas through both photoionizing and heating the gas. The second and third columns show the HI gas phase diagram in the inner and outer halo. The HI gas in both the inner and outer halos is also reduced, indicating once more that the influence of local stellar radiation on the HI gas spreads throughout the whole halo.

The second panel of Figure~\ref{IonNTPhase} shows the $n-T$ phase diagram of the OVI gas mass in the NoRad (left) and StarRad (right) galaxies. The OVI gas mainly resides in the low-density, high-temperature region in both the NoRad and StarRad galaxies. The difference of the OVI gas mass in this region is negligible in both types of simulated galaxies, indicating that OVI in this region is dominated by the collisional ionization process. However, compared to the NoRad galaxies, the StarRad galaxies have the OVI gas in the region where $10^{-3}\,{\rm cm^{-3}}\la n\la 10^{-1}\,{\rm cm^{-3}}$ and $T<10^{5}\,{\rm K}$. Since the typical ionization temperature of the OVI is $10^{5-6}\,{\rm K}$ if only collisional ionization is considered, the OVI gas that appears in the low-temperature region in the StarRad galaxies should be photoionized.
This result implies that local stellar radiation can increase the OVI gas mass. However, we do not see a corresponding increase in the radial profile of the OVI column density. This is because the OVI gas is still dominated by hot gas. Since the hot gas has a large covering fraction, the column density of the OVI should be dominated by the hot gas. The high-density, low-temperature OVI gas on the other hand is distributed in both the inner and outer halos. Since high-density, low-temperature gas exists preferentially in the inner halo, OVI is also more prevalent in the inner halo. 

The third panel of Figure~\ref{IonNTPhase} shows the $n-T$ phase diagram of the MgII gas mass in the NoRad (left) and StarRad (right) galaxies. The MgII gas resides in a similar region in the phase diagram as the HI gas, but it occupies a smaller  region in the diagram than the HI gas. This is because both the MgII and HI gas trace the cold gas, but the HI gas can also trace the cold gas accreted from the IGM, which is not yet polluted with metals. Similar to the HI gas, MgII is also reduced in the StarRad galaxies compared to the NoRad galaxies. The reduction is both due to heating and photoionizing effects. The reduction of the MgII gas in the StarRad galaxies is more significant than that of the HI gas, because the MgII fraction is more sensitive to temperature than the HI gas. The second and third panels show the MgII gas phase diagram in the inner and outer halo. Here the MgII gas mass is reduced in both the inner and outer halos, similar to the HI gas.

\section{Discussion}\label{sec:discussion}

\subsection{Comparison with previous work}\label{sec:comparison}

In the following we briefly discuss three other literature studies that have investigated the ``long-range'' impact of local stellar radiation on galaxy formation and evolution in a comparable fashion as done here \citep{kannan14, obreja19, hopkins20}. In fact, \citet{kannan14} and \citet{obreja19} have used a numerical approach similar to the one we employ in this work, whereas \citet{hopkins20} used both the {\small LEBRON} method and the M1 method for radiative transfer. {\small LEBRON} is a ray-based scheme for radiative transfer, which is conceptually still quite close to the method used in this work, whereas the M1 method refers to the low-order closure used in a moment-based radiative transfer approach. We now discuss the similarities and differences between the findings of these three studies and the current work. 

\subsubsection{Comparison with \citet{kannan14}}

\citet{kannan14} studied the impact of local stellar radiation on MW-like galaxies in the MUGS suite, which is a cosmological zoom-in simulation project. \citet{kannan14} employed the same numerical approach as our current work to calculate the radiation field. However, they identified star particles younger than 10 Myr as young stars and used the SED of the stellar population with constant SFR at 10 Myr. In the current work, we identify star particles borne more recently than 100 Myr as young stars, which implies a higher radiation flux from young stars. 

In \citet{kannan14}, it was found that both the SFR and the disk gas mass of the galaxy with local stellar radiation were reduced by a factor of 2. The amount of gas with temperature $T\sim 10^{4}~{\rm K}$ was strongly suppressed. The peak of the rotation curve at the center was also strongly suppressed. In general, the effect of local stellar radiation in \citet{kannan14} is similar to but stronger than that found in the present work.

Since we identify more stars as young stars, the radiation flux from young stars should be larger than in \citet{kannan14}, and the effects should in principle also be stronger. However, our results show weaker effects than those in \citet{kannan14}. This may be due to the star formation model. In \citet{kannan14}, instead of using an effective EOS to determine the thermal state of the star-forming gas, they directly resolve the cold gas, and this gas can also be affected by local stellar radiation, potentially explaining that star formation was  more significantly affected by radiation in their model.

\subsubsection{Comparison with \citet{obreja19}}

\citet{obreja19} have implemented a similar scheme of local stellar radiation as \citet{kannan14}, but in the NIHAO suite, which is another cosmological zoom-in simulation project. Their implementation is based on \citet{kannan14} and \citet{kannan16}. In comparison to our current work, they also include radiation from the hot gas. The escape fraction in their work is set to 5\% at all frequencies, which is different from \citet{kannan14} and the present work. Such a setup reduces the total amount of ionizing photons.

In \citet{obreja19}, very similar conclusions as  in \citet{kannan14} and the current work were reached for MW-like galaxies, although the effects were weaker. Since ionizing photons are more abundant in the current work because a larger fraction of stars are considered young, and because the escape fraction is assumed to be higher at certain frequencies, the effects of local stellar radiation would have been expected to be more significant, contrary to what we have found.

\subsubsection{Comparison with \citet{hopkins20}}

\citet{hopkins20} studied the impacts of different components of radiation and different methods of radiative transfer on the evolution of galaxies with stellar masses ranging from $10^{4}$ to $10^{11}{\rm M_{\odot}}$. Unlike in our work, the radiation pressure of the continuous spectrum was also included. Their SED of stellar radiation was calculated by {\small STARBURST99}, and the SED of the star particles was evolved with stellar age. However, since their SED of young stellar radiation was calculated by {\small STARBURST99}, they lacked a certain amount of ionized photons coming from the SNRs and HXMBs. They used a ray-based {\small LEBRON} method and a momentum-based M1 method to calculate radiative transfer. The {\small LEBRON} method is similar to the method we implement in this work, but they also included local extinction calculated with the Sobolev approximation as well as radiation pressure.

We focus on a comparison with results for their m12i and m12m galaxies, which are MW-like galaxies. For the time evolution of the stellar mass of m12i and m12m, different from the result presented in this work and \citet{obreja19}, the local stellar radiation did not have an impact on the stellar mass, both for the {\small LERBON} and M1 methods. For the gas in the $n-T$ phase diagram of m12i, the total amount of gas mass lower than $10^4\,{\rm K}$ was reduced, which is qualitatively consistent with the results of \citet{obreja19} and our work. However, the degree of gas reduction in \citet{hopkins20} is nevertheless weaker than found by \citet{obreja19} and in our current work.

In \citet{hopkins20}, there may be two reasons for the quite weak impact of the stellar radiation feedback. First, since \citet{hopkins20} resolve GMCs directly, they do not need to assume an escape fraction of the radiation from the star-forming region. The real escape fraction of radiation from the star-forming region may be lower than what  we have used in the current work. Second, as mentioned above, the radiation from star-forming regions lacks some ionized photons from SNRs and HXMBs, which may make the effects of local stellar radiation become weaker.

\subsection{Other possible radiation sources}

There are various sources of extreme ultraviolet (EUV) and soft X-ray photons that could potentially impact the properties of galaxies and the CGM, as noted in recent studies \citep{US2018}. Here, we focus on two potentially significant contributors: super soft sources (SSS) and the warm-hot interstellar medium (ISM), which emits EUV and soft X-ray photons.

SSS arise from radiation emitted during accretion onto white dwarfs, while the warm-hot ISM emission stems from radiation generated by shocked ISM due to supernovae or stellar winds from OB stars. We utilize the spectral energy distributions (SEDs) presented in \citet{US2018} and compare them with those used in our present work. Given the uncertainty in these SEDs, we incorporate two different models for SSS (with temperatures of $3\times10^5$ K and $5\times10^5$ K) and consider the upper and lower bounds of radiation intensity from the warm-hot ISM as reported in \citet{US2018}.

\begin{figure}
   \includegraphics[width=0.45\textwidth]{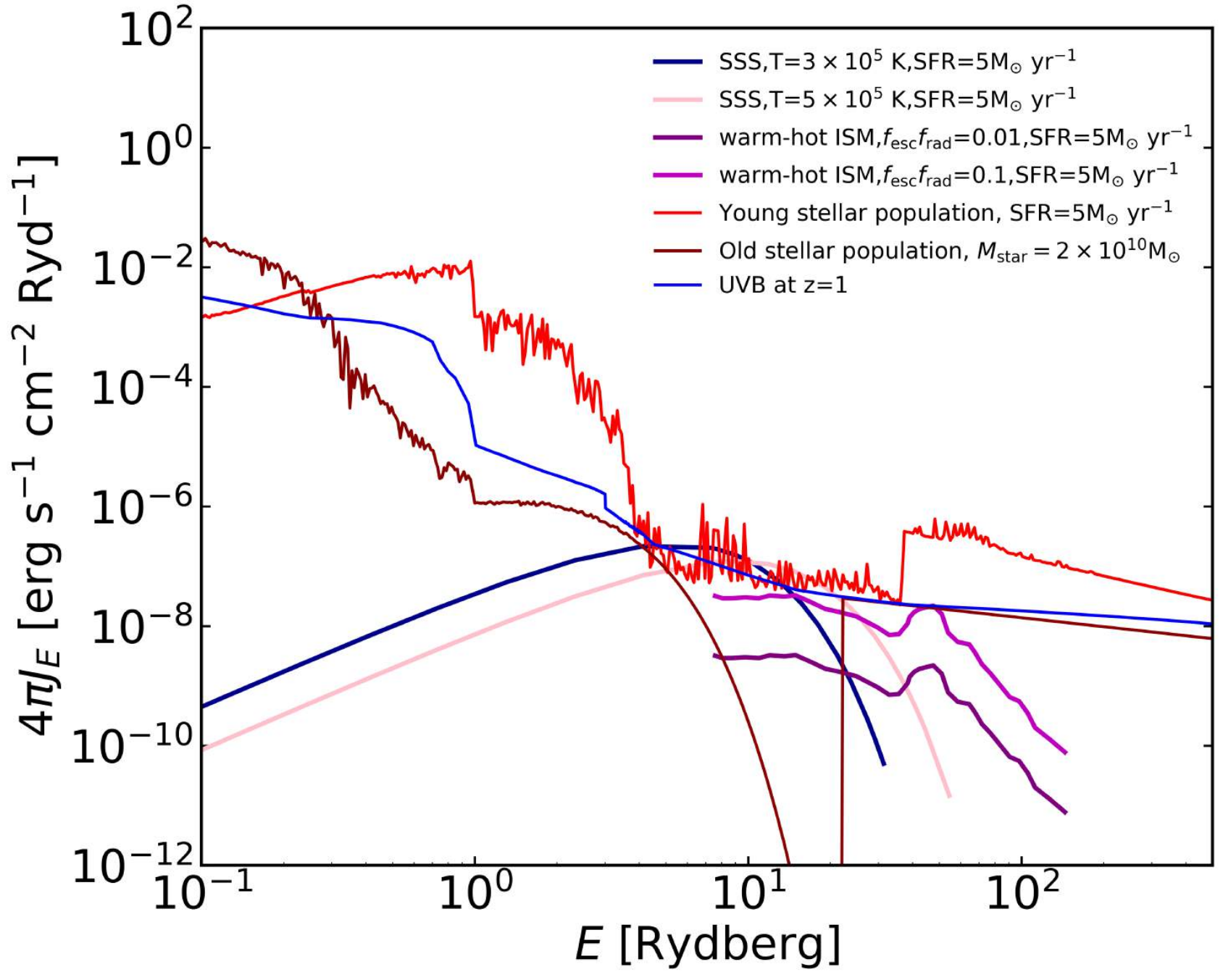}
   \caption{The intensities of SSS (two different models, $T=3\times10^5\, {\rm K}$ and $T=5\times10^5\, {\rm K}$) and warm-hot ISM ($f_{\rm esc}f_{\rm rad}=0.01$ and $f_{\rm esc}f_{\rm rad}=0.1$) are shown with a star formation rate (SFR) of 5 ${\rm M_{\odot}~yr^{-1}}$, calculated at a distance of 50 kpc from the sources, which is around $0.5\,R_{\rm vir}$ of the simulated halos at $z=1$. Intensity values are derived from \citet{US2018}. The intensities of the young stellar population with SFR=5 ${\rm M_{\odot}yr^{-1}}$ at 200 kpc, the old stellar population with ${\rm M_{\mathrm{star}}=2\times10^{10}\,\mathrm{M_{\odot}}}$ at 200 kpc, and the UVB intensity at $z=1$ are also presented, for comparison.}
    \label{sed_new}
\end{figure}

Figure~\ref{sed_new} displays the intensities of SSS and warm-hot ISM (with $f_{\rm esc}f_{\rm rad}=0.01$ and $f_{\rm esc}f_{\rm rad}=0.1$) at a distance of 200 kpc from the radiation sources, assuming a star formation rate (SFR) of 1 ${\rm M_{\odot}~yr^{-1}}$. These intensities are compared with those of the young stellar population (SFR=1 ${\rm M_{\odot}yr^{-1}}$), the old stellar population (${M_{\mathrm{star}}=10^{11}\mathrm{M_{\odot}}}$), and the ultraviolet background (UVB) at $z=0$ for context.

Regarding the intensity from the warm-hot ISM compared to the young stellar population, we find that the former is lower, even at the upper limit of the estimated radiation from the warm-hot ISM. This suggests that the warm-hot ISM does not significantly add to the radiation observed from the young stellar population.

For the SSS intensity, it is comparable to that of the young stellar population. However, since SSS radiation peaks at approximately 1 Gyr after star formation, there is minimal overlap in time with the radiation from the young stellar population, which typically dominates during the first 100 Myr after star formation. Nonetheless, SSS could potentially extend the period of the local radiation impact beyond this initial phase. Investigating the effects of SSS will thus be an important task  of our future work.

\subsection{Possible non-linear coupling of AGN feedback and local stellar radiation} \label{sec:combined}

In comparison with previous studies, our investigation also incorporates BH growth and consequently AGN feedback. The interaction between AGN feedback and local stellar radiation is complex: an increase in AGN feedback typically leads to a decrease in the star formation rate (SFR), thereby reducing local stellar radiation. Conversely, lower local stellar radiation limits the photoionization of gas, potentially suppressing BH accretion. Thus, local stellar radiation may indirectly influence star formation by regulating BH accretion and subsequent AGN feedback.

To directly assess the impact of local stellar radiation on star formation, we conducted re-simulations of the Au3 galaxy with and without local stellar radiation, while maintaining fixed BH accretion rates at specific redshifts. The BH accretion rate function we prescribed for this experiment is:
\begin{equation}
    \dot{M}_{\rm BH} =
    \begin{cases}
        10^{-1.8}\dot{M}_{\rm Edd} & \text{if } z>5 \\
        10^{-0.8-0.4(z-2.5)}\dot{M}_{\rm Edd}  & \text{if } 5>=z>2.5\\
        10^{z-3.3}\dot{M}_{\rm Edd}  & \text{if } z<2.5
    \end{cases}
\end{equation}

\begin{figure*}
   \includegraphics[width=\textwidth]{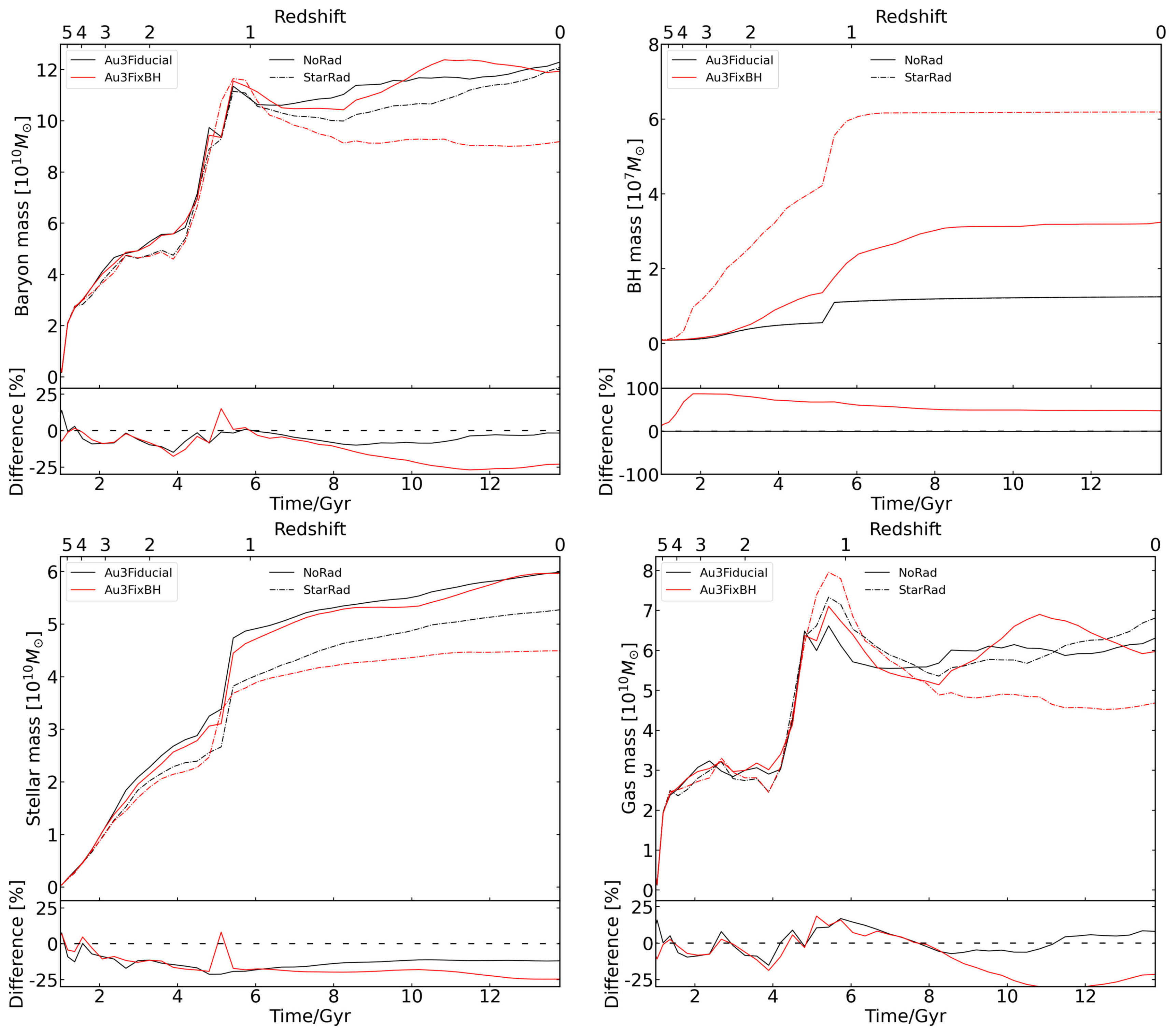}
   \caption{{\it Upper panels}: The time evolution of the total baryonic mass, the BH mass, the stellar mass, and the gas mass from $z=5.5$ to $z=0$ in our simulations of Au3 galaxy with self-regulated BH growth and fixed BH growth. The solid lines give the results from the NoRad simulations, while the dashed lines represent the results from the StarRad simulations. {\it Bottom panels}: The relative difference between the simulations with and without stellar radiation. The NoRad simulations are treated as here as fiducial reference simulations.}
    \label{JoinFixBH}
\end{figure*}

We refer to the simulations with self-regulated BH growth as Fiducial simulations, and to the ones with the fixed BH growth as FixBH simulations. Figure~\ref{JoinFixBH} compares the baryonic mass, the BH mass, the stellar mass and the gas mass evolution from $z=5.5$ to $z=0$. Inspection of the time evolution of the stellar mass in the lower-left panel shows that the stellar mass in the StarRad-FixBH simulation is $\sim$ 10\% lower than in the NoRad-FixBH simulation, implying that the local stellar radiation has a direct effect on the star formation independent of the AGN feedback.

We next consider the evolution of other baryonic components. The upper-left panel of Figure~\ref{JoinFixBH}  displays the evolution of the baryonic mass and the relative difference of the Fiducial and FixBH simulations of the Au3 galaxy. The large difference of the baryonic mass after $z\la1$ disappears. Since the galaxy experiences a merger event at $z\sim1.5$, this result indicates that this huge variation of the baryonic mass within the DM halo is due to the AGN feedback during the merger. The upper right panel shows the evolution of the BH mass. The BH mass in the Fidicial simulations is $\sim$2-4 times larger than in the FixBH simulations. However, we can see that there is no significant difference of the baryonic mass and the stellar mass in the NoRad simulation between the Fiducial and FixBH models. Since higher BH mass indicates a higher AGN output over the evolutionary history, this result indicates that the AGN feedback has only a mild affect on the star formation of this galaxy.

Interestingly, in the StarRad-Fiducial simulation the evolution of the baryonic mass and the gas mass is lower than in the other three simulations after the merger event at $z\la1$. Comparing  also the stellar mass in the StarRad-Fiducial and StarRad-FixBH simulations, we can infer that the further suppression of star formation in the StarRad-Fiducial simulation is due to the lower baryonic mass/gas mass compared to the one in the StarRad-FixBH simulation, which is due to the AGN feedback during the merger event. Overall we conclude that the AGN feedback alone hardly affects the star formation in this $\sim 10^{12}~{\rm M_{\odot}}$ halo despite the merger event, which is consistent with previous work about the quenching of the galaxies with the TNG model \citep{zinger20}. However, when taking the local stellar radiation into account, the AGN feedback during the merger event becomes amplified since the local stellar radiation can prevent the gas from forming star, making it instead available for fueling the BH. The AGN feedback will then become stronger, allowing it to reduce the gas fraction in the simulation more effectively, and ultimately suppressing the star formation more strongly.

\section{Conclusions} \label{sec:conclusions}

In this work, we have studied the impacts of local stellar radiation on MW-like galaxies taken from the Auriga project, which are cosmological zoom-in simulations performed with the moving-mesh code {\small AREPO}. Our numerical implementation for calculating the local stellar radiation field closely follows the method presented in \citet{kannan14}, where the local stellar radiation field is calculated in an optically thing approximation by using the octree method. In particular, we assume that a certain fraction of the radiation emanating from a star-forming region will be absorbed internally, but once the photons have escaped to outside the star forming region the radiation from old stars is not absorbed when it propagates. The spectrum of the local stellar radiation is obtained  using a BPASS v2.3 model, which is a stellar population synthesis code that accounts also for the evolution of binary stars. Since the simulation results exhibit stochastic fluctuations and significant system-to-system variations, we compute five Auriga galaxies and simulate them each with (`StarRad' simulations) and without stellar radiation (`NoRad' models), allowing us to obtain robust average trends for their differences. Our main findings can be summarized as follows:

\begin{itemize}

\item By comparing the differences in the time evolution of different baryon components in the NoRad and StarRad galaxies, we find that local stellar radiation does hardly affect the total baryonic mass in the five simulated systems, implying that it cannot prevent or modify the gas accretion into the halo. However, the growth of the central BH mass is enhanced, while the growth of the stellar mass is suppressed. We find that the BH mass in our StarRad galaxies is $\sim40\%$ higher than in NoRad galaxies, whereas the stellar mass is $\sim 25\%$ lower. The associated suppression of star formation occurs at all redshifts, but the enhancement of the BH mass growth happens predominantly at high redshift. In fact, at low redshift, the inclusion of local stellar radiation leads to a reduction of BH growth.

\item  The suppression of star formation by local stellar radiation happens because some gas is prevented from forming stars by being heated  slightly above $\sim 10^{4}\,{\rm K}$, which is the equilibrium temperature in the presence of the local stellar radiation field, but it is not because gas is ejected outside of the galaxy. The enhancement of the BH growth can be partially understood as a consequence of converting less gas to stars, leaving more gaseous fuel for flowing into the galaxy center at high redshift. However, at low redshift, this channel is becoming less effective, which may be in part related to the higher specific angular momentum of the corresponding gas. In addition, the heating effect of the local stellar radiation trends to increase the gas temperature and decreases the density near the BH, slowing BH accretion.
  
\item As a consequence of the modified star formation, maps of the stellar light are redder for our StarRad galaxies than for the NoRad galaxies at low redshift. By comparing the average radial profiles of the young and old stellar surface density, we find that the surface density of old stars in the StarRad galaxies is only slightly lower than in the StarRad galaxies, whereas the surface density of young stars in the StarRad galaxies is significantly lower than in NoRad galaxies. In general, the half-mass radius of the total stars in our StarRad galaxies is slightly larger than that in NoRad galaxies, while there is no obvious trend for the young stars. 

\item The surface density distribution of the total gas in both NoRad and StarRad galaxies is similar, while the surface density distribution of the HI gas in StarRad galaxies is significantly lower than the NoRad galaxies, which confirms that local stellar radiation does not significantly affect gas inflow onto the disk, but it can heat and/or ionize the cold gas in the disk.

\item By comparing the rotation curve of five NoRad galaxies and StarRad galaxies, we find that local stellar radiation can suppress the peak of the rotation curve, implying that this feedback can contribute to alleviating the overcooling problem, albeit only weakly. This influence on the rotation curve is due to the reduction of old stellar mass within 5~kpc, and originates in the suppression effect of star formation at early times.

\item The young stellar component is the dominated factor for the feedback influence of the local stellar radiation. The ionization parameter of the young stellar radiation for the cold CGM gas is $\sim 0.01-10$, whereas it is $\sim 10$ for the hot CGM gas. Local stellar radiation can significantly affect the thermal state of cold ($T<10^4\,{\rm K}$) and cool ($10^4\,{\rm K}<T<10^5~{\rm K}$) CGM gas. As a result, the cold CGM gas is significantly reduced even outside $0.3\,R_{\rm vir}$ due to the heating effect of local stellar radiation.

\item Local stellar radiation can also reduce the HI and MgII column density of CGM gas within $0.4\,R_{\rm vir}$, while it does not have a significant impact on the OVI column density in our simulations. This result arises from a combination of heating and photoionization effects due to local stellar radiation. For HI and MgII that trace the cold gas, the local stellar radiation can reduce the column density both by reducing the total amount of cold gas by heating, and by reducing the ion fraction for the gas at certain temperatures by photoionization. Since the cold gas mainly resides in the inner region, HI and MgII experience a more significant reduction in the inner region. While the OVI gas traces the hot gas, however our model of local stellar radiation is too soft to heat and photoionize the hot gas. Thus, the OVI column density does not show significant differences between the NoRad and StarRad galaxies. 

\end{itemize}

  The findings presented above demonstrate that local stellar radiation acts as a mild preventive feedback and clearly has important quantitative impact on the formation and evolution of MW-like galaxies, consistent with earlier findings. In particular, this radiative feedback needs to be considered when calculating the physical properties of CGM gas. This form of radiative feedback should therefore be included by default in future cosmological simulations of galaxy formation, something that has largely not been done thus far.

  We note, however, that the model for local radiation considered here is not complete and still calls for many improvements in future work. With respect to the radiation sources, \citet{kannan16} has shown that the radiation from the hot gas can suppress the star formation of satellites in the massive halo, while \citet{Oppenheimer18} and \citet{nelson18} show that the flicker of AGN radiation can help photoionize the OVI in CGM. \citet{US2018} also proposed many components of the local stellar radiation like SSS and warm-hot ISM that can possibly affect the properties of the ISM and CGM. These radiation sources have not been included in the current work. Also, in our numerical implementation of local radiation, the self-shielding of high-density clouds is not yet included, and the critical value of the ionization parameter we used for stopping star formation is quite ad-hoc and poorly constrained. The quantitative impacts of the local radiation field we have found here  provide a strong motivation to address these points in future work.

\section*{Acknowledgements}

The authors would like to thank the anonymous referee for detailed, insightful and constructive feedback that improved the quality of the manuscript. The authors acknowledge support by a Max Planck Partner group between the Shanghai Astronomical Observatory and the Max Planck Institute for Astrophysics (MPA). Computations were performed on the Freya compute cluster of MPA, operated by the Max Planck Computing and Data Facility.

\section*{Data Availability}

The data of this study are available from the corresponding author upon reasonable request.



\bibliographystyle{mnras}
\bibliography{ref} 

\begin{thebibliography}{}
\makeatletter
\relax
\def\mn@urlcharsother{\let\do\@makeother \do\$\do\&\do\#\do\^\do\_\do\%\do\~}
\def\mn@doi{\begingroup\mn@urlcharsother \@ifnextchar [ {\mn@doi@} {\mn@doi@[]}}
\def\mn@doi@[#1]#2{\def\@tempa{#1}\ifx\@tempa\@empty \href {http://dx.doi.org/#2} {doi:#2}\else \href {http://dx.doi.org/#2} {#1}\fi \endgroup}
\def\mn@eprint#1#2{\mn@eprint@#1:#2::\@nil}
\def\mn@eprint@arXiv#1{\href {http://arxiv.org/abs/#1} {{\tt arXiv:#1}}}
\def\mn@eprint@dblp#1{\href {http://dblp.uni-trier.de/rec/bibtex/#1.xml} {dblp:#1}}
\def\mn@eprint@#1:#2:#3:#4\@nil{\def\@tempa {#1}\def\@tempb {#2}\def\@tempc {#3}\ifx \@tempc \@empty \let \@tempc \@tempb \let \@tempb \@tempa \fi \ifx \@tempb \@empty \def\@tempb {arXiv}\fi \@ifundefined {mn@eprint@\@tempb}{\@tempb:\@tempc}{\expandafter \expandafter \csname mn@eprint@\@tempb\endcsname \expandafter{\@tempc}}}

\bibitem[\protect\citeauthoryear{{Anderson}, {Bregman}  \& {Dai}}{{Anderson} et~al.}{2013}]{anderson13}
{Anderson} M.~E.,  {Bregman} J.~N.,   {Dai} X.,  2013, \mn@doi [\apj] {10.1088/0004-637X/762/2/106}, \href {https://ui.adsabs.harvard.edu/abs/2013ApJ...762..106A} {762, 106}

\bibitem[\protect\citeauthoryear{{Babul} \& {Rees}}{{Babul} \& {Rees}}{1992}]{babul92}
{Babul} A.,  {Rees} M.~J.,  1992, \mn@doi [\mnras] {10.1093/mnras/255.2.346}, \href {https://ui.adsabs.harvard.edu/abs/1992MNRAS.255..346B} {255, 346}

\bibitem[\protect\citeauthoryear{{Balogh}, {Pearce}, {Bower}  \& {Kay}}{{Balogh} et~al.}{2001}]{balogh01}
{Balogh} M.~L.,  {Pearce} F.~R.,  {Bower} R.~G.,   {Kay} S.~T.,  2001, \mn@doi [\mnras] {10.1111/j.1365-2966.2001.04667.x}, \href {https://ui.adsabs.harvard.edu/abs/2001MNRAS.326.1228B} {326, 1228}

\bibitem[\protect\citeauthoryear{{Benson}, {Lacey}, {Baugh}, {Cole}  \& {Frenk}}{{Benson} et~al.}{2002}]{benson02}
{Benson} A.~J.,  {Lacey} C.~G.,  {Baugh} C.~M.,  {Cole} S.,   {Frenk} C.~S.,  2002, \mn@doi [\mnras] {10.1046/j.1365-8711.2002.05387.x}, \href {https://ui.adsabs.harvard.edu/abs/2002MNRAS.333..156B} {333, 156}

\bibitem[\protect\citeauthoryear{{Borrow}, {Kannan}, {Garaldi}, {Smith}, {Vogelsberger}, {Pakmor}, {Springel}  \& {Hernquist}}{{Borrow} et~al.}{2023}]{borrow23}
{Borrow} J.,  {Kannan} R.,  {Garaldi} E.,  {Smith} A.,  {Vogelsberger} M.,  {Pakmor} R.,  {Springel} V.,   {Hernquist} L.,  2023, \mn@doi [\mnras] {10.1093/mnras/stad2523}, \href {https://ui.adsabs.harvard.edu/abs/2023MNRAS.525.5932B} {525, 5932}

\bibitem[\protect\citeauthoryear{{Cantalupo}}{{Cantalupo}}{2010}]{cantalupo10}
{Cantalupo} S.,  2010, \mn@doi [\mnras] {10.1111/j.1745-3933.2010.00806.x}, \href {https://ui.adsabs.harvard.edu/abs/2010MNRAS.403L..16C} {403, L16}

\bibitem[\protect\citeauthoryear{{Cen} \& {Ostriker}}{{Cen} \& {Ostriker}}{1992}]{cen92}
{Cen} R.,  {Ostriker} J.,  1992, \mn@doi [\apj] {10.1086/171482}, \href {https://ui.adsabs.harvard.edu/abs/1992ApJ...393...22C} {393, 22}

\bibitem[\protect\citeauthoryear{{Cen} \& {Ostriker}}{{Cen} \& {Ostriker}}{1993}]{cen93}
{Cen} R.,  {Ostriker} J.~P.,  1993, \mn@doi [\apj] {10.1086/173321}, \href {https://ui.adsabs.harvard.edu/abs/1993ApJ...417..404C} {417, 404}

\bibitem[\protect\citeauthoryear{{Cervi{\~n}o}, {Mas-Hesse}  \& {Kunth}}{{Cervi{\~n}o} et~al.}{2002}]{cervino02}
{Cervi{\~n}o} M.,  {Mas-Hesse} J.~M.,   {Kunth} D.,  2002, \mn@doi [\aap] {10.1051/0004-6361:20020785}, \href {https://ui.adsabs.harvard.edu/abs/2002A&A...392...19C} {392, 19}

\bibitem[\protect\citeauthoryear{{Ceverino}, {Klypin}, {Klimek}, {Trujillo-Gomez}, {Churchill}, {Primack}  \& {Dekel}}{{Ceverino} et~al.}{2014}]{ceverino2014}
{Ceverino} D.,  {Klypin} A.,  {Klimek} E.~S.,  {Trujillo-Gomez} S.,  {Churchill} C.~W.,  {Primack} J.,   {Dekel} A.,  2014, \mn@doi [\mnras] {10.1093/mnras/stu956}, \href {https://ui.adsabs.harvard.edu/abs/2014MNRAS.442.1545C} {442, 1545}

\bibitem[\protect\citeauthoryear{{Chabrier}}{{Chabrier}}{2003}]{chabrier03}
{Chabrier} G.,  2003, \mn@doi [\pasp] {10.1086/376392}, \href {https://ui.adsabs.harvard.edu/abs/2003PASP..115..763C} {115, 763}

\bibitem[\protect\citeauthoryear{{Crain} et~al.,}{{Crain} et~al.}{2015}]{crain15}
{Crain} R.~A.,  et~al., 2015, \mn@doi [\mnras] {10.1093/mnras/stv725}, \href {https://ui.adsabs.harvard.edu/abs/2015MNRAS.450.1937C} {450, 1937}

\bibitem[\protect\citeauthoryear{{Das} et~al.,}{{Das} et~al.}{2024}]{das2024}
{Das} S.,  et~al., 2024, \mn@doi [\mnras] {10.1093/mnras/stad3892}, \href {https://ui.adsabs.harvard.edu/abs/2024MNRAS.52710358D} {527, 10358}

\bibitem[\protect\citeauthoryear{{Dav{\'e}}, {Angl{\'e}s-Alc{\'a}zar}, {Narayanan}, {Li}, {Rafieferantsoa}  \& {Appleby}}{{Dav{\'e}} et~al.}{2019}]{dave19}
{Dav{\'e}} R.,  {Angl{\'e}s-Alc{\'a}zar} D.,  {Narayanan} D.,  {Li} Q.,  {Rafieferantsoa} M.~H.,   {Appleby} S.,  2019, \mn@doi [\mnras] {10.1093/mnras/stz937}, \href {https://ui.adsabs.harvard.edu/abs/2019MNRAS.486.2827D} {486, 2827}

\bibitem[\protect\citeauthoryear{{Efstathiou}}{{Efstathiou}}{1992}]{efstathiou92}
{Efstathiou} G.,  1992, \mn@doi [\mnras] {10.1093/mnras/256.1.43P}, \href {https://ui.adsabs.harvard.edu/abs/1992MNRAS.256P..43E} {256, 43P}

\bibitem[\protect\citeauthoryear{{Eldridge}, {Stanway}, {Xiao}, {McClelland}, {Taylor}, {Ng}, {Greis}  \& {Bray}}{{Eldridge} et~al.}{2017}]{eldridge17}
{Eldridge} J.~J.,  {Stanway} E.~R.,  {Xiao} L.,  {McClelland} L.~A.~S.,  {Taylor} G.,  {Ng} M.,  {Greis} S.~M.~L.,   {Bray} J.~C.,  2017, \mn@doi [\pasa] {10.1017/pasa.2017.51}, \href {https://ui.adsabs.harvard.edu/abs/2017PASA...34...58E} {34, e058}

\bibitem[\protect\citeauthoryear{{Faucher-Gigu{\`e}re}}{{Faucher-Gigu{\`e}re}}{2020}]{faucher20}
{Faucher-Gigu{\`e}re} C.-A.,  2020, \mn@doi [\mnras] {10.1093/mnras/staa302}, \href {https://ui.adsabs.harvard.edu/abs/2020MNRAS.493.1614F} {493, 1614}

\bibitem[\protect\citeauthoryear{{Faucher-Gigu{\`e}re}, {Lidz}, {Zaldarriaga}  \& {Hernquist}}{{Faucher-Gigu{\`e}re} et~al.}{2009}]{faucher09}
{Faucher-Gigu{\`e}re} C.-A.,  {Lidz} A.,  {Zaldarriaga} M.,   {Hernquist} L.,  2009, \mn@doi [\apj] {10.1088/0004-637X/703/2/1416}, \href {https://ui.adsabs.harvard.edu/abs/2009ApJ...703.1416F} {703, 1416}

\bibitem[\protect\citeauthoryear{{Ferland}, {Korista}, {Verner}, {Ferguson}, {Kingdon}  \& {Verner}}{{Ferland} et~al.}{1998}]{ferland98}
{Ferland} G.~J.,  {Korista} K.~T.,  {Verner} D.~A.,  {Ferguson} J.~W.,  {Kingdon} J.~B.,   {Verner} E.~M.,  1998, \mn@doi [\pasp] {10.1086/316190}, \href {https://ui.adsabs.harvard.edu/abs/1998PASP..110..761F} {110, 761}

\bibitem[\protect\citeauthoryear{{Fielding} et~al.,}{{Fielding} et~al.}{2020}]{fielding2020}
{Fielding} D.~B.,  et~al., 2020, \mn@doi [\apj] {10.3847/1538-4357/abbc6d}, \href {https://ui.adsabs.harvard.edu/abs/2020ApJ...903...32F} {903, 32}

\bibitem[\protect\citeauthoryear{{Finlator}, {Dav{\'e}}  \& {{\"O}zel}}{{Finlator} et~al.}{2011}]{finlator11}
{Finlator} K.,  {Dav{\'e}} R.,   {{\"O}zel} F.,  2011, \mn@doi [\apj] {10.1088/0004-637X/743/2/169}, \href {https://ui.adsabs.harvard.edu/abs/2011ApJ...743..169F} {743, 169}

\bibitem[\protect\citeauthoryear{{Governato} et~al.,}{{Governato} et~al.}{2004}]{governato2004}
{Governato} F.,  et~al., 2004, \mn@doi [\apj] {10.1086/383516}, \href {https://ui.adsabs.harvard.edu/abs/2004ApJ...607..688G} {607, 688}

\bibitem[\protect\citeauthoryear{{Grand} et~al.,}{{Grand} et~al.}{2017}]{grand17}
{Grand} R. J.~J.,  et~al., 2017, \mn@doi [\mnras] {10.1093/mnras/stx071}, \href {https://ui.adsabs.harvard.edu/abs/2017MNRAS.467..179G} {467, 179}

\bibitem[\protect\citeauthoryear{{Grand}, {Fragkoudi}, {G{\'o}mez}, {Jenkins}, {Marinacci}, {Pakmor}  \& {Springel}}{{Grand} et~al.}{2024}]{Grand2024}
{Grand} R. J.~J.,  {Fragkoudi} F.,  {G{\'o}mez} F.~A.,  {Jenkins} A.,  {Marinacci} F.,  {Pakmor} R.,   {Springel} V.,  2024, \mn@doi [arXiv e-prints] {10.48550/arXiv.2401.08750}, \href {https://ui.adsabs.harvard.edu/abs/2024arXiv240108750G} {p. arXiv:2401.08750}

\bibitem[\protect\citeauthoryear{{Haardt} \& {Madau}}{{Haardt} \& {Madau}}{1996}]{haardt96}
{Haardt} F.,  {Madau} P.,  1996, \mn@doi [\apj] {10.1086/177035}, \href {https://ui.adsabs.harvard.edu/abs/1996ApJ...461...20H} {461, 20}

\bibitem[\protect\citeauthoryear{{Haardt} \& {Madau}}{{Haardt} \& {Madau}}{2012}]{haardt12}
{Haardt} F.,  {Madau} P.,  2012, \mn@doi [\apj] {10.1088/0004-637X/746/2/125}, \href {https://ui.adsabs.harvard.edu/abs/2012ApJ...746..125H} {746, 125}

\bibitem[\protect\citeauthoryear{{Haiman}}{{Haiman}}{2016}]{haiman16}
{Haiman} Z.,  2016, in {Mesinger} A.,  ed.,  Astrophysics and Space Science Library Vol. 423, Understanding the Epoch of Cosmic Reionization: Challenges and Progress. p.~1 (\mn@eprint {arXiv} {1511.01125})

\bibitem[\protect\citeauthoryear{{Haiman}, {Thoul}  \& {Loeb}}{{Haiman} et~al.}{1996}]{haiman96}
{Haiman} Z.,  {Thoul} A.~A.,   {Loeb} A.,  1996, \mn@doi [\apj] {10.1086/177343}, \href {https://ui.adsabs.harvard.edu/abs/1996ApJ...464..523H} {464, 523}

\bibitem[\protect\citeauthoryear{{Hopkins}, {Quataert}  \& {Murray}}{{Hopkins} et~al.}{2012}]{hopkins12}
{Hopkins} P.~F.,  {Quataert} E.,   {Murray} N.,  2012, \mn@doi [\mnras] {10.1111/j.1365-2966.2012.20593.x}, \href {https://ui.adsabs.harvard.edu/abs/2012MNRAS.421.3522H} {421, 3522}

\bibitem[\protect\citeauthoryear{{Hopkins} et~al.,}{{Hopkins} et~al.}{2018}]{hopkins18}
{Hopkins} P.~F.,  et~al., 2018, \mn@doi [\mnras] {10.1093/mnras/sty1690}, \href {https://ui.adsabs.harvard.edu/abs/2018MNRAS.480..800H} {480, 800}

\bibitem[\protect\citeauthoryear{{Hopkins}, {Grudi{\'c}}, {Wetzel}, {Kere{\v{s}}}, {Faucher-Gigu{\`e}re}, {Ma}, {Murray}  \& {Butcher}}{{Hopkins} et~al.}{2020}]{hopkins20}
{Hopkins} P.~F.,  {Grudi{\'c}} M.~Y.,  {Wetzel} A.,  {Kere{\v{s}}} D.,  {Faucher-Gigu{\`e}re} C.-A.,  {Ma} X.,  {Murray} N.,   {Butcher} N.,  2020, \mn@doi [\mnras] {10.1093/mnras/stz3129}, \href {https://ui.adsabs.harvard.edu/abs/2020MNRAS.491.3702H} {491, 3702}

\bibitem[\protect\citeauthoryear{{Hummels} et~al.,}{{Hummels} et~al.}{2019}]{hummels19}
{Hummels} C.~B.,  et~al., 2019, \mn@doi [\apj] {10.3847/1538-4357/ab378f}, \href {https://ui.adsabs.harvard.edu/abs/2019ApJ...882..156H} {882, 156}

\bibitem[\protect\citeauthoryear{{Ji} et~al.,}{{Ji} et~al.}{2020}]{ji20}
{Ji} S.,  et~al., 2020, \mn@doi [\mnras] {10.1093/mnras/staa1849}, \href {https://ui.adsabs.harvard.edu/abs/2020MNRAS.496.4221J} {496, 4221}

\bibitem[\protect\citeauthoryear{{Johnson}, {Chen}  \& {Mulchaey}}{{Johnson} et~al.}{2015}]{Johnson2015}
{Johnson} S.~D.,  {Chen} H.-W.,   {Mulchaey} J.~S.,  2015, \mn@doi [\mnras] {10.1093/mnras/stv553}, \href {https://ui.adsabs.harvard.edu/abs/2015MNRAS.449.3263J} {449, 3263}

\bibitem[\protect\citeauthoryear{{Kannan} et~al.,}{{Kannan} et~al.}{2014}]{kannan14}
{Kannan} R.,  et~al., 2014, \mn@doi [\mnras] {10.1093/mnras/stt2098}, \href {https://ui.adsabs.harvard.edu/abs/2014MNRAS.437.2882K} {437, 2882}

\bibitem[\protect\citeauthoryear{{Kannan}, {Vogelsberger}, {Stinson}, {Hennawi}, {Marinacci}, {Springel}  \& {Macci{\`o}}}{{Kannan} et~al.}{2016}]{kannan16}
{Kannan} R.,  {Vogelsberger} M.,  {Stinson} G.~S.,  {Hennawi} J.~F.,  {Marinacci} F.,  {Springel} V.,   {Macci{\`o}} A.~V.,  2016, \mn@doi [\mnras] {10.1093/mnras/stw463}, \href {https://ui.adsabs.harvard.edu/abs/2016MNRAS.458.2516K} {458, 2516}

\bibitem[\protect\citeauthoryear{{Katz}, {Weinberg}  \& {Hernquist}}{{Katz} et~al.}{1996}]{katz96}
{Katz} N.,  {Weinberg} D.~H.,   {Hernquist} L.,  1996, \mn@doi [\apjs] {10.1086/192305}, \href {https://ui.adsabs.harvard.edu/abs/1996ApJS..105...19K} {105, 19}

\bibitem[\protect\citeauthoryear{{Loeb} \& {Barkana}}{{Loeb} \& {Barkana}}{2001}]{loeb01}
{Loeb} A.,  {Barkana} R.,  2001, \mn@doi [\araa] {10.1146/annurev.astro.39.1.19}, \href {https://ui.adsabs.harvard.edu/abs/2001ARA&A..39...19L} {39, 19}

\bibitem[\protect\citeauthoryear{{Navarro} \& {Benz}}{{Navarro} \& {Benz}}{1991}]{navarro91}
{Navarro} J.~F.,  {Benz} W.,  1991, \mn@doi [\apj] {10.1086/170590}, \href {https://ui.adsabs.harvard.edu/abs/1991ApJ...380..320N} {380, 320}

\bibitem[\protect\citeauthoryear{{Nelson} et~al.,}{{Nelson} et~al.}{2018}]{nelson18}
{Nelson} D.,  et~al., 2018, \mn@doi [\mnras] {10.1093/mnras/sty656}, \href {https://ui.adsabs.harvard.edu/abs/2018MNRAS.477..450N} {477, 450}

\bibitem[\protect\citeauthoryear{{Nelson} et~al.,}{{Nelson} et~al.}{2020}]{nelson20}
{Nelson} D.,  et~al., 2020, \mn@doi [\mnras] {10.1093/mnras/staa2419}, \href {https://ui.adsabs.harvard.edu/abs/2020MNRAS.498.2391N} {498, 2391}

\bibitem[\protect\citeauthoryear{{Obreja}, {Macci{\`o}}, {Moster}, {Udrescu}, {Buck}, {Kannan}, {Dutton}  \& {Blank}}{{Obreja} et~al.}{2019}]{obreja19}
{Obreja} A.,  {Macci{\`o}} A.~V.,  {Moster} B.,  {Udrescu} S.~M.,  {Buck} T.,  {Kannan} R.,  {Dutton} A.~A.,   {Blank} M.,  2019, \mn@doi [\mnras] {10.1093/mnras/stz2639}, \href {https://ui.adsabs.harvard.edu/abs/2019MNRAS.490.1518O} {490, 1518}

\bibitem[\protect\citeauthoryear{{Okamoto}, {Gao}  \& {Theuns}}{{Okamoto} et~al.}{2008}]{okamoto08}
{Okamoto} T.,  {Gao} L.,   {Theuns} T.,  2008, \mn@doi [\mnras] {10.1111/j.1365-2966.2008.13830.x}, \href {https://ui.adsabs.harvard.edu/abs/2008MNRAS.390..920O} {390, 920}

\bibitem[\protect\citeauthoryear{{Oppenheimer}, {Segers}, {Schaye}, {Richings}  \& {Crain}}{{Oppenheimer} et~al.}{2018}]{Oppenheimer18}
{Oppenheimer} B.~D.,  {Segers} M.,  {Schaye} J.,  {Richings} A.~J.,   {Crain} R.~A.,  2018, \mn@doi [\mnras] {10.1093/mnras/stx2967}, \href {https://ui.adsabs.harvard.edu/abs/2018MNRAS.474.4740O} {474, 4740}

\bibitem[\protect\citeauthoryear{{Peeples} et~al.,}{{Peeples} et~al.}{2019}]{peeples19}
{Peeples} M.~S.,  et~al., 2019, \mn@doi [\apj] {10.3847/1538-4357/ab0654}, \href {https://ui.adsabs.harvard.edu/abs/2019ApJ...873..129P} {873, 129}

\bibitem[\protect\citeauthoryear{{Pillepich} et~al.,}{{Pillepich} et~al.}{2018}]{pillepich18}
{Pillepich} A.,  et~al., 2018, \mn@doi [\mnras] {10.1093/mnras/stx2656}, \href {https://ui.adsabs.harvard.edu/abs/2018MNRAS.473.4077P} {473, 4077}

\bibitem[\protect\citeauthoryear{{Planck Collaboration} et~al.,}{{Planck Collaboration} et~al.}{2014}]{Planck2014}
{Planck Collaboration} et~al., 2014, \mn@doi [\aap] {10.1051/0004-6361/201321591}, \href {https://ui.adsabs.harvard.edu/abs/2014A&A...571A..16P} {571, A16}

\bibitem[\protect\citeauthoryear{{Qu} \& {Bregman}}{{Qu} \& {Bregman}}{2018}]{qu18}
{Qu} Z.,  {Bregman} J.~N.,  2018, \mn@doi [\apj] {10.3847/1538-4357/aaafd4}, \href {https://ui.adsabs.harvard.edu/abs/2018ApJ...856....5Q} {856, 5}

\bibitem[\protect\citeauthoryear{{Quinn}, {Katz}  \& {Efstathiou}}{{Quinn} et~al.}{1996}]{quinn96}
{Quinn} T.,  {Katz} N.,   {Efstathiou} G.,  1996, \mn@doi [\mnras] {10.1093/mnras/278.4.L49}, \href {https://ui.adsabs.harvard.edu/abs/1996MNRAS.278L..49Q} {278, L49}

\bibitem[\protect\citeauthoryear{{Raymond} \& {Smith}}{{Raymond} \& {Smith}}{1977}]{raymond1977}
{Raymond} J.~C.,  {Smith} B.~W.,  1977, \mn@doi [\apjs] {10.1086/190486}, \href {https://ui.adsabs.harvard.edu/abs/1977ApJS...35..419R} {35, 419}

\bibitem[\protect\citeauthoryear{{Rees} \& {Ostriker}}{{Rees} \& {Ostriker}}{1977}]{Rees77}
{Rees} M.~J.,  {Ostriker} J.~P.,  1977, \mn@doi [MNRAS] {10.1093/mnras/179.4.541}, \href {https://ui.adsabs.harvard.edu/abs/1977MNRAS.179..541R} {179, 541}

\bibitem[\protect\citeauthoryear{{Ro{\v{s}}kar}, {Teyssier}, {Agertz}, {Wetzstein}  \& {Moore}}{{Ro{\v{s}}kar} et~al.}{2014}]{Roskar14}
{Ro{\v{s}}kar} R.,  {Teyssier} R.,  {Agertz} O.,  {Wetzstein} M.,   {Moore} B.,  2014, \mn@doi [\mnras] {10.1093/mnras/stu1548}, \href {https://ui.adsabs.harvard.edu/abs/2014MNRAS.444.2837R} {444, 2837}

\bibitem[\protect\citeauthoryear{{Scannapieco} et~al.,}{{Scannapieco} et~al.}{2012}]{scannapieco12}
{Scannapieco} C.,  et~al., 2012, \mn@doi [\mnras] {10.1111/j.1365-2966.2012.20993.x}, \href {https://ui.adsabs.harvard.edu/abs/2012MNRAS.423.1726S} {423, 1726}

\bibitem[\protect\citeauthoryear{{Shapiro} \& {Giroux}}{{Shapiro} \& {Giroux}}{1987}]{shapiro87}
{Shapiro} P.~R.,  {Giroux} M.~L.,  1987, \mn@doi [\apjl] {10.1086/185015}, \href {https://ui.adsabs.harvard.edu/abs/1987ApJ...321L.107S} {321, L107}

\bibitem[\protect\citeauthoryear{{Silk}}{{Silk}}{1977}]{silk77}
{Silk} J.,  1977, \mn@doi [\apj] {10.1086/154972}, \href {https://ui.adsabs.harvard.edu/abs/1977ApJ...211..638S} {211, 638}

\bibitem[\protect\citeauthoryear{{Springel}}{{Springel}}{2010}]{springel10}
{Springel} V.,  2010, \mn@doi [\mnras] {10.1111/j.1365-2966.2009.15715.x}, \href {https://ui.adsabs.harvard.edu/abs/2010MNRAS.401..791S} {401, 791}

\bibitem[\protect\citeauthoryear{{Springel} \& {Hernquist}}{{Springel} \& {Hernquist}}{2003}]{springel03}
{Springel} V.,  {Hernquist} L.,  2003, \mn@doi [\mnras] {10.1046/j.1365-8711.2003.06206.x}, \href {https://ui.adsabs.harvard.edu/abs/2003MNRAS.339..289S} {339, 289}

\bibitem[\protect\citeauthoryear{{Stern}, {Hennawi}, {Prochaska}  \& {Werk}}{{Stern} et~al.}{2016}]{stern16}
{Stern} J.,  {Hennawi} J.~F.,  {Prochaska} J.~X.,   {Werk} J.~K.,  2016, \mn@doi [\apj] {10.3847/0004-637X/830/2/87}, \href {https://ui.adsabs.harvard.edu/abs/2016ApJ...830...87S} {830, 87}

\bibitem[\protect\citeauthoryear{{Stinson}, {Bailin}, {Couchman}, {Wadsley}, {Shen}, {Nickerson}, {Brook}  \& {Quinn}}{{Stinson} et~al.}{2010}]{stinson10}
{Stinson} G.~S.,  {Bailin} J.,  {Couchman} H.,  {Wadsley} J.,  {Shen} S.,  {Nickerson} S.,  {Brook} C.,   {Quinn} T.,  2010, \mn@doi [\mnras] {10.1111/j.1365-2966.2010.17187.x}, \href {https://ui.adsabs.harvard.edu/abs/2010MNRAS.408..812S} {408, 812}

\bibitem[\protect\citeauthoryear{{Strawn} et~al.,}{{Strawn} et~al.}{2021}]{strawn21}
{Strawn} C.,  et~al., 2021, \mn@doi [\mnras] {10.1093/mnras/staa3972}, \href {https://ui.adsabs.harvard.edu/abs/2021MNRAS.501.4948S} {501, 4948}

\bibitem[\protect\citeauthoryear{{Suresh}, {Rubin}, {Kannan}, {Werk}, {Hernquist}  \& {Vogelsberger}}{{Suresh} et~al.}{2017}]{Suresh17}
{Suresh} J.,  {Rubin} K. H.~R.,  {Kannan} R.,  {Werk} J.~K.,  {Hernquist} L.,   {Vogelsberger} M.,  2017, \mn@doi [\mnras] {10.1093/mnras/stw2499}, \href {https://ui.adsabs.harvard.edu/abs/2017MNRAS.465.2966S} {465, 2966}

\bibitem[\protect\citeauthoryear{{Thoul} \& {Weinberg}}{{Thoul} \& {Weinberg}}{1996}]{thoul96}
{Thoul} A.~A.,  {Weinberg} D.~H.,  1996, \mn@doi [\apj] {10.1086/177446}, \href {https://ui.adsabs.harvard.edu/abs/1996ApJ...465..608T} {465, 608}

\bibitem[\protect\citeauthoryear{{Tumlinson} et~al.,}{{Tumlinson} et~al.}{2011}]{tumlinson11}
{Tumlinson} J.,  et~al., 2011, \mn@doi [Science] {10.1126/science.1209840}, \href {https://ui.adsabs.harvard.edu/abs/2011Sci...334..948T} {334, 948}

\bibitem[\protect\citeauthoryear{{Upton Sanderbeck}, {McQuinn}, {D'Aloisio}  \& {Werk}}{{Upton Sanderbeck} et~al.}{2018}]{US2018}
{Upton Sanderbeck} P.~R.,  {McQuinn} M.,  {D'Aloisio} A.,   {Werk} J.~K.,  2018, \mn@doi [\apj] {10.3847/1538-4357/aaeff2}, \href {https://ui.adsabs.harvard.edu/abs/2018ApJ...869..159U} {869, 159}

\bibitem[\protect\citeauthoryear{{Vogelsberger} et~al.,}{{Vogelsberger} et~al.}{2014}]{vogelsberger14}
{Vogelsberger} M.,  et~al., 2014, \mn@doi [\mnras] {10.1093/mnras/stu1536}, \href {https://ui.adsabs.harvard.edu/abs/2014MNRAS.444.1518V} {444, 1518}

\bibitem[\protect\citeauthoryear{{Wadsley}, {Stadel}  \& {Quinn}}{{Wadsley} et~al.}{2004}]{wadsley04}
{Wadsley} J.~W.,  {Stadel} J.,   {Quinn} T.,  2004, \mn@doi [\na] {10.1016/j.newast.2003.08.004}, \href {https://ui.adsabs.harvard.edu/abs/2004NewA....9..137W} {9, 137}

\bibitem[\protect\citeauthoryear{{Wang}, {Mao}, {Xiang}  \& {Yuan}}{{Wang} et~al.}{2009}]{wang09}
{Wang} L.,  {Mao} J.,  {Xiang} S.,   {Yuan} Y.~F.,  2009, \mn@doi [\aap] {10.1051/0004-6361/200809916}, \href {https://ui.adsabs.harvard.edu/abs/2009A&A...494..817W} {494, 817}

\bibitem[\protect\citeauthoryear{{Wang}, {Dutton}, {Stinson}, {Macci{\`o}}, {Penzo}, {Kang}, {Keller}  \& {Wadsley}}{{Wang} et~al.}{2015}]{wang15}
{Wang} L.,  {Dutton} A.~A.,  {Stinson} G.~S.,  {Macci{\`o}} A.~V.,  {Penzo} C.,  {Kang} X.,  {Keller} B.~W.,   {Wadsley} J.,  2015, \mn@doi [\mnras] {10.1093/mnras/stv1937}, \href {https://ui.adsabs.harvard.edu/abs/2015MNRAS.454...83W} {454, 83}

\bibitem[\protect\citeauthoryear{{Weinberg}, {Hernquist}  \& {Katz}}{{Weinberg} et~al.}{1997}]{weinberg97}
{Weinberg} D.~H.,  {Hernquist} L.,   {Katz} N.,  1997, \mn@doi [\apj] {10.1086/303683}, \href {https://ui.adsabs.harvard.edu/abs/1997ApJ...477....8W} {477, 8}

\bibitem[\protect\citeauthoryear{{Weinberger} et~al.,}{{Weinberger} et~al.}{2018}]{weinberger17}
{Weinberger} R.,  et~al., 2018, \mn@doi [\mnras] {10.1093/mnras/sty1733}, \href {https://ui.adsabs.harvard.edu/abs/2018MNRAS.479.4056W} {479, 4056}

\bibitem[\protect\citeauthoryear{{Werk}, {Prochaska}, {Thom}, {Tumlinson}, {Tripp}, {O'Meara}  \& {Peeples}}{{Werk} et~al.}{2013}]{Werk2013}
{Werk} J.~K.,  {Prochaska} J.~X.,  {Thom} C.,  {Tumlinson} J.,  {Tripp} T.~M.,  {O'Meara} J.~M.,   {Peeples} M.~S.,  2013, \mn@doi [\apjs] {10.1088/0067-0049/204/2/17}, \href {https://ui.adsabs.harvard.edu/abs/2013ApJS..204...17W} {204, 17}

\bibitem[\protect\citeauthoryear{{Werk} et~al.,}{{Werk} et~al.}{2014}]{werk14}
{Werk} J.~K.,  et~al., 2014, \mn@doi [\apj] {10.1088/0004-637X/792/1/8}, \href {https://ui.adsabs.harvard.edu/abs/2014ApJ...792....8W} {792, 8}

\bibitem[\protect\citeauthoryear{{White} \& {Frenk}}{{White} \& {Frenk}}{1991}]{white91}
{White} S. D.~M.,  {Frenk} C.~S.,  1991, \mn@doi [\apj] {10.1086/170483}, \href {https://ui.adsabs.harvard.edu/abs/1991ApJ...379...52W} {379, 52}

\bibitem[\protect\citeauthoryear{{White} \& {Rees}}{{White} \& {Rees}}{1978}]{white78}
{White} S.~D.~M.,  {Rees} M.~J.,  1978, \mn@doi [MNRAS] {10.1093/mnras/183.3.341}, \href {https://ui.adsabs.harvard.edu/abs/1978MNRAS.183..341W} {183, 341}

\bibitem[\protect\citeauthoryear{{Wiersma}, {Schaye}  \& {Smith}}{{Wiersma} et~al.}{2009}]{wiersma09}
{Wiersma} R. P.~C.,  {Schaye} J.,   {Smith} B.~D.,  2009, \mn@doi [\mnras] {10.1111/j.1365-2966.2008.14191.x}, \href {https://ui.adsabs.harvard.edu/abs/2009MNRAS.393...99W} {393, 99}

\bibitem[\protect\citeauthoryear{{Woods}}{{Woods}}{2015}]{woods15}
{Woods} R.,  2015, PhD thesis, McMaster University, Canada

\bibitem[\protect\citeauthoryear{{Zinger} et~al.,}{{Zinger} et~al.}{2020}]{zinger20}
{Zinger} E.,  et~al., 2020, \mn@doi [\mnras] {10.1093/mnras/staa2607}, \href {https://ui.adsabs.harvard.edu/abs/2020MNRAS.499..768Z} {499, 768}

\bibitem[\protect\citeauthoryear{{van de Voort}, {Springel}, {Mandelker}, {van den Bosch}  \& {Pakmor}}{{van de Voort} et~al.}{2019}]{vdv19}
{van de Voort} F.,  {Springel} V.,  {Mandelker} N.,  {van den Bosch} F.~C.,   {Pakmor} R.,  2019, \mn@doi [\mnras] {10.1093/mnrasl/sly190}, \href {https://ui.adsabs.harvard.edu/abs/2019MNRAS.482L..85V} {482, L85}

\makeatother
\end{thebibliography}







\bsp	
\label{lastpage}
\end{document}